\pdfoutput=1
\documentclass[a4paper,12pt]{article}
\usepackage{natbib,epsfig, graphics, lscape, longtable,
rotating, geometry, threeparttable,amsmath, multirow,tabularx,setspace,amssymb}

\usepackage{enumerate}

\usepackage{amsmath,mathtools,amsthm, float}

\usepackage{titlesec}

\usepackage{appendix}
\usepackage[bottom]{footmisc}

\usepackage{graphicx}
\usepackage[centerlast]{caption}
\usepackage{subcaption}
\usepackage{placeins}

\usepackage{booktabs}

\usepackage{pdflscape}
\usepackage{fancyhdr}
\usepackage{adjustbox}
\usepackage[absolute,overlay]{textpos}

\usepackage{hyperref}
\hypersetup{
    colorlinks=true,
    linkcolor=black,
    citecolor=black,
    filecolor=black,
    urlcolor=black
}

\usepackage{tikz}
\usetikzlibrary{calc,arrows,automata,shapes.misc,shapes.arrows,chains,matrix,positioning,scopes,decorations.pathmorphing,shadows,shapes,snakes,fadings}
\usetikzlibrary{chains, positioning, shapes.symbols}
\tikzset{start/.style = {signal, fill=#1,minimum height=12mm,draw=white,
                 font=\footnotesize,text=black,inner sep=4pt,
                 signal pointer angle=150, on chain},
  cont/.style = {start=#1, signal from=west},
  end/.style = {start=#1, signal from=west, signal pointer angle=150}}
\tikzset{
    invisible/.style={opacity=.5},
    visible on/.style={alt={#1{}{invisible}}},
    highlight on/.style={alt={#1{fill=title.fg!25!white}{fill=gray!20!white}}},
    alt/.code args={<#1>#2#3}{%
      \alt<#1>{\pgfkeysalso{#2}}{\pgfkeysalso{#3}}
    },
  }

\newcommand{\sym}[1]{\rlap{$^{#1}$}}

\newtheorem{proposition}{Proposition}

\newtheorem{result}{Result}

\newtheorem{hypothesis}{Hypothesis}

\renewcommand{\footnotesize}{\small}

\renewcommand{\thesection}{\Roman{section}}
\renewcommand{\thesubsection}{\thesection.\Alph{subsection}}

\usepackage{dsfont}
\usepackage{threeparttable}
\newenvironment{mytablenotes}
  {\begin{tablenotes}[flushleft]\vspace{3pt}\setlength\labelsep{0pt}
\setstretch{.95}\begin{footnotesize}}
  {\end{footnotesize}\end{tablenotes}}

\usepackage{changepage}

\usepackage{etoolbox}
\AtBeginEnvironment{proof}{\setstretch{1.5}}

\begin{document}

\title{\vspace{-1cm} Choosing Complexity: \\ Cognitive Ability, Overconfidence, and Task Choice
\vspace{0.5cm}\thanks{We are grateful for helpful comments from Victoria Prowse, seminar participants, and in private conversations. We thank the Vernon Smith Experimental Economics Laboratory at Purdue University for hosting our experiment and the Daniels School of Business for providing internal funding.}
}

\date{}

\author{
\large{David Gill \thanks{Purdue University, Department of Economics; dgill.econ@gmail.com}}\\
\large{Zhongheng Qiao \thanks{Purdue University, Department of Economics; qiao50@purdue.edu}}\\
}

\maketitle

\thispagestyle{empty}

\vspace{-0.5cm}
\begin{center}
This version: November 9, 2026 \\

\end{center}

\vspace{0.1cm}

\begin{abstract}
\begin{adjustwidth}{-2mm}{-2mm}
\setstretch{1.3}
\noindent

Workers and organizations routinely choose between complex tasks that differ in their success-payment tradeoff. We study how cognitive ability and overconfidence shape this choice of task complexity, both theoretically and experimentally. Consistent with our model, we find that more cognitively able individuals choose more complex tasks. Conditional on cognitive ability, we find that more overconfident individuals also choose more complex tasks. In about 75\% of the decisions, subjects pick their payoff-maximizing task, and higher cognitive ability is weakly associated with more efficient choices of task complexity. But errors are strikingly one-sided: those who fail to choose optimally pick tasks that are too complex. Our findings suggest that organizations should pair task menus and incentives for choosing more complex tasks with information and feedback that help workers assess their ability and probability of success. This preserves the benefits of self-selection while reducing costly overreach.

\vspace{1cm}

\textbf{Keywords}: \textit{Cognitive Ability, Complexity, Task Choice, Choice Efficiency, Overconfidence, Experiment}\\

\textbf{JEL Classification}: \textit{C91, D03, D83, D91, J24}
\end{adjustwidth}
\end{abstract}

\newpage

\newpage
\pagenumbering{arabic}

\section{Introduction}\label{Sect:Intro}\

Workers and organizations routinely choose how complex a task to undertake, and the choice has real consequences for pay and productivity. Simpler tasks are more likely to succeed, while more complex tasks offer higher rewards when they do succeed. A manager may choose between a simple and a complex pricing strategy. The complex strategy delivers higher profits when executed well, but it has more moving parts and is easier to misapply. A lawyer may choose between routine small cases and high-stakes cases. High-stakes cases are harder to win, but they bring larger rewards when won.

Cognitive ability is a powerful predictor of labor-market outcomes and of the quality of economic decisions \citep{burks2009cognitive, fe2022cognitive, deming2025skills}. Field evidence links managerial ability to performance in complex environments \citep{goldfarb2011thinks, hortaccsu2019does, han2024minds}, and lab evidence shows that cognitive ability affects strategic sophistication \citep{carpenter2013cognitive, gill2016cognitive, gill2025cognitive}. Yet when workers can choose how complex a task to undertake, we know little about who selects into more complex tasks, whether those choices are well calibrated to their cognitive ability, and how beliefs about cognitive ability shape choices of task complexity.

In this paper, we provide theoretical and experimental evidence on how cognitive ability and overconfidence shape choice between complex tasks that differ in their success-payment tradeoff. We address three research questions. First, do more cognitively able people choose more complex tasks? Second, conditional on cognitive ability, do more overconfident people choose more complex tasks? Third, do more cognitively able people choose their optimal task more often? Briefly, our contributions are three-fold. First, we extend the literature on choice of complexity by showing that cognitive ability and overconfidence predict choice of more complex tasks. Second, we find some weak evidence that cognitive ability predicts efficiency of complexity choices, complementing prior work on cognition and decision quality. Third, we document a sharp asymmetry: subjects who err tend to overreach by choosing tasks that are too complex.

We develop a model in which a decision-maker chooses how complex a task to attempt. The probability of making a mistake depends on her cognitive ability and on the complexity of the chosen task. Higher cognitive ability lowers the mistake probability at every level of complexity, and higher complexity raises the mistake probability at every level of cognitive ability. The effect of complexity on the mistake probability can itself depend on cognitive ability, with greater cognitive ability weakening the impact of complexity. The success payment (that is, the reward for success) rises with the complexity of the chosen task, while the failure payment is constant. The decision-maker therefore trades off a higher success payment on a more complex task against a higher mistake probability on that task. The model directly yields our first hypothesis: more cognitively able individuals choose more complex tasks.

We then augment the model with two behavioral assumptions motivated by prior empirical work, and each assumption delivers a further hypothesis. First, we assume that the decision-maker's belief about her own cognitive ability helps to shape her choice. This assumption yields our second hypothesis: more overconfident individuals choose more complex tasks. Hypothesis 2 is motivated by prior evidence, specifically prior work that links overconfidence and decision making. For example, \citet{camerer1999overconfidence} and \citet{larkin2012incentive} document that overconfidence predicts, respectively, excess entry into competitions and the selection of riskier incentive schemes.\footnote{In \citet{koszegi2006ego}'s theoretical model, overconfidence drives the choice of more ambitious tasks via self-image concerns. \citet{burks2013overconfidence} provide evidence linking overconfidence about relative cognitive ability to social signaling.} Second, we assume that identifying the optimal level of task complexity is itself cognitively demanding. This assumption yields our third hypothesis: more cognitively able individuals choose their payoff-maximizing task more frequently. Hypothesis 3 is also motivated by prior evidence, specifically prior work that links cognition and the quality of decisions. For example, \citet{agarwal2013cognitive} and \citet{ambuehl2022evaluating} document that cognitive ability and deliberative competence, respectively, link to the quality of financial decisions.

We test these hypotheses in a pre-registered laboratory experiment.\footnote{The experiment was pre-registered on the Open Science Framework (\texttt{osf.io/8dbrw}). We pre-registered our model, experimental design, and hypotheses, including the definition of overconfidence.} To bring the model to the laboratory, we construct two pairs of tasks adapted from \citet{oprea2020makes}. In each pair, the subject chooses between a less complex task and a more complex task to implement. As in the model, the success payment rises with the complexity of the chosen task, while the failure payment is constant. We use rule implementation tasks from \citet{oprea2020makes}, in which subjects implement finite automata of varying complexity. We measure cognitive ability using the 11-item matrix-reasoning test from the International Cognitive Ability Resource \citep{dworak2021using}. Matrix reasoning is a leading measure of fluid intelligence and has been used in economics by, for example, \citet{gill2016cognitive} and \citet{proto2019intelligence, proto2022intelligence}. We elicit each subject's belief about her own score in the matrix-reasoning test and define overconfidence as the difference between her belief and her score.\footnote{We use `overconfidence' in the sense of overestimation of one's own cognitive ability, following the taxonomy in \citet{moore2008trouble}. Like us, \citet{tekleselassie2025feedback} use a matrix-reasoning test and elicit subjects' beliefs about their own scores to measure overconfidence, and find an asymmetry: underconfident job seekers intensify search after feedback about their relative ability, while overconfident ones do not adjust their behavior.}

The first two hypotheses are strongly supported by the data. From the choice between the less complex and the more complex task in each pair of tasks, we obtain our first two results. First, higher cognitive ability predicts the choice of more complex tasks. Second, conditional on cognitive ability, higher overconfidence also predicts the choice of more complex tasks. Given that each subject attempts only the task she chooses, evaluating the optimality of that choice raises a selection problem. We therefore structurally estimate a model that accounts for endogenous selection, derive subjects' task-specific success probabilities, and use these estimates to identify each subject's optimal task. We find that in about 75\% of the decisions, subjects choose optimally; and those who fail to choose optimally are undercautious, choosing the more complex task when the less complex task had a higher expected payoff. We also find some weak evidence that cognitive ability is associated with more efficient choices of task complexity. In addition, we find that cognitive ability predicts realized earnings, as implied by our model.

Our primary contribution is to the literature on choice of task complexity. We study a setting in which subjects choose between two tasks that are both complex but differ in complexity and in the payment conditional on success. This design creates a direct tradeoff: the more complex task offers a higher success payment, but is expected to have a lower success probability. We link this tradeoff to measured cognitive ability and overconfidence, and show that both predict selection into the more complex task. \citet{oprea2020makes} measures task complexity using elicited willingness to pay to avoid a complex task in favor of a zero-complexity task (typing the same letter repeatedly), mistake rates, and implementation time, and finds that all three measures are influenced by cognitive ability. However, \citet{oprea2020makes} does not study choice among tasks of varying levels of complexity with a tradeoff between success probability and success payment. Related experiments (e.g., \citealp{niederle2008gender, ou2021effect}) study choice between a hard, higher-paid task and an easy, lower-paid task, focusing on gender differences in task choice. However, they do not focus on how cognitive ability affects this process. Our paper instead connects cognitive ability and overconfidence to choice of task complexity and further explores the optimality of task choice.\footnote{Other work considers choice of complexity without focusing on the link to cognitive ability or overconfidence. Examples include choice of: decision-making procedures (e.g., \citealp{banovetz2023complexity, kendall2024complexity}); strategies in games (e.g., \citealp{romero2018constructing, romero2019evolution, salant2026complexity}); choice sets or lotteries (e.g., \citealp{sonsino2002complexity,burton2012self, moffatt2015heterogeneity}); and disclosures or policies (e.g., \citealp{jin2022complex, foarta2025common}). Closest to our angle, \citet{carvalho2024complexity} study how a composite decision-making skill measure (based on numeracy, financial literacy, and consistency with GARP) predicts choice between a complex portfolio and a simple outside option.}

We also contribute to the literature on cognitive ability and decision making in complex settings. We consider the link between cognitive ability and the optimality of task choice.\footnote{Cognitive ability is also associated with strategic behavior and sophistication in environments where subjects do not choose task complexity (e.g., \citealp{burks2009cognitive, carpenter2013cognitive, fehr2016knows, gill2016cognitive, proto2019intelligence, lambrecht2024intelligence, gill2025cognitive, gill2026identifying}).} Because we observe each subject's task success or failure only for the task she chooses, we estimate subject-specific success probabilities using a model that accounts for endogenous selection. This allows us to identify each subject's payoff-maximizing task and measure the efficiency of her task choice. We find some weak evidence that cognitive ability is associated with choice efficiency, and we also find that cognitive ability predicts earnings. Related papers connect cognitive ability to errors in complex decisions. \citet{guan2025beyond} study information demand through incentivized rankings of information structures. They find that subjects sometimes rank less instrumentally valuable structures above more valuable alternatives, and that cognitive ability predicts the optimality of these rankings. \citet{brocas2019consistency} find that working-memory and IQ scores correlate with GARP consistency in complex bundle choices involving three distinct goods, but not in simpler bundle choices involving only two goods. \citet{puri2025simplicity} finds that lotteries with more possible outcomes are valued less, especially by participants with lower cognitive ability. Relative to these studies, our paper examines the question of whether cognitive ability shapes the optimality of subjects' complexity choices.\footnote{A broader literature studies how complexity affects decisions in various settings, such as lottery choice and the evaluation of risky prospects (e.g., \citealp{huck1999risk, mador2000complexity, de2024caution, oprea2024decisions}); Bayesian updating (e.g., \citealp{charness2005optimal}); intertemporal choice (e.g., \citealp{enke2025complexity}); contingent reasoning (e.g., \citealp{park2025complexity}); effort provision (e.g., \citealp{abeler2025incentive}); strategic interactions (e.g., \citealp{devetag2008playing, gill2023strategic, zhou2024complexity}); and subjective perceptions of complexity (e.g., \citealp{agranov2025complex}). See also \citet{gabaix2024complexity} for a general framework for the complexity of economic decisions.}

Our findings suggest that organizations and managers should actively design task menus, incentives for choosing more complex tasks, and feedback structures to account for (i) how workers choose task complexity and sometimes overreach by choosing tasks that are too complex, and (ii) how workers' task choices link to both their cognitive ability and confidence. In our setting, self-selection in task complexity often works: in about 75\% of the decisions subjects choose their payoff-maximizing task. At the same time, the mistakes are asymmetric. When subjects fail to choose optimally, they always choose the more complex task when the less complex task has the higher expected payoff. This pattern maps naturally to workplaces in which employees volunteer for stretch projects, take on high-stakes client accounts, apply for promotion into more complex roles, or choose between narrower and more complex analytical assignments. Each case involves a choice of task complexity: the organization may benefit when able workers choose more complex tasks, because those tasks are more valuable when completed successfully. At the same time, workers may not have accurate beliefs about their own ability or about their probability of success on tasks. The implication is therefore not that organizations should remove ambitious options, but that they should design task menus carefully. Rewards for completing more complex tasks can help attract workers who are able to perform them, but they may also encourage overambitious workers to choose tasks for which the expected payoff is lower. Organizations and managers can respond by pairing the choice of more complex tasks with clear information and feedback: historical success rates for comparable tasks, low-stakes trial assignments before full responsibility, decision aids that make the payoff tradeoff salient, and post-project reviews that distinguish poor task choice from poor task execution. These features can help preserve the benefits of self-selection while helping workers learn when ambition is productive, when a more complex task is worth attempting, and when overconfidence might push them to be too ambitious.

The paper proceeds as follows. Section \ref{Sect:Model} presents the model. Section \ref{Sect:Design} describes the experimental design. Section \ref{Sect:Hypothesis} develops the hypotheses. Section \ref{Sect:Testing} tests the hypotheses. Section \ref{Sect:Conclusion} concludes.

\clearpage

\section{Model}\label{Sect:Model}\

We consider a model in which the probability of making a mistake in a task is determined by two factors: task complexity and the decision-maker's cognitive ability. The mistake probability depends on task complexity and cognitive ability in two intuitive ways.
\begin{itemize}
    \item The mistake probability \textit{increases} with task complexity. For any given level of cognitive ability, implementing a more complex task is more likely to produce an error.
    \item The mistake probability \textit{decreases} with cognitive ability. For any given task, a more cognitively able decision-maker is less likely to make a mistake.
\end{itemize}

Let \(a_i\) denote the cognitive ability of decision-maker \(i\) and \(c_t\) denote the complexity level of task \(t\). A greater \(a_i\) represents higher cognitive ability, while a greater \(c_t\) represents a higher complexity level. \(P_m(a_i, c_t)\) denotes the probability that decision-maker \(i\) makes a mistake when implementing task \(t\). The functional form should satisfy the following requirements:
\begin{enumerate}[(1)]
    \item \( \frac{\partial P_m(a_i, c_t)}{\partial c_t} > 0 \): The mistake probability \textit{increases} with the complexity of the task.
    \item \( \frac{\partial P_m(a_i, c_t)}{\partial a_i} < 0 \): The mistake probability \textit{decreases} with the decision-maker's cognitive ability.
\end{enumerate}

\subsection{Mistake Probability Function}

We use the following functional form of the probability of making a mistake, \(P_m(a_i, c_t)\):
\begin{equation}
    P_m(a_i, c_t) = \alpha + \beta_1 a_i + \beta_2 c_t + \beta_3 a_i c_t
    \label{eqn:linear-mistake-probability-new}
\end{equation}
where \(\alpha\) is a constant term. \(\beta_1\), \(\beta_2\), and \(\beta_3\) are coefficients that quantify the impact of the decision-maker's cognitive ability (\(a_i\)) and the task's complexity (\(c_t\)). Without loss of generality, \(a_i\) and \(c_t\) are both normalized to \([0,1]\). We impose the following restrictions:

\begin{enumerate}[\bfseries{A}-1.]
    \item $\frac{\partial P_m}{\partial c_t} = \beta_2 + \beta_3 a_i > 0, \ \forall a_i \in [0,1]$.
    This assumption ensures that an increase in task complexity \(c_t\) increases the mistake probability, for any given cognitive ability.
    \item $\frac{\partial P_m}{\partial a_i} = \beta_1 + \beta_3 c_t < 0, \ \forall c_t \in [0,1]$. This assumption ensures that an increase in cognitive ability \(a_i\) decreases the mistake probability, for any given task complexity.
    \item $\frac{\partial^2 P_m(a_i,c_t)}{\partial c_t \partial a_i} = \beta_3 \leq 0$. This assumption ensures that task complexity has a weakly smaller effect on the mistake probability of a more cognitively able decision-maker. This assumption is a sufficient but not necessary condition for Propositions 1 and 2, which continue to hold when \(\beta_3\) is not too positive.

    \item Probability bounds: $P_m(a_i, c_t) \in [0, 1], \ \forall (a_i,c_t) \in [0,1]^2$.
\end{enumerate}

We estimate the model freely to verify these restrictions, and all the restrictions are satisfied. See \ref{Sect:Model-Estimation} for details.

\subsection{Optimal Task Complexity and Cognitive Ability}

Consider the maximization problem in which the decision-maker chooses a task \(t\) with complexity level \(c_t \in [0,1]\) to maximize her expected payoff. The decision-maker receives a success payment \(w_H(c_t)\) if she successfully implements the task and a failure payment \(w_L\) if she makes a mistake. \(w_H(c_t)\) increases with task complexity, while \(w_L\) remains constant:
\begin{equation}
   w_H(c_t) = w_L + k \cdot c_t
\end{equation}
where \(k > 0\) indicates the rate at which the success payment increases with complexity.

The decision-maker's objective is to choose task complexity to maximize her expected payoff \(\pi(a_i, c_t)\):
\begin{equation}
    \begin{aligned}
        \max\limits_{c_t \in [0,1]} \quad \pi(a_i, c_t) &= P_m(a_i, c_t) \cdot w_L + \big( 1 - P_m(a_i, c_t) \big) \cdot w_H(c_t) \\
        &= P_m(a_i, c_t)\cdot w_L + \big(1 -  P_m(a_i, c_t) \big) \cdot (w_L +k \cdot c_t) \\
        &= w_L + k \cdot c_t \cdot \big( 1 - P_m(a_i, c_t) \big)  \\
    \end{aligned}
    \label{eqn:linear-maximization-problem-new}
\end{equation}

Our main result describes how optimal task complexity varies with cognitive ability.
\begin{proposition}[Optimal Complexity]
    As the decision-maker's cognitive ability increases, the optimal level of task complexity weakly increases.
    \label{prop:optimal-complexity}
\end{proposition}

\begin{proof}
    See \ref{subsec:proof-prop1}.
\end{proof}

The intuition is as follows. Higher cognitive ability lowers the mistake probability at every level of complexity. Because the success premium (difference between success and failure payments) is larger for more complex tasks, the gain in expected payoff from a lower mistake probability is larger on more complex tasks, making them relatively more attractive as cognitive ability rises. Under Assumption A-3 ($\beta_3 \leq 0$), higher cognitive ability also weakly slows the rate at which the mistake probability rises with complexity, reinforcing this effect. A more cognitively able decision-maker therefore chooses a weakly higher level of task complexity. The result continues to hold as long as $\beta_3$ is not sufficiently positive.

Equivalently, in terms of the success probability ($1-P_m$), the decision-maker's problem is analogous to a monopolist's choice of a price on a linear demand curve, with complexity playing the role of price and the success probability playing the role of demand. The comparative static mirrors the standard result that an upward shift in linear demand raises the optimal price, as long as the demand curve does not steepen too much as it shifts up.

In our experiment, subjects make a binary choice between a less complex task and a more complex task. Although Proposition \ref{prop:optimal-complexity} is stated for the continuous case, its logic extends naturally to the binary case:
\begin{proposition}[Binary Choice of Task Complexity]
    When the decision-maker faces a binary choice of task complexity, her optimal choice can switch only from the less complex task to the more complex task as her cognitive ability increases.
    \label{prop:binary-complexity}
\end{proposition}

\begin{proof}
    See \ref{subsec:proof-prop2}.
\end{proof}

Our setup also carries a welfare implication: under optimal choice of complexity, expected payoff $\pi^* (a_i )\equiv \max_{c_t \in [0,1]}\pi(a_i,c_t )$ strictly increases in cognitive ability. A decision-maker with higher cognitive ability can choose the same complexity as a less able decision-maker. The replicated choice yields strictly higher expected payoff by A-2 (since optimal complexity is always strictly positive from the proof of Proposition \ref{prop:optimal-complexity}), and re-optimizing can only raise expected payoff further. We summarize this implication:
\begin{proposition}[Expected Payoff]
    When the decision-maker chooses task complexity optimally, her expected payoff strictly increases in cognitive ability.
    \label{prop:expected-payoff}
\end{proposition}

\clearpage

\section{Experimental Design}\label{Sect:Design}\

\subsection{Procedures}

We collected experimental data from 307 student subjects at the Vernon Smith Experimental Economics Laboratory (VSEEL) at Purdue University in November 2025 (Purdue IRB-2024-1586). The experiment is pre-registered on the Open Science Framework (\texttt{osf.io/8dbrw}).\footnote{We pre-registered our plan to run sessions until we reached 300 subjects. We also collected experimental data from 103 subjects across three pilot sessions before we submitted our pre-registration. In these pilot sessions, subjects did not choose task complexity (see \ref{Sect:Pilot} for details).\label{fn:pilot}} We drew subjects from the VSEEL subject pool (administered using ORSEE; \citealp{greiner2015subject}). The sample is balanced by gender, with 50.5\% female. The experiment was programmed in oTree \citep{chen2016otree}. At the beginning of the experiment, subjects electronically provided consent to participate. We provided the instructions to each subject on her computer screen.

We ran 11 sessions, each lasting approximately 30 minutes. Procedures were identical across sessions, although session sizes varied due to fluctuations in the show-up rate. Subjects earned \$18.62 on average, including the \$5.00 show-up fee, with total payments ranging from \$7.50 to \$30.50.

Each session consisted of three parts. In Part 1, subjects completed a test of cognitive ability followed by a belief elicitation task about their cognitive ability. Part 2 consisted of two rounds. In each round, subjects saw a pair of rule implementation tasks that differed in their levels of complexity. Subjects first chose one task from the pair, and then worked on the task they had chosen. In Part 3, subjects completed a brief demographic questionnaire that asked about gender and age.

\subsection{Pairs of Tasks with Choice of Complexity}\label{Sect:Main-task}

\begin{figure}[h!]
    \centering
    \includegraphics[width=1.0\textwidth]{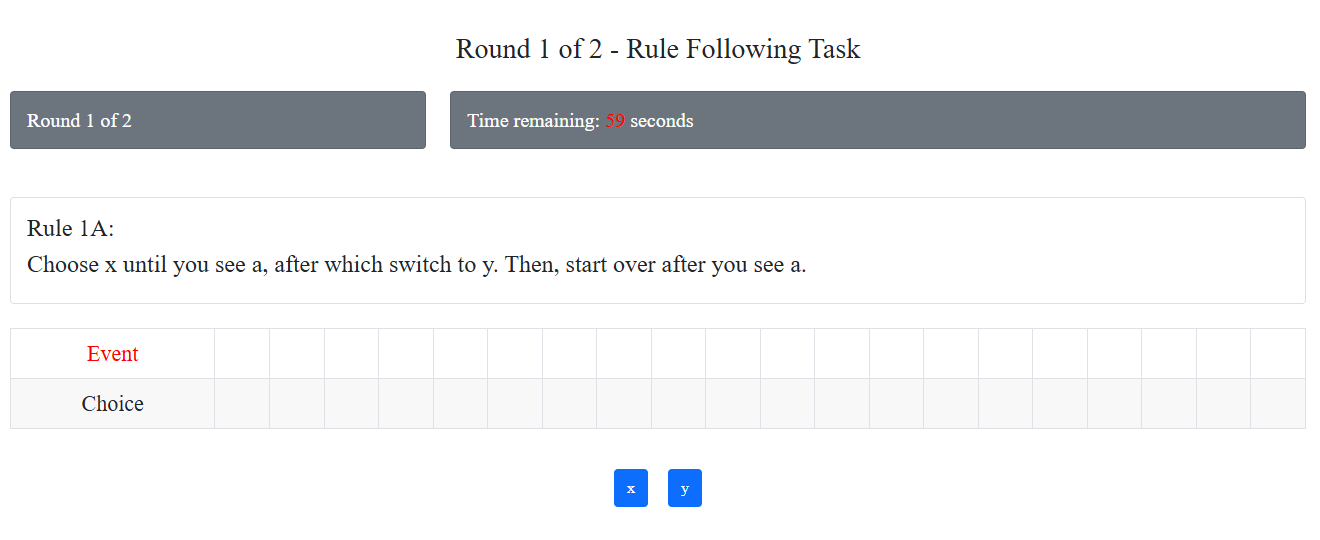}
    \caption{Experiment Interface Screenshot - Rule Implementation Task}
    \label{fig:screenshot-task1}
\end{figure}

\begin{figure}[h!]
    \centering
    \includegraphics[width=1.0\textwidth]{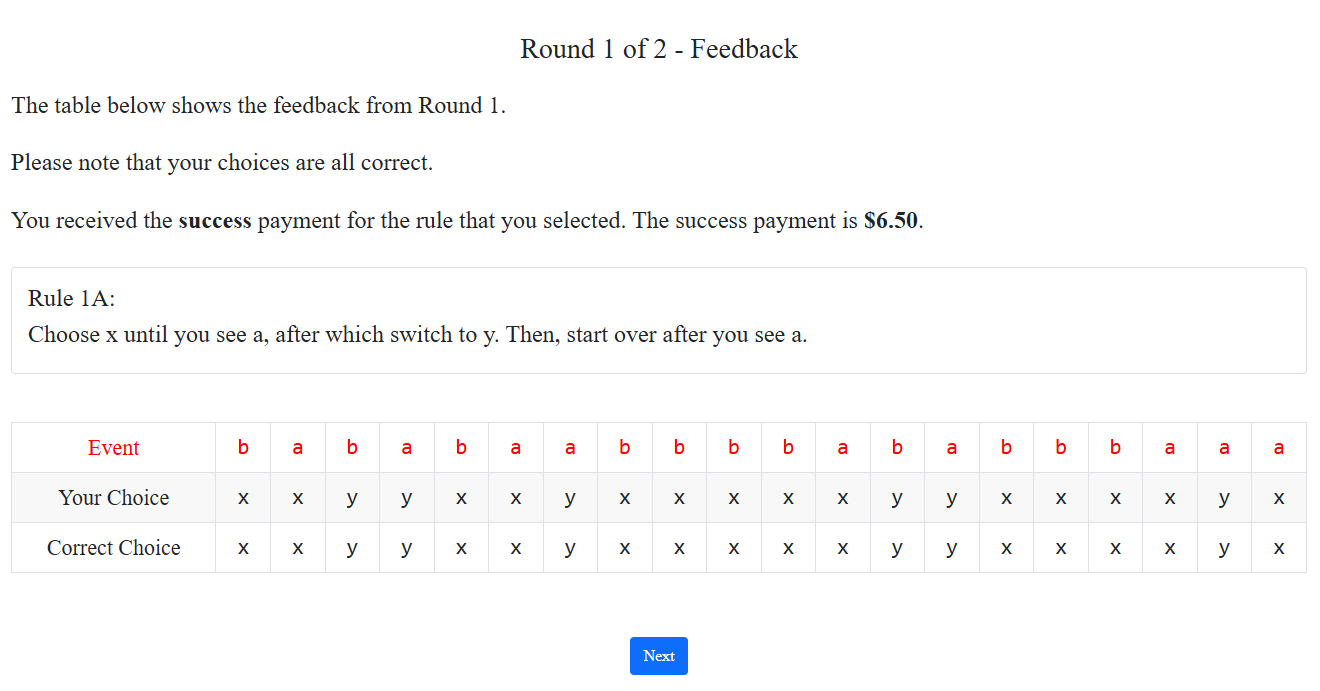}
    \caption{Experiment Interface Screenshot - Feedback Screen}
    \label{fig:screenshot-task2}
\end{figure}

We use rule implementation tasks from \citet{oprea2020makes}. In each task, the subject follows a rule, formally described as a finite automaton with a particular number of states and transitions. Figures \ref{fig:screenshot-task1} and \ref{fig:screenshot-task2} provide an example rule implementation task and feedback screen. Appendix B provides a full set of screenshots from the experiment. Specifically, each task consists of 20 ticks. In each tick, the subject first types a letter and then observes a letter randomly chosen by the computer. The subject is given a rule that specifies which letter to type in the first tick and which letter to type in response to each randomly chosen letter. The rule is displayed on the screen throughout the task. Each task has a time constraint of 60 seconds.\footnote{\citet{oprea2020makes} does not impose a time constraint. We chose 60 seconds to generate the variation in mistake rates that our analysis relies on.} We define task success as correctly typing the sequence of 20 letters within the 60-second time constraint, and task failure (which corresponds to a mistake in the model) as any outcome that does not meet this criterion. The subject receives the success payment if and only if she correctly types all 20 letters within the time constraint, and otherwise receives the failure payment. At the end of each task, the subject observes the correct sequence alongside her own responses. This feedback screen includes the subject’s payment for the round and is displayed until the subject chooses to continue, with no time constraint.\footnote{A possible concern with using rule implementation tasks is that mistakes are driven mainly by differences in effort rather than cognitive ability. However, \citet[Appendix B.3]{oprea2020makes} argues against an important role for effort, noting that: failures in high-mistake rules often happen far into the task, and time spent on failed tasks is longer than on successful ones. Furthermore, the per-task incentives in our experiment are substantially higher than in \citet{oprea2020makes}.}

\vspace{2cm}

\clearpage

\vspace{1cm}

We select four tasks from \citet{oprea2020makes} that span a range of complexity levels and organize them into two pairs, each containing a less complex task and a more complex task. Pair 1 consists of 2S-1Ta (less complex) and 3S-3T (more complex). Pair 2 consists of 2S-2T (less complex) and Sequenceable (more complex).\footnote{We use the same task names as \citet{oprea2020makes}. For example, 2S-1Ta refers to an automaton with two states, one transition, and one absorbing state (Sequenceable has four states and transitions). Subjects see the exact on-screen text that \citet{oprea2020makes} uses to describe each task. We preserve Oprea's original instructions, except for the additions of the choice of task and the time constraint.} Tasks are labeled A or B in the interface: within each pair, the task labeled B is always more complex than the one labeled A. Across the two pairs, 2S-1Ta is less complex than 2S-2T, and 3S-3T is less complex than Sequenceable. Therefore, Pair 1 is less complex overall than Pair 2. The ordering of complexity is confirmed by the mistake rates in \citet{oprea2020makes}, and by the mistake rates in our pilot sessions (without choice of task complexity).

\begin{figure}[h!]
    \centering
    \includegraphics[width=1.0\textwidth]{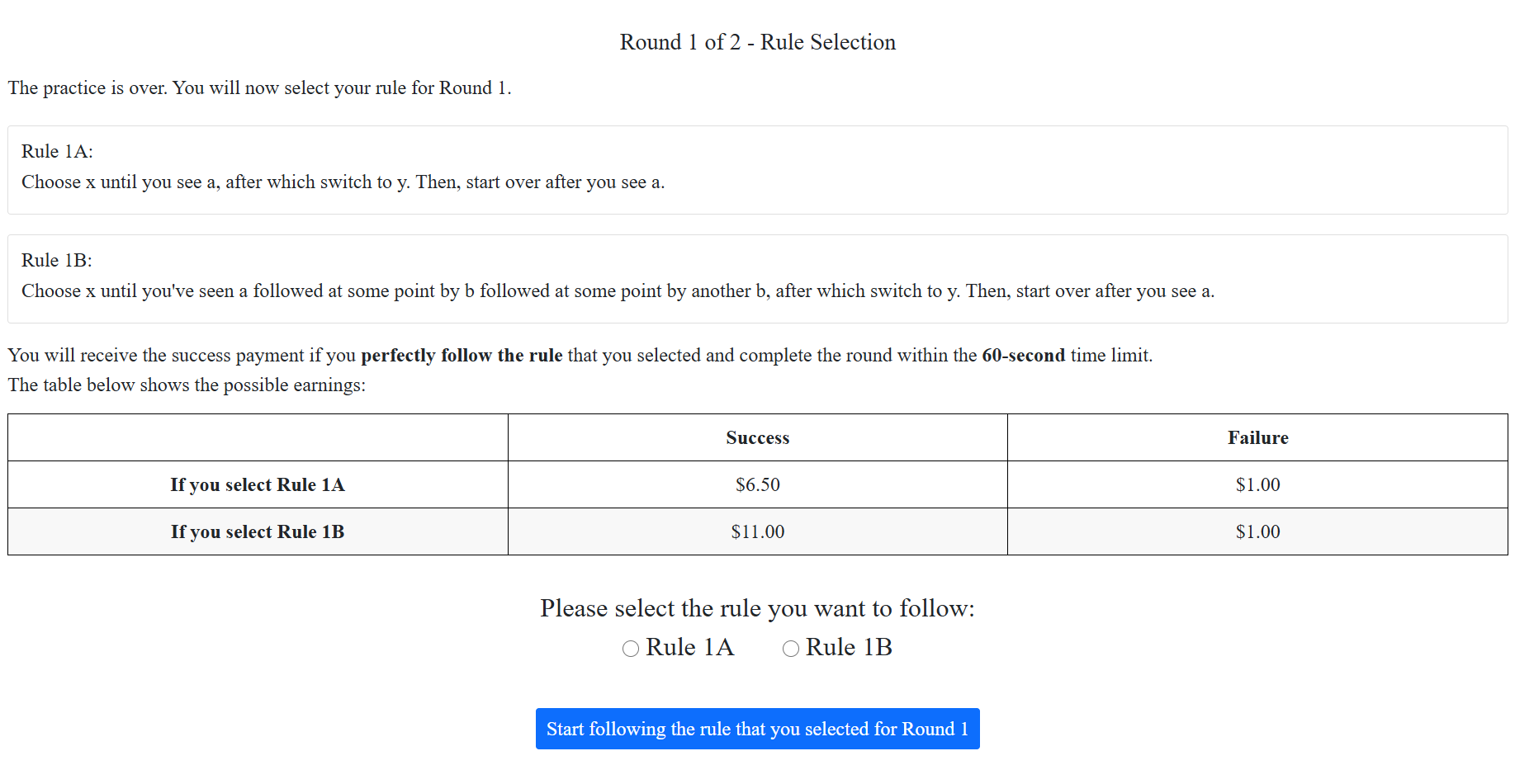}
    \caption{Experiment Interface Screenshot - Round 1 Choice Page}
    \label{fig:screenshot-selection1}
\end{figure}

\begin{figure}[htbp]
    \centering
    \includegraphics[width=1.0\textwidth]{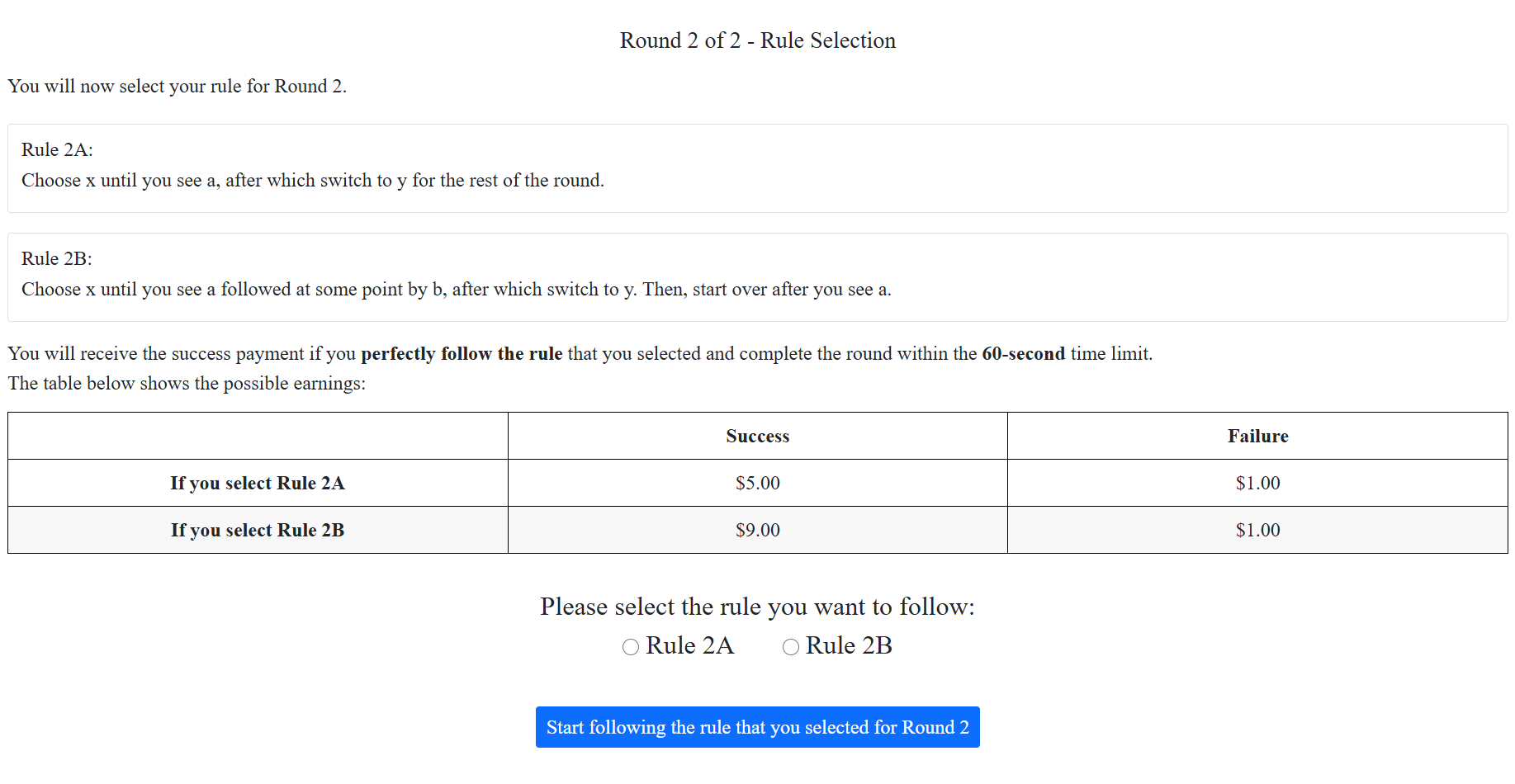}
    \caption{Experiment Interface Screenshot - Round 2 Choice Page}
    \label{fig:screenshot-selection2}
\end{figure}

In each of two rounds, the subject is presented with one pair and chooses one of the two tasks to complete. The order of the two pairs across rounds is randomized at the subject level. The tasks are prefixed with the round number in the interface (e.g., 1A and 1B in Round 1); because pair order is randomized, the digit prefix indicates the round rather than the pair. Figures \ref{fig:screenshot-selection1} and \ref{fig:screenshot-selection2} provide an example of the choice pages in both rounds, with Pair 2 appearing first in this example.

\clearpage

\vspace{0.5cm}

\vspace{1cm}

Table \ref{tab:payment} summarizes the payment structure. The more complex task in each pair offers a higher success payment, creating a tradeoff between the reward from success and the probability of success. The failure payment is \$1.00 for all tasks. As Pair 1 is relatively less complex than Pair 2, the success payment for each task in Pair 1 is lower than that for its counterpart in Pair 2. This payment structure reflects our model, in which the failure payment is constant, and the success payment rises with complexity.

\begin{table}[h!]
    \centering
    \caption{Payment Structure}
    \label{tab:payment}
    \resizebox{\textwidth}{!}{%
    \begin{threeparttable}
    \begin{tabular}{llcc}
        \toprule
        & Task & Success Payment & Failure Payment \\
        \midrule
        \noalign{\vspace{5pt}}
        \multirow{2}{*}{Pair 1} & 2S-1Ta (less complex) & \$5.00 & \$1.00 \\[5pt]
                                & 3S-3T (more complex)  & \$9.00 & \$1.00 \\[10pt]
        \multirow{2}{*}{Pair 2} & 2S-2T (less complex)       & \$6.50 & \$1.00 \\[5pt]
                                & Sequenceable (more complex) & \$11.00 & \$1.00 \\[5pt]
        \bottomrule
    \end{tabular}
    \begin{mytablenotes}
        \item \textit{Notes:} Task names follow \citet{oprea2020makes}. We paid subjects for each pair.
    \end{mytablenotes}
    \end{threeparttable}
    }
\end{table}

\vspace{0.25cm}

After subjects had read the instructions and previewed the payment structure and the on-screen text descriptions of the four tasks used in Pair 1 and Pair 2, they worked on five unpaid practice tasks. The practice lets subjects learn the structure of the rule implementation tasks and helps them understand the complexity of the four tasks in Pair 1 and Pair 2. All subjects worked on the same five practice tasks without making any choice over tasks.\footnote{The five practice tasks are 3S-1Ta, 4S-1Ta, 3S-2T, 4S-2T, and 4S-3T, all from \citet{oprea2020makes}. No practice task matches any of the four tasks in Pair 1 and Pair 2 in both the number of states and the number of transitions.}

\subsection{Measurement of Cognitive Ability and Overconfidence}\label{Sect:Tests}

To measure cognitive ability, subjects completed the 11-item matrix-reasoning test from the International Cognitive Ability Resource (\citealp{dworak2021using}; \href{https://icar-project.com}{https://icar-project.com}). This test is an open-source analogue of Raven's Progressive Matrices \citep{raven2000raven}, and has been used in economics by, e.g., \citet{gill2024beliefs} and \citet{gill2025cognitive}. The test requires subjects to identify missing elements that complete visual patterns: Appendix B (page 3) provides a sample item. Matrix reasoning captures fluid intelligence (i.e., logical or analytical intelligence), and is a leading measure of cognitive ability that correlates strongly with general intelligence \textit{g} \citep[Box~1]{gray2004neurobiology} and with performance on other complex cognitive tasks \citep{carpenter1990one}. Subjects earned \$0.30 for each correct answer and had a 10-minute time limit to complete all 11 items. Since our matrix-reasoning test is an open-source analogue of Raven’s Matrices, we refer to the number of correct answers in the matrix-reasoning test as the subject's Raven score, which can range from 0 to 11. In the regressions reported in Section \ref{Sect:Testing}, we normalize the Raven score to \([0, 1]\).

After completing the test, subjects reported their belief about the number of items they solved correctly. Subjects received a \$0.20 fixed payment and earned an additional \$2.00 if their reported belief matched their Raven score exactly. We define overconfidence as the difference between a subject's reported belief and her Raven score, as pre-registered. A positive value indicates that the subject overestimates her cognitive ability, while a negative value indicates underconfidence. Subjects were informed only at the end of the experiment about whether their belief matched their Raven score.

\section{Development of Hypotheses}\label{Sect:Hypothesis}\

We draw on the theoretical analysis from Section \ref{Sect:Model} to develop three hypotheses. All three hypotheses were pre-registered before the data were collected.

Our first hypothesis follows from Propositions \ref{prop:optimal-complexity} and \ref{prop:binary-complexity}. The model predicts that the optimal complexity level increases with cognitive ability. More cognitively able subjects succeed more often at any given task, and so they face a more favorable tradeoff between the higher success payment from a more complex task and its lower probability of success.

\begin{hypothesis}
    Subjects with higher cognitive ability are more likely to choose more complex tasks.
    \label{hyp:ability-choice}
\end{hypothesis}

\vspace{0.5cm}

Our second hypothesis concerns the role of overconfidence. Recall from Section \ref{Sect:Tests} that we define overconfidence as the difference between a subject's belief about her Raven score and her actual Raven score, with a negative value indicating underconfidence. To allow overconfidence to affect choice, we adopt the behavioral assumption that a subject's belief about her own cognitive ability helps to shape her choice of task complexity. Under this assumption, our second hypothesis is that, conditional on cognitive ability, higher overconfidence raises chosen complexity.

\begin{hypothesis}
    Conditional on cognitive ability, subjects with higher overconfidence are more likely to choose more complex tasks.
    \label{hyp:confidence-choice}
\end{hypothesis}

\vspace{0.5cm}

Our third hypothesis concerns the efficiency of task choice. In our model, a decision-maker chooses the task complexity that maximizes her expected payoff, so the model itself predicts no variation in optimality. We augment the model with the behavioral assumption that identifying the optimal level of task complexity is cognitively demanding. To make the optimal choice, a subject must understand the tradeoff between the success payment and the probability of success, and we further assume that higher cognitive ability is associated with a greater probability of correctly evaluating this tradeoff. Under these assumptions, the optimality of the choice improves with cognitive ability, so we expect more cognitively able subjects to choose their optimal task more frequently.

\begin{hypothesis}
    Subjects with higher cognitive ability are more likely to choose task complexity optimally.
    \label{hyp:ability-optimality}
\end{hypothesis}

\section{Testing of Hypotheses}\label{Sect:Testing}\

We report descriptive statistics and regression results that test the three hypotheses developed in Section \ref{Sect:Hypothesis}. Section \ref{Sect:Descriptive} describes the distributions of cognitive ability and overconfidence. Section \ref{Sect:Choice} tests Hypotheses \ref{hyp:ability-choice} and \ref{hyp:confidence-choice}, and Section \ref{Sect:Optimality} tests Hypothesis \ref{hyp:ability-optimality}.

\subsection{Descriptive Statistics}\label{Sect:Descriptive}

\begin{figure}[h!]
    \centering
    \includegraphics[width=0.8\linewidth]{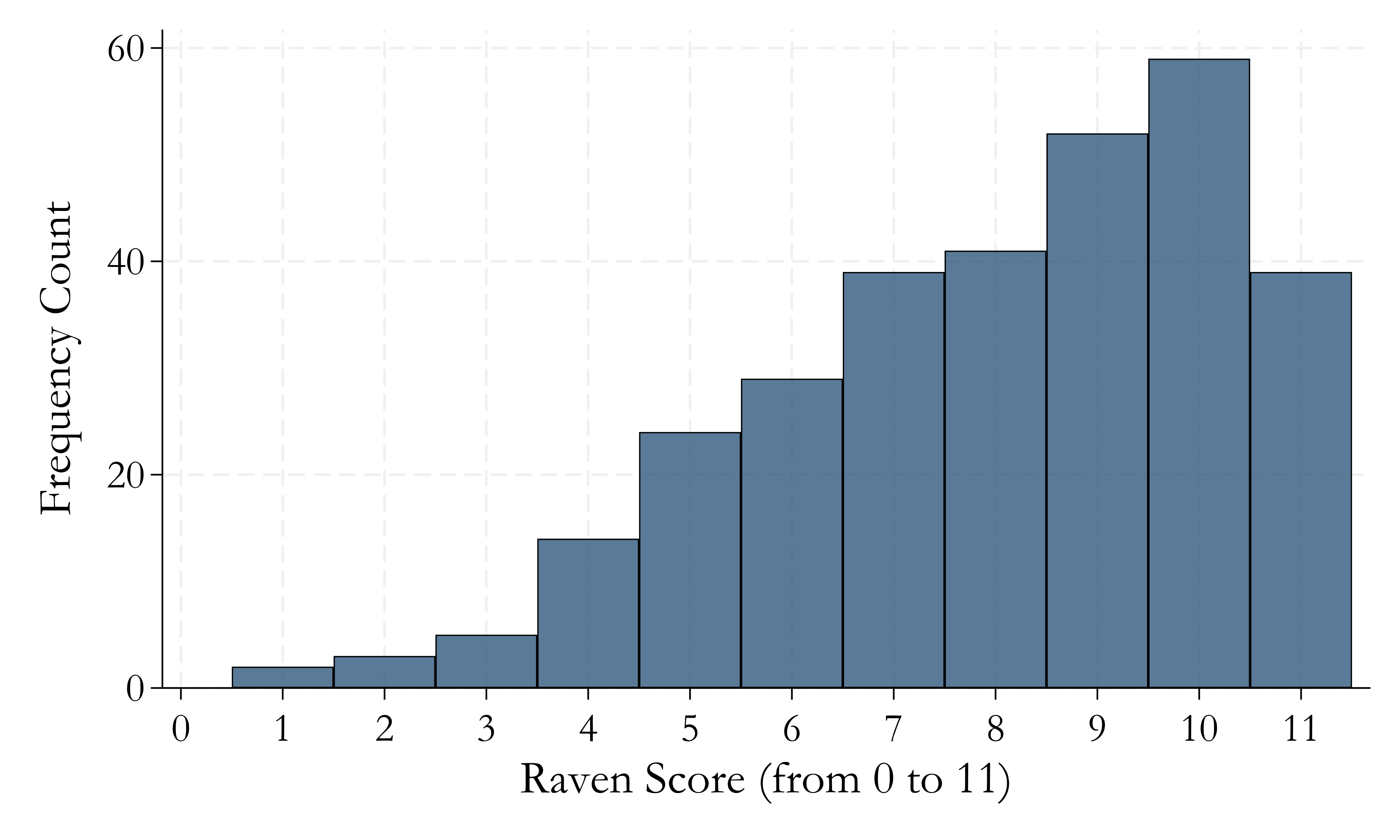}
    \caption{Distribution of Raven Score}
    \label{fig:raven}
\end{figure}

We begin by describing the distributions of cognitive ability and overconfidence. Figure \ref{fig:raven} shows the distribution of Raven scores. The average Raven score is 8.0 out of a maximum of 11, with observed scores ranging from 1 to 11. The distribution is concentrated at high scores, and most subjects score between 5 and 11.

\begin{figure}[h!]
    \centering
    \includegraphics[width=0.8\linewidth]{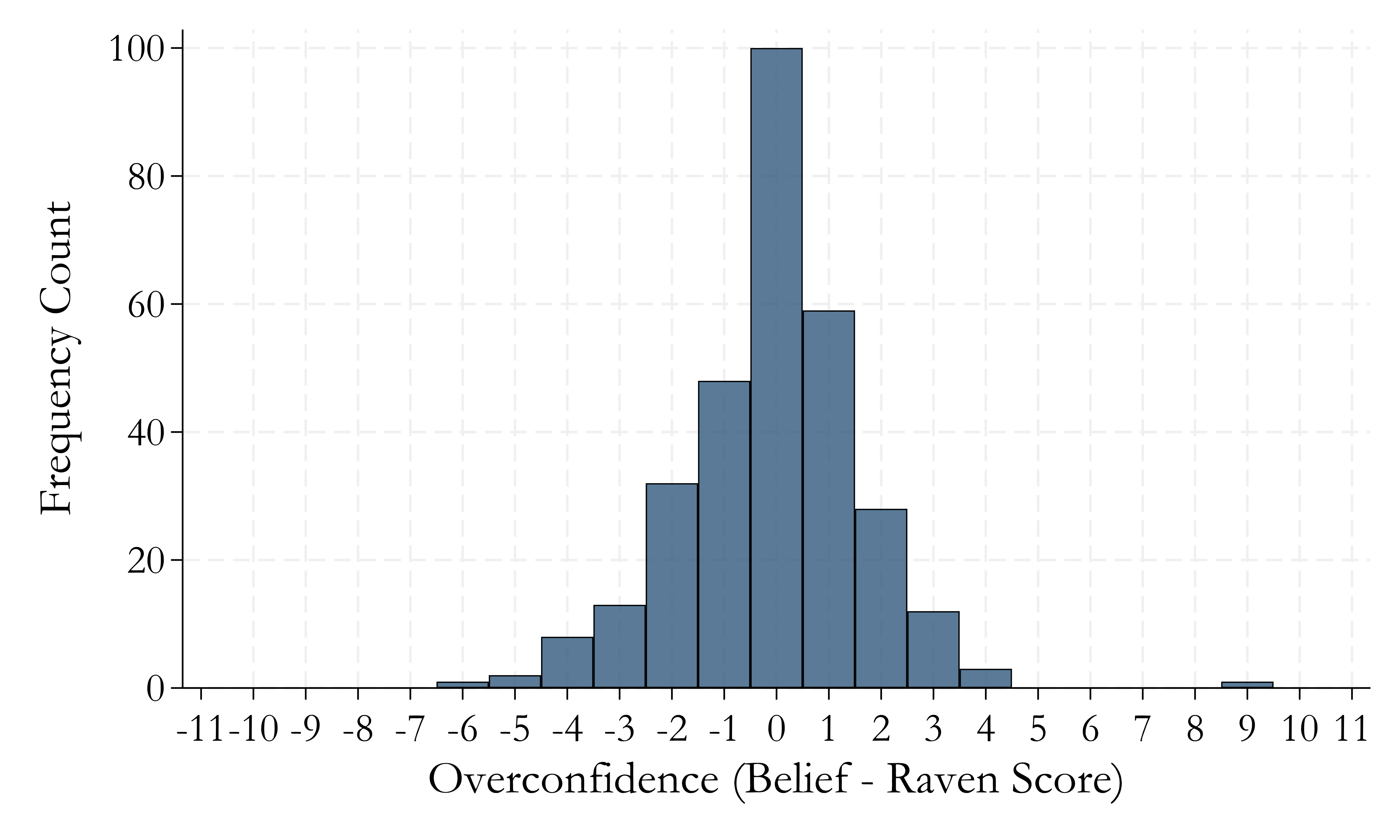}
    \caption{Distribution of Overconfidence}
    \label{fig:confidence}
\end{figure}

Figure \ref{fig:confidence} shows the distribution of overconfidence, measured as the difference between a subject's belief and her actual Raven score. The distribution is roughly symmetric around zero, and most subjects have overconfidence values between $-2$ and $+2$. One subject reports an overconfidence value of $+9$, well above the rest of the distribution. The main results are unaffected when we remove this outlier.\footnote{Our regression results on choice of complexity, earnings, and optimality using our choice efficiency measure are robust when we remove this outlier. See Tables \ref{tab:choice-outlier}--\ref{tab:optimality-outlier} in \ref{Sect:Additional-Analysis}.}

\vspace{0.5cm}

\begin{figure}[h!]
\centering
\begin{subfigure}{0.48\textwidth}
    \centering
    \includegraphics[width=\textwidth]{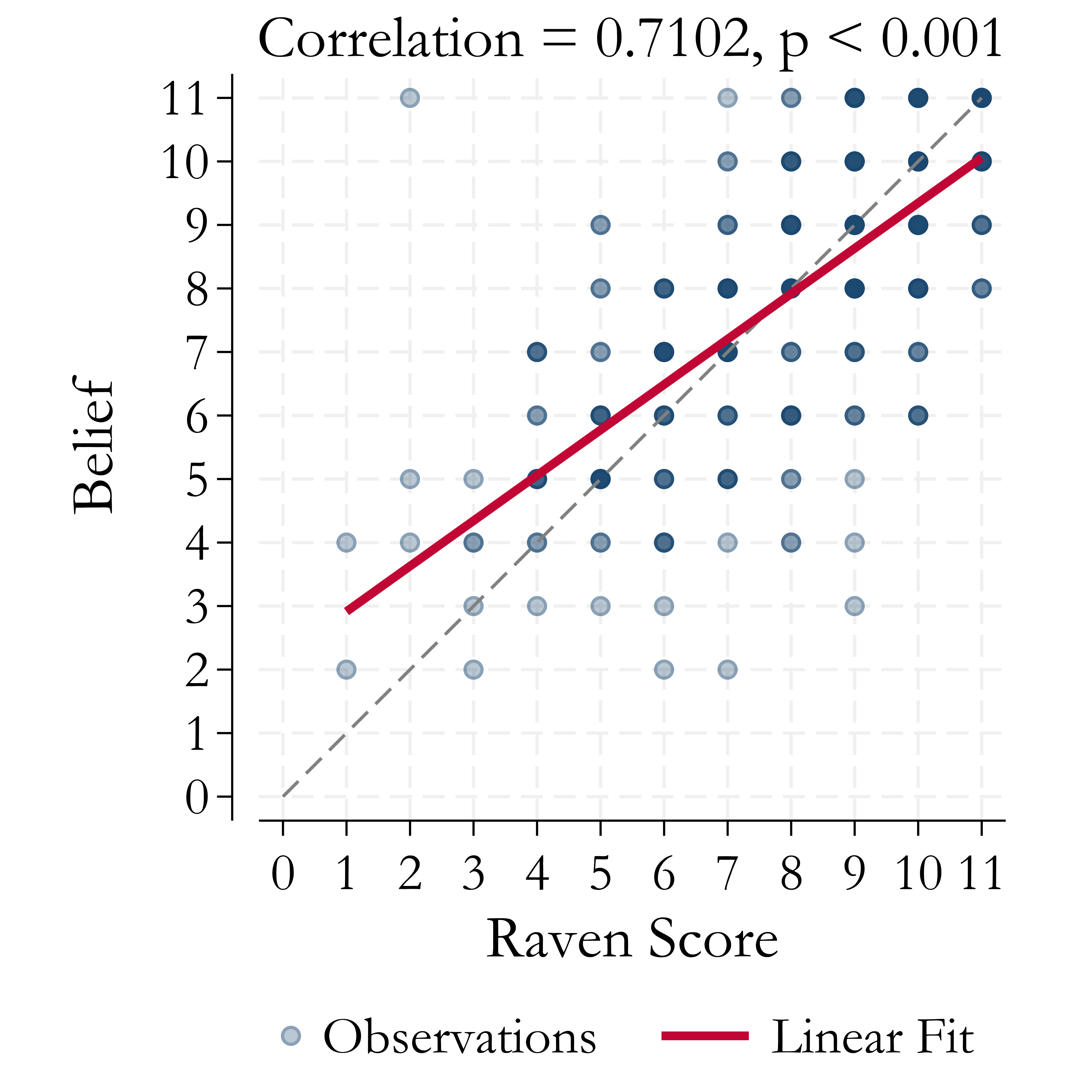}
    \caption{Raven Score and Belief}
    \label{fig:corr-raven}
\end{subfigure}
\hfill
\begin{subfigure}{0.48\textwidth}
    \centering
    \includegraphics[width=\textwidth]{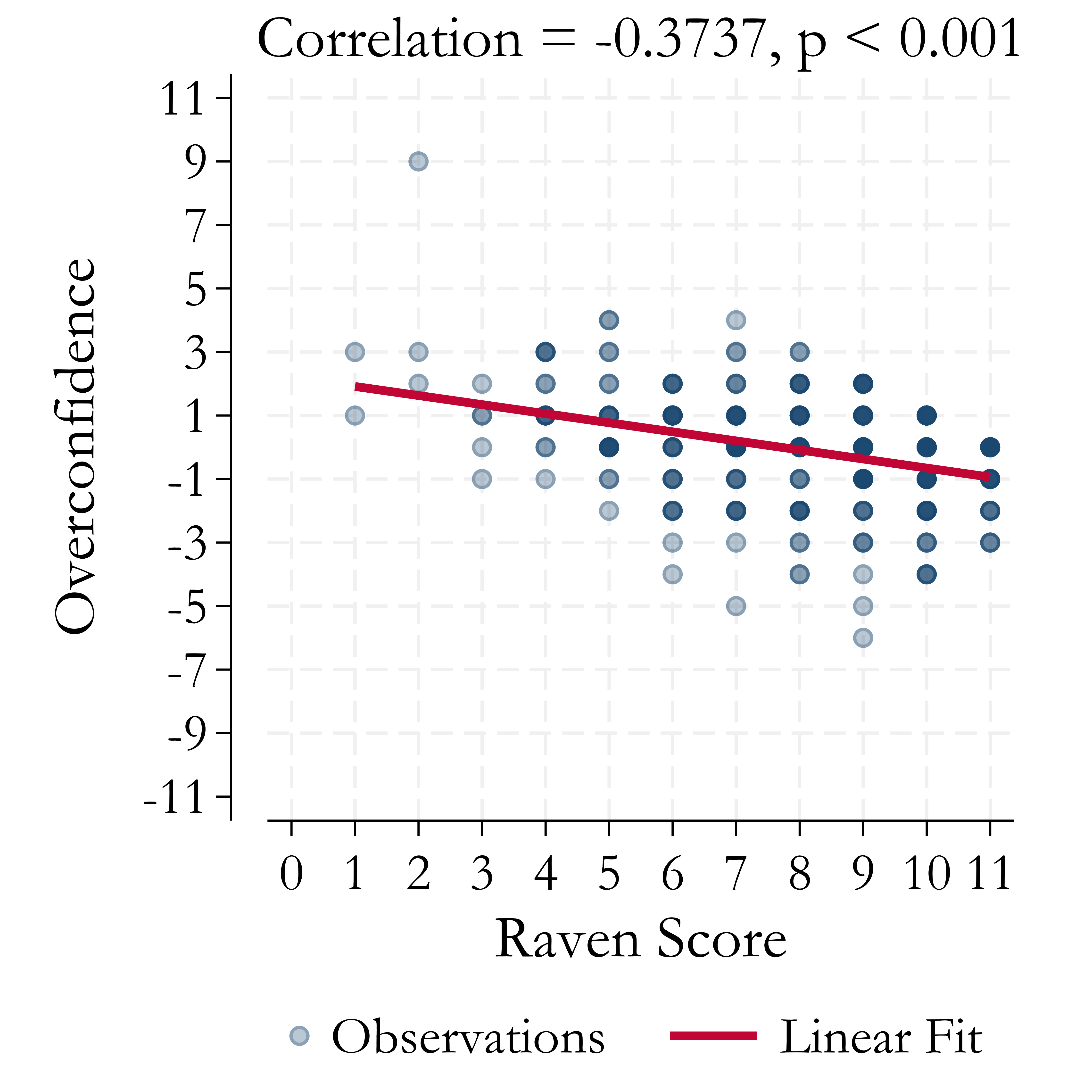}
    \caption{Raven Score and Overconfidence}
    \label{fig:corr-raven-error}
\end{subfigure}
\caption{Correlation Between Raven Score, Belief, and Overconfidence}
\end{figure}

\clearpage

Figure \ref{fig:corr-raven} shows the correlation between subjects' Raven scores and their beliefs. The correlation between the two is 0.71 ($p < 0.001$). Subjects with low cognitive ability tend to overestimate their ability, while subjects with high cognitive ability tend to underestimate their ability. Figure \ref{fig:corr-raven-error} shows the correlation between Raven score and overconfidence, which is defined as the difference between a subject's belief and her score. The correlation between the two is $-0.37$ ($p < 0.001$).

In all regressions reported below, we normalize the Raven score to $[0, 1]$ by dividing by 11, and we normalize overconfidence to $[-1, 1]$ by dividing the difference between belief and score by 11.

\subsection{Choice of Task Complexity and Earnings}\label{Sect:Choice}

Table \ref{tab:choice} reports OLS regressions of task choice on cognitive ability and overconfidence. The dependent variable is a binary indicator equal to one if the subject chose the more complex task within a pair of tasks. Columns (1)--(2) pool both pairs, columns (3)--(4) report results for Pair 1, and columns (5)--(6) report results for Pair 2.

We find strong support for Hypothesis \ref{hyp:ability-choice}: Raven score predicts the choice of the more complex task in both pairs. Subjects with higher Raven scores choose the more complex task substantially more often. An increase in the normalized Raven score from 0 to 1 is associated with a 36 percentage-point higher probability of choosing the more complex task, on average. This positive effect is robust whether we pool the data (Column 1 of Table \ref{tab:choice}) or analyze each pair separately (Columns 3 and 5 of Table \ref{tab:choice}).

\begin{result}
    As predicted by our model, more cognitively able subjects choose more complex tasks.
\end{result}

We also find strong support for Hypothesis \ref{hyp:confidence-choice}: overconfidence predicts the choice of the more complex task in both pairs. Adding overconfidence to Columns 2, 4, and 6 of Table \ref{tab:choice}, we find that, conditional on cognitive ability, more overconfident subjects (or less underconfident subjects) choose the more complex task substantially more often. As shown in Figure \ref{fig:confidence}, most subjects (87\%) have raw overconfidence scores in the range $[-2,2]$. An increase in overconfidence over this range is associated with a 21 percentage-point higher probability of choosing the more complex task, on average ($0.584 \times \frac{4}{11} \approx 0.212$).\footnote{Controlling for overconfidence also raises the coefficient on Raven score in all specifications. This follows mechanically from the positive correlation between overconfidence and choice of the more complex task, together with the negative correlation between Raven score and overconfidence.}

\begin{result}
    Controlling for cognitive ability, more overconfident subjects choose more complex tasks.
\end{result}

\vspace{0.1cm}

\begin{table}[h!]
    \centering
    \caption{Effect of Cognitive Ability and Overconfidence on Complexity Choice}
    \label{tab:choice}
    \small
    \begin{threeparttable}
    \begingroup
    \setlength{\tabcolsep}{4pt}
    {\def\sym#1{\ifmmode^{#1}\else\(^{#1}\)\fi}
\begin{tabular}{l*{3}{c}*{3}{c}*{2}{c}}
\toprule
\multirow{2}{*}{DV: Choose More Complex Task} & \multicolumn{2}{c}{Pooled} && \multicolumn{2}{c}{Pair 1} && \multicolumn{2}{c}{Pair 2} \\
\cmidrule(lr){2-3} \cmidrule(lr){5-6} \cmidrule(lr){8-9}
                    &\multicolumn{1}{c}{(1)}         &\multicolumn{1}{c}{(2)}         &&\multicolumn{1}{c}{(3)}         &\multicolumn{1}{c}{(4)}         &&\multicolumn{1}{c}{(5)}         &\multicolumn{1}{c}{(6)}         \\
\midrule
Raven Score $[0,1]$   &       0.363\sym{***}&       0.551\sym{***}&&       0.297\sym{**} &       0.494\sym{***}&&       0.428\sym{***}&       0.609\sym{***}\\
                    &     (0.125)         &     (0.132)         &&     (0.147)         &     (0.160)         &&     (0.143)         &     (0.149)         \\
Overconfidence $[-1,1]$      &                     &       0.584\sym{***}&&                     &       0.608\sym{***}&&                     &       0.559\sym{***}\\
                    &                     &     (0.171)         &&                     &     (0.186)         &&                     &     (0.187)         \\
\midrule
Observations        &         614         &         614         &&         307         &         307         &&         307         &         307         \\
Mean of DV          &       0.511         &       0.511         &&       0.531         &       0.531         &&       0.492         &       0.492         \\
\bottomrule
\end{tabular}
}

    \endgroup
    \begin{mytablenotes}
        \item \textit{Notes:} OLS regressions. The dependent variable is a binary indicator equal to one if the subject chose the more complex task within a pair of tasks. Raven score is normalized to $[0, 1]$. Overconfidence is defined as the difference between belief and Raven score, normalized to $[-1, 1]$. All specifications include categorical controls for gender, age, and the order of the two pairs. Columns (1)--(2) additionally include a pair fixed effect. Heteroskedasticity-robust standard errors clustered at the subject level are reported in parentheses. $^{***}$, $^{**}$, and $^{*}$ denote significance at the 1\%, 5\%, and 10\% levels (two-sided tests).
    \end{mytablenotes}
    \end{threeparttable}
\end{table}

\vspace{0.5cm}

\begin{table}[h!]
    \centering
    \caption{Effect of Cognitive Ability and Overconfidence on Earnings}
    \label{tab:payoff}
    \small
    \begin{threeparttable}
    \begingroup
    \setlength{\tabcolsep}{4pt}
    {\def\sym#1{\ifmmode^{#1}\else\(^{#1}\)\fi}
\begin{tabular}{l*{3}{c}*{3}{c}*{2}{c}}
\toprule
\multirow{2}{*}{DV: Earnings} & \multicolumn{2}{c}{Pooled} && \multicolumn{2}{c}{Pair 1} && \multicolumn{2}{c}{Pair 2} \\
\cmidrule(lr){2-3} \cmidrule(lr){5-6} \cmidrule(lr){8-9}
                    &\multicolumn{1}{c}{(1)}         &\multicolumn{1}{c}{(2)}         &&\multicolumn{1}{c}{(3)}         &\multicolumn{1}{c}{(4)}         &&\multicolumn{1}{c}{(5)}         &\multicolumn{1}{c}{(6)}         \\
\midrule
Raven Score $[0,1]$   &       2.894\sym{***}&       3.573\sym{***}&&       2.581\sym{***}&       2.537\sym{***}&&       3.206\sym{***}&       4.609\sym{***}\\
                    &     (0.796)         &     (0.893)         &&     (0.842)         &     (0.928)         &&     (1.133)         &     (1.261)         \\
Overconfidence $[-1,1]$      &                     &       2.101\sym{*}  &&                     &      -0.136         &&                     &       4.338\sym{**} \\
                    &                     &     (1.210)         &&                     &     (1.082)         &&                     &     (1.700)         \\
\midrule
Observations        &         614         &         614         &&         307         &         307         &&         307         &         307         \\
Mean of DV          &       5.184         &       5.184         &&       5.130         &       5.130         &&       5.238         &       5.238         \\
\bottomrule
\end{tabular}
}

    \endgroup
    \begin{mytablenotes}
        \item \textit{Notes:} OLS regressions. The dependent variable is the subject's realized earnings in dollars (success or failure payment for the task chosen from a pair of tasks). Raven score is normalized to $[0, 1]$. Overconfidence is defined as the difference between belief and Raven score, normalized to $[-1, 1]$. All specifications include categorical controls for gender, age, and the order of the two pairs. Columns (1)--(2) additionally include a pair fixed effect. Heteroskedasticity-robust standard errors clustered at the subject level are reported in parentheses. $^{***}$, $^{**}$, and $^{*}$ denote significance at the 1\%, 5\%, and 10\% levels (two-sided tests).
    \end{mytablenotes}
    \end{threeparttable}
\end{table}

Our model also predicts that earnings rise with cognitive ability (Proposition \ref{prop:expected-payoff}). Table \ref{tab:payoff} reports OLS regressions of earnings on cognitive ability and overconfidence. The dependent variable is the subject's realized earnings, in dollars. We find that Raven score predicts earnings in both pairs. Controlling for cognitive ability, overconfidence predicts earnings in Pair 2, but not in Pair 1: our model does not make a prediction about the effect of overconfidence on earnings, but any association may reflect a productive dimension of ability that is correlated with overconfidence but not captured by the Raven score.

\subsection{Optimality of Task Choice}\label{Sect:Optimality}

We now turn to Hypothesis \ref{hyp:ability-optimality}. To compute each subject's optimal choice in a pair of tasks that vary in complexity, we need to estimate her probability of success on each of the two tasks. We cannot simply regress observed task success on observed individual characteristics because each subject's success or failure is observed only on the task she chose, and unobserved individual characteristics can shift both task choice and task success. For example, suppose an unobserved trait, such as task-specific skill, raises both the probability that a subject chooses the more complex task and her probability of succeeding at it. Subjects who choose the more complex task then have higher unobserved task-specific skill on average. A naive regression of observed success on observed characteristics among subjects who chose the more complex task therefore overestimates the success probability that an average subject would have on the more complex task.

\vspace{0.25cm}

\begin{table}[h!]
    \centering
    \caption{Average Success Probabilities}
    \label{tab:avg_success}
    \begin{threeparttable}
    \begin{tabular}{llcc}
        \toprule
        & \multirow{2}{*}{Task} & Main Experiment    & Pilot Sessions   \\
        &                       & (With task choice) & (Without task choice) \\
        \midrule
        \noalign{\vspace{5pt}}
        \multirow{2}{*}{Pair 1} & 2S-1Ta (less complex) & 97.9\% & 98.1\% \\[5pt]
                                & 3S-3T (more complex)  & 54.0\% & 51.5\% \\[10pt]
        \multirow{2}{*}{Pair 2} & 2S-2T (less complex)       & 52.6\% & 75.7\% \\[5pt]
                                & Sequenceable (more complex) & 56.3\% & 43.2\% \\[5pt]
        \bottomrule
    \end{tabular}
    \begin{mytablenotes}
        \item \textit{Notes:} We report the average success probabilities using all observations of the four tasks. In the pilot sessions (mentioned in footnote \ref{fn:pilot}), subjects did not choose task complexity (see \ref{Sect:Pilot} for details).
    \end{mytablenotes}
    \end{threeparttable}
\end{table}

\vspace{0.25cm}

Table \ref{tab:avg_success} suggests strong endogenous selection into tasks. The table reports average success probabilities on each of the four tasks in the main experiment, where subjects chose task complexity, and in the pilot sessions, where subjects completed all the tasks without any choice. Specifically, in the pilot sessions there is a large difference in success rates between the less complex and more complex tasks in Pair 2, which disappears when subjects choose the task in the main experiment.

\clearpage

To account for endogenous selection, we estimate a random-effects probit model that leverages the panel structure of our experiment. \ref{Sect:esr} provides the full details of the model and estimation. We observe each subject choosing a task in each of two pairs, and we observe her outcome on the chosen task. The model jointly estimates the choice equation for each pair and the success equation for each task in both pairs, with a subject-level random effect that enters all equations and captures unobserved individual traits that can shift both task choice and task success. Table \ref{tab:esr-coefficients} reports estimated coefficients of the random-effects probit model. We find that the loadings on the random effect are statistically significant in both the choice and the success equations (\(p < 0.01\)), so unobserved traits have an effect on task choice and task success.\footnote{We follow, e.g., \citet{heckman2005structural} in using a latent factor structure to account for endogenous selection (see also \citealp{aakvik2005estimating}, who use a random-effects model to capture the latent factor). See \citet{heckman2006effects} for an application to a setting with multiple discrete choices and outcomes. We estimate our random-effects model by maximum likelihood with Gauss-Hermite quadrature, following, e.g., \citet{rabe2005maximum}.}

\vspace{0.5cm}

\begin{table}[htbp]
    \centering
    \caption{Random-Effects Probit Estimates}
    \label{tab:esr-coefficients}
    \begin{threeparttable}
    {\def\sym#1{\ifmmode^{#1}\else\(^{#1}\)\fi}
\begin{tabular}{l*{4}{c}*{2}{c}}
\toprule
\multirow{2}{*}{DV} &&& \multicolumn{1}{c}{Choose More Complex Task} &&& \multicolumn{1}{c}{Task Success} \\
\cmidrule(lr){4-4} \cmidrule(lr){7-7}
                    &&&\multicolumn{1}{c}{(1)}         &&&\multicolumn{1}{c}{(2)}         \\
\midrule
Raven Score $[0,1]$  &&& 2.672\sym{***} & && 2.992\sym{***} \\
                     &&& (0.714) & && (0.910) \\
Overconfidence $[-1,1]$  &&& 2.891\sym{***} & && 2.621\sym{***} \\
                     &&& (0.933) & && (0.977) \\
Loading on $\eta_i$ (random effect) &&& 1.527\sym{***} & && 1.610\sym{***} \\
                     &&& (0.226) & && (0.355) \\
\midrule
Observations        &&&         614         &&&         614         \\
Mean of DV          &&&       0.511         &&&       0.645         \\
\bottomrule
\end{tabular}
}

    \begin{mytablenotes}
        \item \textit{Notes:} We estimate the random-effects probit model using maximum likelihood (specifically using the gsem command in Stata with 20-point Gauss-Hermite quadrature for numerical integration). As described in \ref{Sect:esr}, we estimate a restricted model where coefficients are common across the two choice equations, and also across the four success equations, except for six equation-specific intercepts (the unrestricted model does not converge). Raven score is normalized to $[0, 1]$. Overconfidence is defined as the difference between belief and Raven score, normalized to $[-1, 1]$. All equations include categorical controls for gender, age, and the order of the two pairs. Bootstrap standard errors from 500 replications, in which we draw 307 subjects with replacement and re-estimate the random-effects probit model, are reported in parentheses (see footnote \ref{fn:bootstrap} for details). $^{***}$, $^{**}$, and $^{*}$ denote significance at the 1\%, 5\%, and 10\% levels (two-sided tests).
    \end{mytablenotes}
    \end{threeparttable}
\end{table}

\vspace{0.5cm}

After we estimate the random-effects probit model, we use the model estimates to derive predicted success probabilities for each subject on each of the two tasks in each pair. The model has one random effect for each subject, which enters all equations but is unobserved. To derive the predicted success probabilities, we predict each subject's random effect twice. To derive the subject's predicted success probabilities in Pair 1, we compute her first predicted random effect, which is the empirical Bayes posterior mean of her random effect given the model estimates, her task choices in both pairs, and her observed success on the chosen task only in Pair 2. To derive her predicted success probabilities in Pair 2, we compute her second predicted random effect, which uses her observed success on the chosen task in Pair 1 instead of Pair 2 (see \ref{Sect:esr-probability} for details).\footnote{The two predicted random effects are highly correlated, with a Pearson correlation of 0.89.} So, holding the model estimates fixed, we derive the subject's predicted success probabilities in each pair without using her success in that pair. Otherwise, the predicted success probabilities would be fitted in part to her own success or failure in the pair, and using them to evaluate the optimality of her choice in that pair would be circular.

\vspace{0.25cm}

Subsequently, we use the predicted success probabilities and our payment structure (see Table \ref{tab:payment}) to compute the expected payoff of each task for each subject, and we define two measures of choice optimality. \textit{Efficiency} is the ratio of the expected payoff from the chosen task to the expected payoff from the optimal task; efficiency equals one when the subject chooses optimally and is less than one otherwise. \textit{Binary optimality} is a binary indicator equal to one if the subject chose the task with the higher expected payoff. Unlike binary optimality, efficiency weights mistakes by their payoff consequences.\footnote{\ref{Sect:esr-optimality} sets out the construction procedure for both optimality measures.} We also distinguish two types of suboptimal choice: a choice is \textit{undercautious} if the subject chose the more complex task when the less complex task had a higher expected payoff, and \textit{overcautious} if she chose the less complex task when the more complex task had a higher expected payoff. Each decision therefore falls into one of four decision types: an optimal choice of the less complex task, an optimal choice of the more complex task, an undercautious choice, or an overcautious choice.

\vspace{0.25cm}

\begin{figure}[htbp]
    \centering
    \includegraphics[width=0.8\textwidth]{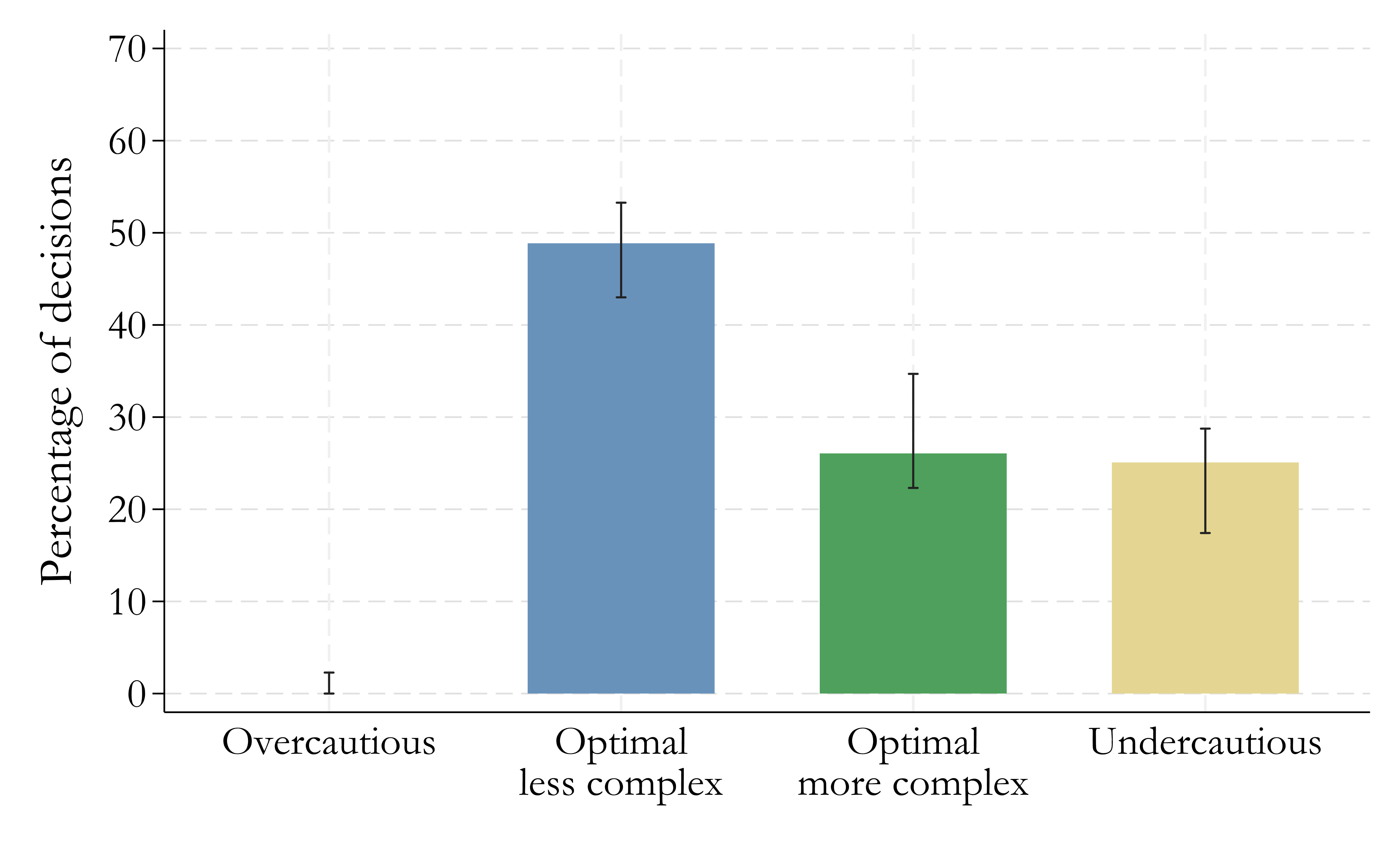}
    \caption{Decision Classification}
    \label{fig:classification}
    \begin{minipage}{0.8\textwidth}
    \vspace{0.25cm}
    \footnotesize \textit{Notes:} Bars report the share of decisions in each category. Error bars show 95\% bootstrap confidence intervals, given by the 2.5th and 97.5th percentiles across 500 bootstrap replications in which we draw 307 subjects with replacement, re-estimate the random-effects probit model, and reclassify all decisions (see footnote \ref{fn:bootstrap} for details).
    \end{minipage}
\end{figure}

\begin{figure}[htbp]
    \centering
    \includegraphics[width=0.8\textwidth]{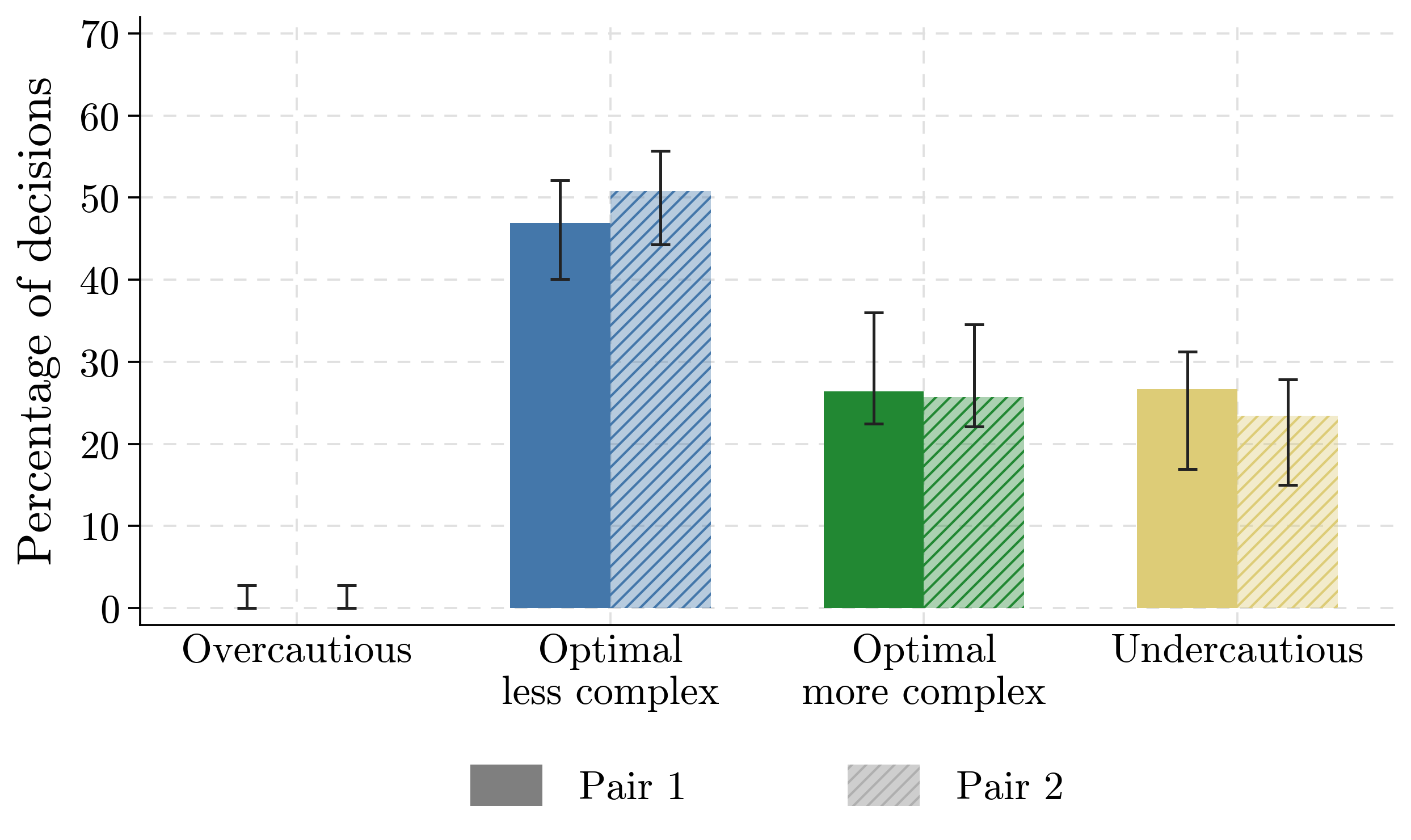}
    \caption{Decision Classification by Pair}
    \label{fig:classification-by-pair}
    \begin{minipage}{0.8\textwidth}
      \vspace{0.25cm}
      \footnotesize \textit{Notes:} Bars report the share of decisions in each category within each pair of tasks. See the notes to Figure \ref{fig:classification} for other details.
      \end{minipage}
\end{figure}

We use a single bootstrap for the random-effects probit model, the predicted random effects and predicted success probabilities, the optimality measures and decision classification, and the regressions of the optimality measures on cognitive ability and overconfidence that we report later in this subsection.\footnote{Specifically, in each of 500 bootstrap replications, we draw 307 subjects at random with replacement, re-estimate the random-effects probit model on the redrawn sample, compute each subject's two predicted random effects and her predicted success probabilities, derive the optimality measures, classify all decisions into decision types, and run the regressions of the optimality measures on cognitive ability and overconfidence. The coefficients and shares that we report are those estimated on the actual sample, and the standard errors that we report are the standard deviations of the estimates across the 500 replications.\label{fn:bootstrap}} Our bootstrap brings the uncertainty in the model estimates downstream. If the downstream analysis ignored this uncertainty, it would tend to understate the standard errors. This bootstrap gives the standard errors in Table \ref{tab:esr-coefficients}, the 95\% bootstrap confidence intervals in Figures \ref{fig:classification} and \ref{fig:classification-by-pair}, and the standard errors in Tables \ref{tab:efficiency} and \ref{tab:optimality}.

\vspace{0.5cm}

\vspace{0.5cm}

Figure \ref{fig:classification} shows the distribution of decision types, pooling the decisions from both pairs. About 75\% of decisions are optimal: in roughly 50\% of decisions subjects optimally choose the less complex task, and in roughly 25\% of decisions they optimally choose the more complex task. Those who fail to choose optimally are undercautious, choosing the more complex task when the less complex task had a higher expected payoff.

\vspace{0.25cm}

Figure \ref{fig:classification-by-pair} shows the distribution of decision types separately for each of the two pairs. In each pair, roughly half of the decisions are optimal choices of the less complex task, and about 25\% of the decisions are optimal choices of the more complex task. Similarly, in both pairs, those who fail to choose optimally pick tasks that are too complex.

\vspace{0.5cm}

\vspace{0.5cm}

Table \ref{tab:efficiency} reports OLS regressions of efficiency on cognitive ability and overconfidence. We find some weak evidence consistent with Hypothesis \ref{hyp:ability-optimality}: the coefficient on Raven score is positive in every specification, so higher cognitive ability is associated with more efficient choice, a continuous measure that accounts for the payoff consequences of mistakes. Only the Pair 1 estimates in columns (3) and (4) are statistically significant, at the 10\% level (two-sided tests).

\begin{table}[h!]
    \centering
    \caption{Effect of Cognitive Ability and Overconfidence on Efficiency}
    \label{tab:efficiency}
    \small
    \begin{threeparttable}
    {\def\sym#1{\ifmmode^{#1}\else\(^{#1}\)\fi}
\begin{tabular}{l*{3}{c}*{3}{c}*{2}{c}}
\toprule
\multirow{2}{*}{DV: Efficiency} & \multicolumn{2}{c}{Pooled} && \multicolumn{2}{c}{Pair 1} && \multicolumn{2}{c}{Pair 2} \\
\cmidrule(lr){2-3} \cmidrule(lr){5-6} \cmidrule(lr){8-9}
                    &\multicolumn{1}{c}{(1)}         &\multicolumn{1}{c}{(2)}         &&\multicolumn{1}{c}{(3)}         &\multicolumn{1}{c}{(4)}         &&\multicolumn{1}{c}{(5)}         &\multicolumn{1}{c}{(6)}         \\
\midrule
Raven Score $[0,1]$  & 0.124 & 0.121 & & 0.172\sym{*} & 0.173\sym{*} & & 0.075 & 0.070 \\
                     & (0.075) & (0.077) & & (0.096) & (0.100) & & (0.082) & (0.088) \\
Overconfidence $[-1,1]$  &  & -0.007 & &  & 0.001 & &  & -0.015 \\
                     &  & (0.092) & &  & (0.104) & &  & (0.103) \\
\midrule
Observations  & 614 & 614 & & 307 & 307 & & 307 & 307 \\
Mean of DV  & 0.912 & 0.912 & & 0.908 & 0.908 & & 0.916 & 0.916 \\
\bottomrule
\end{tabular}
}

    \begin{mytablenotes}
        \item \textit{Notes:} OLS regressions. The dependent variable ``efficiency" is the ratio of the expected payoff from the chosen task to the expected payoff from the optimal task. Expected payoffs are computed using subject-specific success probabilities estimated from the random-effects probit model (see \ref{Sect:esr}). Raven score is normalized to $[0, 1]$. Overconfidence is defined as the difference between belief and Raven score, normalized to $[-1, 1]$. All specifications include categorical controls for gender, age, and the order of the two pairs. Columns (1)--(2) additionally include a pair fixed effect. Bootstrap standard errors from 500 replications are reported in parentheses. In each replication, we draw 307 subjects with replacement, re-estimate the random-effects probit model, derive the optimality measures, and run these regressions (see footnote \ref{fn:bootstrap} for details). $^{***}$, $^{**}$, and $^{*}$ denote significance at the 1\%, 5\%, and 10\% levels (two-sided tests).
    \end{mytablenotes}
    \end{threeparttable}
\end{table}

\begin{table}[h!]
    \centering
    \caption{Effect of Cognitive Ability and Overconfidence on Binary Optimality}
    \label{tab:optimality}
    \small
    \begin{threeparttable}
    {\def\sym#1{\ifmmode^{#1}\else\(^{#1}\)\fi}
\begin{tabular}{l*{3}{c}*{3}{c}*{2}{c}}
\toprule
\multirow{2}{*}{DV: Binary Optimality} & \multicolumn{2}{c}{Pooled} && \multicolumn{2}{c}{Pair 1} && \multicolumn{2}{c}{Pair 2} \\
\cmidrule(lr){2-3} \cmidrule(lr){5-6} \cmidrule(lr){8-9}
                    &\multicolumn{1}{c}{(1)}         &\multicolumn{1}{c}{(2)}         &&\multicolumn{1}{c}{(3)}         &\multicolumn{1}{c}{(4)}         &&\multicolumn{1}{c}{(5)}         &\multicolumn{1}{c}{(6)}         \\
\midrule
Raven Score $[0,1]$  & 0.091 & 0.008 & & 0.158 & 0.076 & & 0.023 & -0.060 \\
                     & (0.125) & (0.134) & & (0.160) & (0.167) & & (0.154) & (0.169) \\
Overconfidence $[-1,1]$  &  & -0.256 & &  & -0.254 & &  & -0.258 \\
                     &  & (0.168) & &  & (0.202) & &  & (0.182) \\
\midrule
Observations  & 614 & 614 & & 307 & 307 & & 307 & 307 \\
Mean of DV  & 0.749 & 0.749 & & 0.733 & 0.733 & & 0.765 & 0.765 \\
\bottomrule
\end{tabular}
}

    \begin{mytablenotes}
        \item \textit{Notes:} OLS regressions. The dependent variable ``binary optimality" is an indicator for choosing the task with the higher expected payoff. Expected payoffs are computed using subject-specific success probabilities estimated from the random-effects probit model (see \ref{Sect:esr}). Raven score is normalized to $[0, 1]$. Overconfidence is defined as the difference between belief and Raven score, normalized to $[-1, 1]$. All specifications include categorical controls for gender, age, and the order of the two pairs. Columns (1)--(2) additionally include a pair fixed effect. Bootstrap standard errors from 500 replications are reported in parentheses. In each replication, we draw 307 subjects with replacement, re-estimate the random-effects probit model, derive the optimality measures, and run these regressions (see footnote \ref{fn:bootstrap} for details). $^{***}$, $^{**}$, and $^{*}$ denote significance at the 1\%, 5\%, and 10\% levels (two-sided tests).
    \end{mytablenotes}
    \end{threeparttable}
\end{table}

\vspace{0.25cm}

Table \ref{tab:optimality} reports OLS regressions of binary optimality on cognitive ability and overconfidence. Binary optimality is a coarse measure that weights all mistakes equally regardless of their payoff consequences. The coefficient on Raven score is positive in every specification except column (6), where it is slightly negative. However, it is not statistically significant in any specification.

\vspace{0.25cm}

\begin{result}
    We find some weak evidence that more cognitively able subjects make more efficient choices of task complexity.
\end{result}

\clearpage

Recall that we have no hypothesis linking overconfidence to optimality. When we add overconfidence in columns (2), (4), and (6) of Table \ref{tab:optimality}, overconfidence is negatively associated with binary optimality, but the association is never statistically significant. The negative association between overconfidence and binary optimality is driven by undercautious overconfident subjects who choose the more complex task when the less complex task offers a higher expected payoff (we know this from Figure \ref{fig:classification}, which shows that all subjects who fail to choose optimally are undercautious). However, in columns (2), (4), and (6) of Table \ref{tab:efficiency}, overconfidence is not associated with efficiency.

\subsection{Summary of Hypothesis Testing}

Taken together, our results support our first two pre-registered hypotheses, and provide some weak support for the third. Cognitive ability predicts the choice of the more complex task in both pairs (Hypothesis \ref{hyp:ability-choice}; Result 1), and overconfidence predicts the choice of the more complex task in both pairs, conditional on cognitive ability (Hypothesis \ref{hyp:confidence-choice}; Result 2). Cognitive ability is also weakly associated with the efficiency of task choice (Hypothesis \ref{hyp:ability-optimality}; Result 3). In addition, consistent with Proposition \ref{prop:expected-payoff}, we find that cognitive ability predicts realized task earnings in both pairs. These findings establish that cognitive ability and overconfidence are independent predictors of the choice of task complexity.

\section{Conclusion}\label{Sect:Conclusion}

Our paper makes three contributions. First, we extend the literature on choice of task complexity by linking cognitive ability and overconfidence jointly to selection between complex tasks that differ in their success-payment tradeoff. Second, we provide some weak evidence that cognitive ability is associated with the efficiency of complexity choices, complementing prior work on cognition and decision quality in fixed-complexity settings. Third, we document a sharp asymmetry: when subjects err, they overreach by choosing tasks that are too complex, with direct implications for how organizations should design feedback over self-selected task menus.

These findings suggest that self-selected task menus can work well, but that organizations should design task menus, incentives for choosing more complex tasks, and feedback structures together. A menu that lets workers choose task complexity can allocate able workers to demanding tasks, and rewards for completing more complex tasks can make those choices attractive. At the same time, workers may misjudge their own ability or their probability of success on more complex tasks. When rewards rise mechanically with task complexity, overambitious workers may therefore choose tasks for which the expected payoff is lower. Organizations can preserve the benefits of self-selection by pairing complex-task choices with clear information and feedback, such as historical success rates for comparable tasks, low-stakes trial assignments before full responsibility, decision aids that make the payoff tradeoff salient, and post-project reviews that distinguish poor task choice from poor task execution.

Our analysis suggests directions for future work. One direction is strategic interaction. In many workplaces, the complexity of tasks chosen by coworkers, competitors, or team members directly affects a worker's own payoffs. Embedding choices of task complexity in cooperative or competitive environments could therefore connect our results to team assignment, promotion systems, and organizational incentive design. A second direction concerns dynamic settings with learning and skill accumulation. People often face repeated decisions about task complexity, observe feedback, and invest in skills, so it would be valuable to understand when feedback corrects overreach and when it instead reinforces miscalibrated confidence.

\clearpage

\renewcommand{\refname}{References}
\fontsize{11}{10}
\begin{spacing}{1.36}
\interlinepenalty=10000
\bibliographystyle{experiment}
\bibliography{Literature}
\end{spacing}

\clearpage
\appendix

\thispagestyle{empty}

\text{}
\vspace{3cm}
\begin{center}

\textbf{\Huge{Appendix A:}}

\vspace{0.5cm}

\textbf{\Huge{Supplemental Appendix}}

\vspace{1cm}

\LARGE{(Intended for Online Publication)}

\end{center}

\clearpage

\setcounter{table}{0}
\renewcommand{\thetable}{A.\arabic{table}}

\setcounter{figure}{0}
\renewcommand{\thefigure}{A.\arabic{figure}}

\setcounter{section}{0}
\renewcommand{\thesection}{Appendix A.\arabic{section}}
\renewcommand{\thesubsection}{Appendix A.\arabic{section}.\arabic{subsection}}

\pagenumbering{arabic}
\renewcommand*{\thepage}{Appendix A, page \arabic{page}}

\setcounter{proposition}{0}
\renewcommand{\theproposition}{A.\arabic{proposition}}

\section{Additional Theoretical Analysis}\label{Sect:Additional-Model}\

\subsection{Proof of Proposition 1}\label{subsec:proof-prop1}

\begin{proof}

First, we record the parameter sign restrictions on $\alpha$, $\beta_1$, $\beta_2$, and $\beta_3$ implied by \textbf{A-1} through \textbf{A-4}. We use these restrictions throughout the proof.

Assumption \textbf{A-2} requires $\beta_1 + \beta_3 c_t < 0$ for all $c_t \in [0, 1]$. Evaluating at $c_t = 0$ yields the necessary sign condition
\begin{equation*}
    \beta_1 < 0.
\end{equation*}

Assumption \textbf{A-1} requires $\beta_2 + \beta_3 a_i > 0$ for all $a_i \in [0, 1]$. Evaluating at $a_i = 0$ yields the necessary sign condition
\begin{equation*}
    \beta_2 > 0.
\end{equation*}

Because \(P_m(a_i, c_t)\) is bilinear and, by \textbf{A-1} and \textbf{A-2}, strictly increasing in \(c_t\) for every \(a_i \in [0,1]\) and strictly decreasing in \(a_i\) for every \(c_t \in [0,1]\), the maximum of \(P_m\) over \([0,1]^2\) is attained at \((a_i, c_t) = (0, 1)\) and the minimum at \((a_i, c_t) = (1, 0)\). Assumption \textbf{A-4} thus reduces to two conditions: $P_m(0, 1) = \alpha + \beta_2 \leq 1$ and $P_m(1, 0) = \alpha + \beta_1 \geq 0$.

Combining $\alpha + \beta_1 \geq 0$ with $\beta_1 < 0$ gives $\alpha \geq -\beta_1 > 0$. Combining $\alpha + \beta_2 \leq 1$ with $\beta_2 > 0$ gives $\alpha \leq 1 - \beta_2 < 1$. Hence, the necessary condition
\begin{equation*}
    \alpha \in (0, 1).
\end{equation*}

Finally, assumption \textbf{A-3} states directly that
\begin{equation*}
    \beta_3 \leq 0.
\end{equation*}

To find the unconstrained optimal complexity level, we derive the first-order and second-order conditions for expected payoff with respect to \(c_t\). Differentiating Equation (\ref{eqn:linear-maximization-problem-new}), we get:
\begin{equation}
    \begin{aligned}
        \frac{\partial \pi}{\partial c_t}  &= k \Big[ 1 - P_m(a_i, c_t) - c_t \frac{\partial P_m(a_i, c_t)}{\partial c_t} \Big] \\
        &= k \Big[ 1 - \big( \alpha + \beta_1 a_i + \beta_2 c_t + \beta_3 a_i c_t \big) - c_t \big( \beta_2 + \beta_3 a_i \big) \Big] \\
        &= k \Big[ 1 - \alpha - \beta_1 a_i - 2\beta_2 c_t - 2\beta_3 a_i c_t \Big]  \\
        & \\
        \frac{\partial^2 \pi}{\partial c_t^2} &= -2k \big( \beta_2 + \beta_3 a_i \big)  \ < 0 \ \text{by} \ \textbf{A-1}
    \end{aligned}
    \label{eqn:linear-single-foc-soc-new}
\end{equation}

To determine the unconstrained optimal complexity level \(c^u\), we set the first-order condition to zero. This yields the following solution for \(c^u\):
\begin{equation}
    c^u = \frac{1 - \alpha - \beta_1 a_i}{2\big(\beta_2 + \beta_3 a_i\big)}
\end{equation}
By \textbf{A-1}, \(\beta_2 + \beta_3 a_i > 0\); and since \(\alpha < 1\) and \(\beta_1 a_i \leq 0\) (from \(\beta_1 < 0\) by \textbf{A-2} and \(a_i \geq 0\)), \(1 - \alpha - \beta_1 a_i > 0\). Hence \( c^u > 0 \). By concavity, the global optimal solution \(c^*\) is
\begin{equation}
    c^* = \min \Big\{ c^u, 1 \Big\}
\end{equation}

To investigate how the unconstrained optimal complexity level $c^u$ is affected by cognitive ability, we differentiate $c^u$ with respect to $a_i$:
\begin{equation}
    \frac{d c^u}{d a_i} = \frac{-\beta_1(2 \beta_2 + 2 \beta_3 a_i) - (1 - \alpha - \beta_1 a_i)(2\beta_3)}{(2 \beta_2 + 2 \beta_3 a_i)^2}
    = \frac{M}{(2\beta_2 + 2\beta_3 a_i)^2}
    \label{eqn:single-complexity-statics}
\end{equation}
where $M \equiv -2\beta_1\beta_2 - 2\beta_3(1-\alpha)$. The denominator is strictly positive by \textbf{A-1}, so $d c^u/d a_i > 0$ if and only if $M > 0$. Because \(\beta_1 < 0\), \(\beta_2 > 0\), \(\alpha \in (0,1)\), and \(\beta_3 \leq 0\), \(M\) is strictly positive.

Therefore, $c^u$ strictly increases in $a_i$. Whenever $c^u(a_i) < 1$, $c^*$ tracks $c^u$ and strictly increases with $a_i$; once $c^u(a_i) \geq 1$, the constraint binds and $c^* = 1$.

The proof uses \textbf{A-3} in one place only, to show that $M > 0$. Therefore, \textbf{A-3} is a sufficient condition for Proposition \ref{prop:optimal-complexity}, and Proposition \ref{prop:optimal-complexity} continues to hold as long as \(\beta_3 < -\beta_1 \beta_2/(1-\alpha)\), so that M remains strictly positive.

\end{proof}

\subsection{Proof of Proposition 2}\label{subsec:proof-prop2}

\begin{proof}

Let \(c_L\) and \(c_H\) denote the complexity levels of the less complex and more complex tasks in the binary choice, with \(0 \leq c_L < c_H \leq 1\). Define the expected-payoff difference between the more complex task and the less complex task as
\begin{equation*}
    \Delta(a_i) \equiv \pi(a_i,c_H)-\pi(a_i,c_L).
\end{equation*}
Using Equation (\ref{eqn:linear-maximization-problem-new}), we get
\begin{equation}
\begin{aligned}
    \Delta(a_i)
    &= k c_H \big(1-P_m(a_i,c_H)\big)
       - k c_L \big(1-P_m(a_i,c_L)\big) \\
    &= k c_H\Big[1-\alpha-\beta_1 a_i-\beta_2 c_H-\beta_3 a_i c_H\Big]
       - k c_L\Big[1-\alpha-\beta_1 a_i-\beta_2 c_L-\beta_3 a_i c_L\Big] \\
    &= k\Big[(c_H-c_L)(1-\alpha-\beta_1 a_i)
       -(\beta_2+\beta_3 a_i)(c_H^2-c_L^2)\Big] \\
    &= k(c_H-c_L)
       \Big[1-\alpha-\beta_2(c_H+c_L)
       -a_i\big(\beta_1+\beta_3(c_H+c_L)\big)\Big].
\end{aligned}
\end{equation}
Differentiating with respect to cognitive ability gives
\begin{equation}
    \frac{d\Delta(a_i)}{d a_i}
    =
    -k(c_H-c_L)\big(\beta_1+\beta_3(c_H+c_L)\big).
\end{equation}
By \textbf{A-2}, we have \(\beta_1<0\) (see proof of Proposition 1). By \textbf{A-3}, we have \(\beta_3\leq 0\). Since \(c_H+c_L > 0\), it follows that
\begin{equation}
    \beta_1+\beta_3(c_H+c_L)<0.
    \label{eqn:binary-complexity}
\end{equation}
Because \(k>0\) and \(c_H-c_L>0\), we therefore have
\begin{equation*}
    \frac{d\Delta(a_i)}{d a_i}>0.
\end{equation*}
Hence, the expected-payoff difference between the more complex task and the less complex task is strictly increasing in cognitive ability, giving the result.

Note that \textbf{A-3} is a sufficient condition for Proposition \ref{prop:binary-complexity}, which continues to hold as long as \(\beta_3 < -\frac{\beta_1}{c_H+c_L}\), so that Equation (\ref{eqn:binary-complexity}) continues to hold.

\end{proof}

\newpage

\section{Random-Effects Probit Model and Estimation}\label{Sect:esr}

This appendix details the six-equation random-effects probit model introduced in Section \ref{Sect:Optimality}. We use the estimated model to compute subject-specific success probabilities on each task, which we then use to construct the optimality measures used in Section \ref{Sect:Optimality}.

\subsection{Specification}\label{Sect:esr-specification}

For subject \(i\) and pair of tasks \(p \in \{1, 2\}\), let \(\text{choice}_{ip} \in \{0, 1\}\) indicate the choice of the less complex or more complex task. For Pair \(p\), let \(\text{success}_{ip}^1 \in \{0, 1\} \) denote the success indicator if subject \(i\) chose the more complex task (observed only when \(\text{choice}_{ip} = 1\)), and let \(\text{success}_{ip}^0 \in \{0, 1\} \) denote the corresponding success indicator if she chose the less complex task (observed only when \(\text{choice}_{ip} = 0\)).

Let \(X_i\) collect the subject's time-invariant covariates: the Raven score, overconfidence, gender, and age. Let \(Z_{ip} \in \{0, 1\} \) be an indicator equal to one if the random order assigned pair $p$ first to subject $i$. Let \(\eta_i\) denote a subject-level latent factor (random effect) capturing unobserved heterogeneity that can affect both the choice of task and success conditional on task choice. We assume $\eta_i \sim \mathcal{N}(0, 1)$ (with unit variance normalized for identification) and independent of \(X_i\) and \(Z_{ip}\).

The random-effects model consists of six probit equations estimated jointly: one choice equation for each pair, and one success equation for each of the four tasks. We estimate a restricted model in which the coefficients and the loading on \(\eta_i\) are shared within the two choice equations and within the four success equations. We allow the intercept to differ across all six equations to reflect the level differences in task choice and in task success.\footnote{In principle each equation could carry its own coefficients and its own loading on \(\eta_i\). We impose these restrictions because the unrestricted model fails to converge.} Superscripts 1 and 0 denote the more complex and the less complex task within a pair, and the superscripts \(\text{Sel}\) and \(\text{Suc}\) denote parameters of the selection (choice) and success equations.

\begin{align}
    \text{choice}_{i1} = 1 \iff& \; \alpha_{1}^{\text{Sel}} + X_i' \gamma^{\text{Sel}} + Z_{i1} \delta^{\text{Sel}} + \lambda^{\text{Sel}} \eta_i + \varepsilon_{i1}^{\text{Sel}} > 0, \\
    \text{choice}_{i2} = 1 \iff& \; \alpha_{2}^{\text{Sel}} + X_i' \gamma^{\text{Sel}} + Z_{i2} \delta^{\text{Sel}} + \lambda^{\text{Sel}} \eta_i + \varepsilon_{i2}^{\text{Sel}} > 0, \\
    \text{success}_{i1}^1 = 1 \iff& \; \alpha_{1}^1 + X_i' \gamma^{\text{Suc}} + Z_{i1} \delta^{\text{Suc}} + \lambda^{\text{Suc}} \eta_i + \varepsilon_{i1}^1 > 0, \\
    \text{success}_{i1}^0 = 1 \iff& \; \alpha_{1}^0 + X_i' \gamma^{\text{Suc}} + Z_{i1} \delta^{\text{Suc}} + \lambda^{\text{Suc}} \eta_i + \varepsilon_{i1}^0 > 0, \\
    \text{success}_{i2}^1 = 1 \iff& \; \alpha_{2}^1 + X_i' \gamma^{\text{Suc}} + Z_{i2} \delta^{\text{Suc}} + \lambda^{\text{Suc}} \eta_i + \varepsilon_{i2}^1 > 0, \\
    \text{success}_{i2}^0 = 1 \iff& \; \alpha_{2}^0 + X_i' \gamma^{\text{Suc}} + Z_{i2} \delta^{\text{Suc}} + \lambda^{\text{Suc}} \eta_i + \varepsilon_{i2}^0 > 0,
\end{align}
where the idiosyncratic shocks \(\varepsilon\) are standard normal, mutually independent, and independent of \(\eta_i\) and \(X_i\).

Table \ref{tab:esr-coefficients} in Section \ref{Sect:Optimality} reports coefficients from the estimated model.

\subsection{Predicted Success Probabilities}\label{Sect:esr-probability}

For each subject \(i\), we predict her random effect \(\eta_i\) twice (obtained in Stata via predict after gsem). To compute her predicted success probabilities in Pair 1, we use her first predicted random effect $\hat{\eta}_{i1}$, which is the empirical Bayes posterior mean of $\eta_i$ given the estimated model parameters, the subject's observed characteristics, the order of the two pairs, her task choices in both pairs, and her success on the chosen task only in Pair 2. To compute her predicted success probabilities in Pair 2, we use her second predicted random effect $\hat{\eta}_{i2}$, which uses her success on the chosen task in Pair 1 instead of Pair 2. So, holding the estimated model parameters fixed, $\hat{\eta}_{i1}$ does not use her success in Pair 1, and $\hat{\eta}_{i2}$ does not use her success in Pair 2 (we do this to avoid circularity, as described in the fourth paragraph of Section~\ref{Sect:Optimality}). Across the 307 subjects, the Pearson correlation between $\hat{\eta}_{i1}$ and $\hat{\eta}_{i2}$ is 0.89 (\(p < 0.001\)). Using the estimated model parameters, the predicted random effects, the subject's observed characteristics, and the order of the two pairs, we then compute predicted success probabilities for each task as:
  \begin{equation}
  \begin{aligned}
      \widehat{\Pr}(\text{success}_{i1}^1 = 1) &= \Phi\!\left( \hat{\alpha}_{1}^1 +  X_i' \hat{\gamma}^{\text{Suc}} + Z_{i1}
  \hat{\delta}^{\text{Suc}} + \hat{\lambda}^{\text{Suc}} \hat{\eta}_{i1} \right)\\
      \widehat{\Pr}(\text{success}_{i1}^0 = 1) &= \Phi\!\left( \hat{\alpha}_{1}^0 +  X_i' \hat{\gamma}^{\text{Suc}} + Z_{i1}
  \hat{\delta}^{\text{Suc}} + \hat{\lambda}^{\text{Suc}} \hat{\eta}_{i1} \right)\\
      \widehat{\Pr}(\text{success}_{i2}^1 = 1) &= \Phi\!\left( \hat{\alpha}_{2}^1 +  X_i' \hat{\gamma}^{\text{Suc}} + Z_{i2}
  \hat{\delta}^{\text{Suc}} + \hat{\lambda}^{\text{Suc}} \hat{\eta}_{i2} \right)\\
      \widehat{\Pr}(\text{success}_{i2}^0 = 1) &= \Phi\!\left( \hat{\alpha}_{2}^0 +  X_i' \hat{\gamma}^{\text{Suc}} + Z_{i2}
  \hat{\delta}^{\text{Suc}} + \hat{\lambda}^{\text{Suc}} \hat{\eta}_{i2} \right)\\
  \end{aligned}
  \label{eq:predicted-prob}
  \end{equation}
where \(\Phi\) is the standard normal CDF.

\subsection{Optimality Measures}\label{Sect:esr-optimality}

As described in Table \ref{tab:payment}, the success payment of the more complex task in Pair 1 is \(\$9.00\) and the success payment of the less complex task in Pair 1 is \(\$5.00\); the success payment of the more complex task in Pair 2 is \(\$11.00\) and the success payment of the less complex task in Pair 2 is \(\$6.50\). The failure payment of all tasks is \(\$1.00\). Let \(\widehat{\mathbb{E}}[\pi_{ip}^1]\) denote the expected payoff of the more complex task in Pair \(p\) for subject \(i\) and let \(\widehat{\mathbb{E}}[\pi_{ip}^0]\) denote the expected payoff of the less complex task in Pair \(p\) for subject \(i\). Based on the predicted success probabilities, the expected payoffs are calculated as follows:
\begin{equation}
\begin{aligned}
    \widehat{\mathbb{E}}[\pi_{i1}^1] &= 9 \cdot \widehat{\Pr}(\text{success}_{i1}^1 = 1) + 1 \cdot \big(1 - \widehat{\Pr}(\text{success}_{i1}^1 = 1)\big) \\
    \widehat{\mathbb{E}}[\pi_{i1}^0] &= 5 \cdot \widehat{\Pr}(\text{success}_{i1}^0 = 1) + 1 \cdot \big(1 - \widehat{\Pr}(\text{success}_{i1}^0 = 1)\big) \\
    \widehat{\mathbb{E}}[\pi_{i2}^1] &= 11 \cdot \widehat{\Pr}(\text{success}_{i2}^1 = 1) + 1 \cdot \big(1 - \widehat{\Pr}(\text{success}_{i2}^1 = 1)\big) \\
    \widehat{\mathbb{E}}[\pi_{i2}^0] &= 6.5 \cdot \widehat{\Pr}(\text{success}_{i2}^0 = 1) + 1 \cdot \big(1 - \widehat{\Pr}(\text{success}_{i2}^0 = 1)\big)
\end{aligned}
\end{equation}
The optimal task for subject \(i\) is the one with the higher predicted expected payoff. In our data, there are no ties: each subject has a unique optimal choice for each pair. The subject's choice is \emph{optimal} if her chosen task matches the optimal task, \emph{undercautious} if she chose the more complex task when the less complex task was optimal, and \emph{overcautious} if she chose the less complex task when the more complex task was optimal. Efficiency for Pair \(p\) is defined as
\begin{equation*}
    \text{efficiency}_{ip} = \frac{\widehat{\mathbb{E}}[\pi_{ip}^{\text{chosen}}]}{\max \Big\{ \widehat{\mathbb{E}}[\pi_{ip}^1], \widehat{\mathbb{E}}[\pi_{ip}^0]\Big\}},
\end{equation*}
where ``chosen'' refers to the task that subject $i$ actually chose. Efficiency equals one when the subject chooses optimally and is less than one otherwise. These are the decision-classification and efficiency measures reported in Section \ref{Sect:Optimality} and Figure \ref{fig:classification}.

\newpage

\section{Pilot Sessions}\label{Sect:Pilot}

The pilot collected data from 103 student subjects across three sessions run at the Vernon Smith Experimental Economics Laboratory (VSEEL) at Purdue University in April 2025.

Each session consisted of three parts. In Part 1, subjects completed a battery of cognitive ability measures that included the 11-item matrix-reasoning test from the International Cognitive Ability Resource \citep{dworak2021using}.\footnote{See Section \ref{Sect:Tests} for test implementation details.} Subjects earned \$0.30 for each correct answer. In Part 2, subjects worked on all 14 rule implementation tasks from \citet{oprea2020makes}, with each task completed twice in randomized order, and subjects earned \$0.70 for each task that they correctly completed. In Part 3, subjects answered a brief demographic questionnaire.

Two features of the pilot deserve explicit emphasis. First, subjects did not choose tasks in the pilot, so task assignment was exogenous. This contrasts with the main experiment, in which subjects choose between a less complex and a more complex task within each pair. Second, the visual presentation and time constraints of the matrix-reasoning test and the rule implementation tasks in the pilot are identical to those in the main experiment, and subjects received the same rule implementation task feedback as in the main experiment.

\clearpage

\section{Model Estimation}\label{Sect:Model-Estimation}

We use the pilot data (see \ref{Sect:Pilot}) to estimate the linear mistake-probability function in Equation~(\ref{eqn:linear-mistake-probability-new}) separately using each of the two task pairs from the main experiment. Pair 1 consists of 2S-1Ta (less complex) and 3S-3T (more complex). Pair 2 consists of 2S-2T (less complex) and Sequenceable (more complex). We rely on the pilot data rather than the main-experiment data because task assignment in the pilot is exogenous, which allows us to estimate the mistake-probability function without selection induced by task choice.\footnote{Subjects face each task twice in the pilot, while subjects face each task at most once in the main experiment. Therefore, to maintain consistency with the main experiment, we use subjects' first attempt at each task to estimate the model.}

Within each pair, we estimate the model using the lower complexity level in the pair ($c_{\min}$) and the higher complexity level in the pair ($c_{\max}$). We estimate the following OLS regression, using an indicator for the more complex task on the right-hand side:

\begin{equation}
    P_m(a_i, c_t) = \alpha + \beta_1 a_i + \beta_2 \mathbf{1}(c_t = c_{\max}) + \beta_3 a_i \mathbf{1}(c_t = c_{\max})
    \label{eqn:indicator}
\end{equation}
The numerical values of $c_{\min}$ and $c_{\max}$ on the $[0,1]$ scale are not identified from our data. Due to linearity, the model is invariant under positive affine transformations (with appropriately rescaled parameters). Thus, we can use an indicator for $c_{\max}$ in Equation~(\ref{eqn:indicator}). We note that our estimates should be interpreted as applying only to the interval from $c_{\min}$ to $c_{\max}$.

We estimate our model using OLS regression without imposing any parameter constraints, and then verify the model's assumptions ex post. For each pair, all four assumptions \textbf{A-1} through \textbf{A-4} are satisfied.

\clearpage

\section{Further Tables}\label{Sect:Additional-Analysis}\

\begin{table}[h!]
    \centering
    \caption{Effect of Cognitive Ability and Overconfidence on Complexity Choice \\ (Excluding the Outlier)}
    \label{tab:choice-outlier}
    \small
    \begin{threeparttable}
     \begingroup
    \setlength{\tabcolsep}{4pt}
    {\def\sym#1{\ifmmode^{#1}\else\(^{#1}\)\fi}
\begin{tabular}{l*{2}{c}*{3}{c}*{3}{c}}
\toprule
\multirow{2}{*}{DV: Choose More Complex Task} & \multicolumn{2}{c}{Pooled} & \multicolumn{3}{c}{Pair 1} & \multicolumn{3}{c}{Pair 2} \\
\cmidrule(lr){2-3} \cmidrule(lr){5-6} \cmidrule(lr){8-9}
                    &\multicolumn{1}{c}{(1)}         &\multicolumn{1}{c}{(2)}         &&\multicolumn{1}{c}{(3)}         &\multicolumn{1}{c}{(4)}         &&\multicolumn{1}{c}{(5)}         &\multicolumn{1}{c}{(6)}         \\
\midrule
Raven Score $[0,1]$   &       0.392\sym{***}&       0.551\sym{***}&&       0.326\sym{**} &       0.494\sym{***}&&       0.459\sym{***}&       0.609\sym{***}\\
                    &     (0.124)         &     (0.132)         &&     (0.147)         &     (0.160)         &&     (0.142)         &     (0.150)         \\
Overconfidence $[-1,1]$     &                     &       0.537\sym{***}&&                     &       0.567\sym{***}&&                     &       0.507\sym{**} \\
                    &                     &     (0.181)         &&                     &     (0.199)         &&                     &     (0.198)         \\
\midrule
Observations        &         612         &         612         &&         306         &         306         &&         306         &         306         \\
Mean of DV          &       0.510         &       0.510         &&       0.529         &       0.529         &&       0.490         &       0.490         \\
\bottomrule
\end{tabular}
}

    \endgroup
    \begin{mytablenotes}
        \item \textit{Notes:} We run the same OLS regressions as in Table \ref{tab:choice}, without including the one subject with overconfidence of $+9$. $^{***}$, $^{**}$, and $^{*}$ denote significance at the 1\%, 5\%, and 10\% levels (two-sided tests).
    \end{mytablenotes}
    \end{threeparttable}
\end{table}

\vspace{2cm}

\begin{table}[h!]
    \centering
    \caption{Effect of Cognitive Ability and Overconfidence on Earnings \\ (Excluding the Outlier)}
    \label{tab:payoff-outlier}
    \small
    \begin{threeparttable}
    \begingroup
    \setlength{\tabcolsep}{4pt}
    {\def\sym#1{\ifmmode^{#1}\else\(^{#1}\)\fi}
\begin{tabular}{l*{2}{c}*{3}{c}*{3}{c}}
\toprule
\multirow{2}{*}{DV: Earnings} & \multicolumn{2}{c}{Pooled} && \multicolumn{2}{c}{Pair 1} && \multicolumn{2}{c}{Pair 2} \\
\cmidrule(lr){2-3} \cmidrule(lr){5-6} \cmidrule(lr){8-9}
                    &\multicolumn{1}{c}{(1)}         &\multicolumn{1}{c}{(2)}         &&\multicolumn{1}{c}{(3)}         &\multicolumn{1}{c}{(4)}         &&\multicolumn{1}{c}{(5)}         &\multicolumn{1}{c}{(6)}         \\
\midrule
Raven Score $[0,1]$   &       2.814\sym{***}&       3.574\sym{***}&&       2.477\sym{***}&       2.538\sym{***}&&       3.151\sym{***}&       4.610\sym{***}\\
                    &     (0.803)         &     (0.894)         &&     (0.849)         &     (0.929)         &&     (1.148)         &     (1.262)         \\
Overconfidence $[-1,1]$     &                     &       2.566\sym{**} &&                     &       0.204         &&                     &       4.929\sym{***}\\
                    &                     &     (1.234)         &&                     &     (1.130)         &&                     &     (1.759)         \\
\midrule
Observations        &         612         &         612         &&         306         &         306         &&         306         &         306         \\
Mean of DV          &       5.198         &       5.198         &&       5.144         &       5.144         &&       5.252         &       5.252         \\
\bottomrule
\end{tabular}
}

    \endgroup
    \begin{mytablenotes}
        \item \textit{Notes:} We run the same OLS regressions as in Table \ref{tab:payoff}, without including the one subject with overconfidence of $+9$. $^{***}$, $^{**}$, and $^{*}$ denote significance at the 1\%, 5\%, and 10\% levels (two-sided tests).
    \end{mytablenotes}
    \end{threeparttable}
\end{table}

\clearpage

\begin{table}[h!]
    \centering
    \caption{Effect of Cognitive Ability and Overconfidence on Efficiency\\ (Excluding the Outlier)}
    \label{tab:efficiency-outlier}
    \small
    \begin{threeparttable}
    {\def\sym#1{\ifmmode^{#1}\else\(^{#1}\)\fi}
\begin{tabular}{l*{3}{c}*{3}{c}*{2}{c}}
\toprule
\multirow{2}{*}{DV: Efficiency} & \multicolumn{2}{c}{Pooled} && \multicolumn{2}{c}{Pair 1} && \multicolumn{2}{c}{Pair 2} \\
\cmidrule(lr){2-3} \cmidrule(lr){5-6} \cmidrule(lr){8-9}
                    &\multicolumn{1}{c}{(1)}         &\multicolumn{1}{c}{(2)}         &&\multicolumn{1}{c}{(3)}         &\multicolumn{1}{c}{(4)}         &&\multicolumn{1}{c}{(5)}         &\multicolumn{1}{c}{(6)}         \\
\midrule
Raven Score $[0,1]$  & 0.109 & 0.119 & & 0.158\sym{*} & 0.169\sym{*} & & 0.060 & 0.068 \\
                     & (0.074) & (0.079) & & (0.095) & (0.098) & & (0.080) & (0.089) \\
Overconfidence $[-1,1]$  &  & 0.033 & &  & 0.037 & &  & 0.029 \\
                     &  & (0.077) & &  & (0.087) & &  & (0.090) \\
\midrule
Observations  & 612 & 612 & & 306 & 306 & & 306 & 306 \\
Mean of DV  & 0.914 & 0.914 & & 0.911 & 0.911 & & 0.918 & 0.918 \\
\bottomrule
\end{tabular}
}

    \begin{mytablenotes}
        \item \textit{Notes:} We run the same OLS regressions as in Table \ref{tab:efficiency}, without including the one subject with overconfidence of $+9$. We re-estimate the random-effects probit model on the remaining 306 subjects, so the predicted success probabilities are computed without the outlier. Bootstrap standard errors from 500 replications are reported in parentheses, following footnote \ref{fn:bootstrap} but drawing 306 subjects in each replication. $^{***}$, $^{**}$, and $^{*}$ denote significance at the 1\%, 5\%, and 10\% levels (two-sided tests).
    \end{mytablenotes}
    \end{threeparttable}
\end{table}

\vspace{2cm}

\begin{table}[h!]
    \centering
    \caption{Effect of Cognitive Ability and Overconfidence on Binary Optimality\\ (Excluding the Outlier)}
    \label{tab:optimality-outlier}
    \small
    \begin{threeparttable}
    {\def\sym#1{\ifmmode^{#1}\else\(^{#1}\)\fi}
\begin{tabular}{l*{3}{c}*{3}{c}*{2}{c}}
\toprule
\multirow{2}{*}{DV: Binary Optimality} & \multicolumn{2}{c}{Pooled} && \multicolumn{2}{c}{Pair 1} && \multicolumn{2}{c}{Pair 2} \\
\cmidrule(lr){2-3} \cmidrule(lr){5-6} \cmidrule(lr){8-9}
                    &\multicolumn{1}{c}{(1)}         &\multicolumn{1}{c}{(2)}         &&\multicolumn{1}{c}{(3)}         &\multicolumn{1}{c}{(4)}         &&\multicolumn{1}{c}{(5)}         &\multicolumn{1}{c}{(6)}         \\
\midrule
Raven Score $[0,1]$  & 0.064 & 0.018 & & 0.133 & 0.095 & & -0.004 & -0.060 \\
                     & (0.129) & (0.140) & & (0.168) & (0.175) & & (0.156) & (0.172) \\
Overconfidence $[-1,1]$  &  & -0.158 & &  & -0.130 & &  & -0.186 \\
                     &  & (0.163) & &  & (0.201) & &  & (0.180) \\
\midrule
Observations  & 612 & 612 & & 306 & 306 & & 306 & 306 \\
Mean of DV  & 0.755 & 0.755 & & 0.742 & 0.742 & & 0.768 & 0.768 \\
\bottomrule
\end{tabular}
}

    \begin{mytablenotes}
        \item \textit{Notes:} We run the same OLS regressions as in Table \ref{tab:optimality}, without including the one subject with overconfidence of $+9$. We re-estimate the random-effects probit model on the remaining 306 subjects, so the predicted success probabilities are computed without the outlier. Bootstrap standard errors from 500 replications are reported in parentheses, following footnote \ref{fn:bootstrap} but drawing 306 subjects in each replication. $^{***}$, $^{**}$, and $^{*}$ denote significance at the 1\%, 5\%, and 10\% levels (two-sided tests).
    \end{mytablenotes}
    \end{threeparttable}
\end{table}

\clearpage

\begin{landscape}

\thispagestyle{empty}

\text{}
\vspace{3cm}
\begin{center}

\textbf{\Huge{Appendix B:}}

\vspace{0.5cm}

\textbf{\Huge{Experimental Screenshots}}

\vspace{1cm}

\LARGE{(Intended for Online Publication)}

\end{center}

\clearpage

\setcounter{table}{0}
\renewcommand{\thetable}{B.\arabic{table}}

\setcounter{figure}{0}
\renewcommand{\thefigure}{B.\arabic{figure}}

\setcounter{section}{0}
\renewcommand{\thesection}{Appendix B.\arabic{section}}

\pagenumbering{arabic}
\renewcommand*{\thepage}{Appendix B, page \arabic{page}}

\renewcommand{\headrulewidth}{0pt}
 \fancypagestyle{landscape}{
        \fancyhf{}
    \fancyfoot[C]{%
            \vspace*{-15.7cm}
            \makebox[\textwidth][c]{\hspace{19cm}\rotatebox{90}{\thepage}}
        }
    }                        \pagestyle{landscape}

\section*{}\label{Screen:Intro}

\begin{textblock*}{\textwidth}(0.8cm,-0.1cm)
 \rotatebox{90}{
    \includegraphics[width=1.85\textwidth]{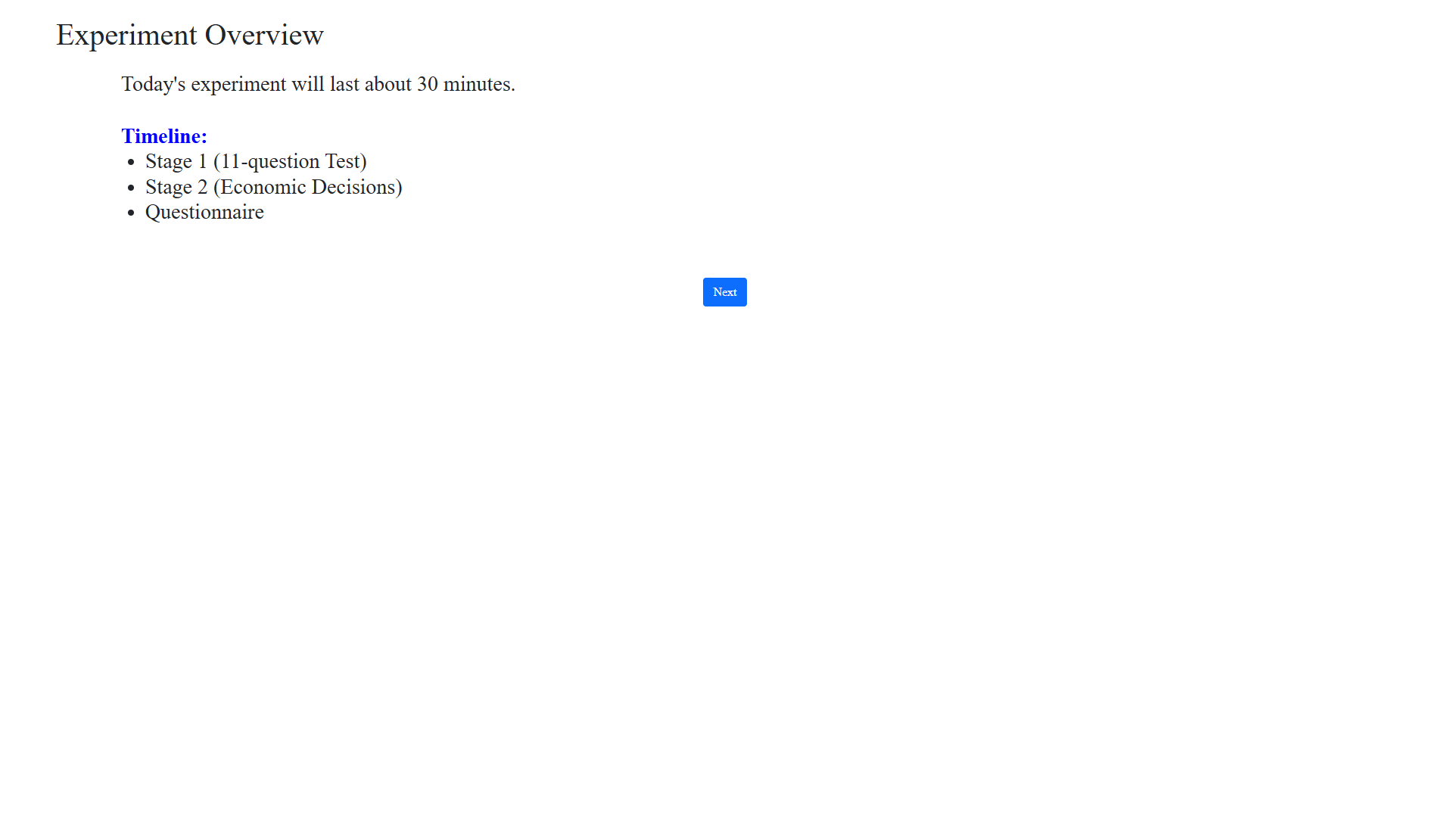}
    }
\end{textblock*}

\clearpage

\section*{}\label{Screen:Section1}

\begin{textblock*}{\textwidth}(0.8cm,-0.1cm)
 \rotatebox{90}{
    \includegraphics[width=1.85\textwidth]{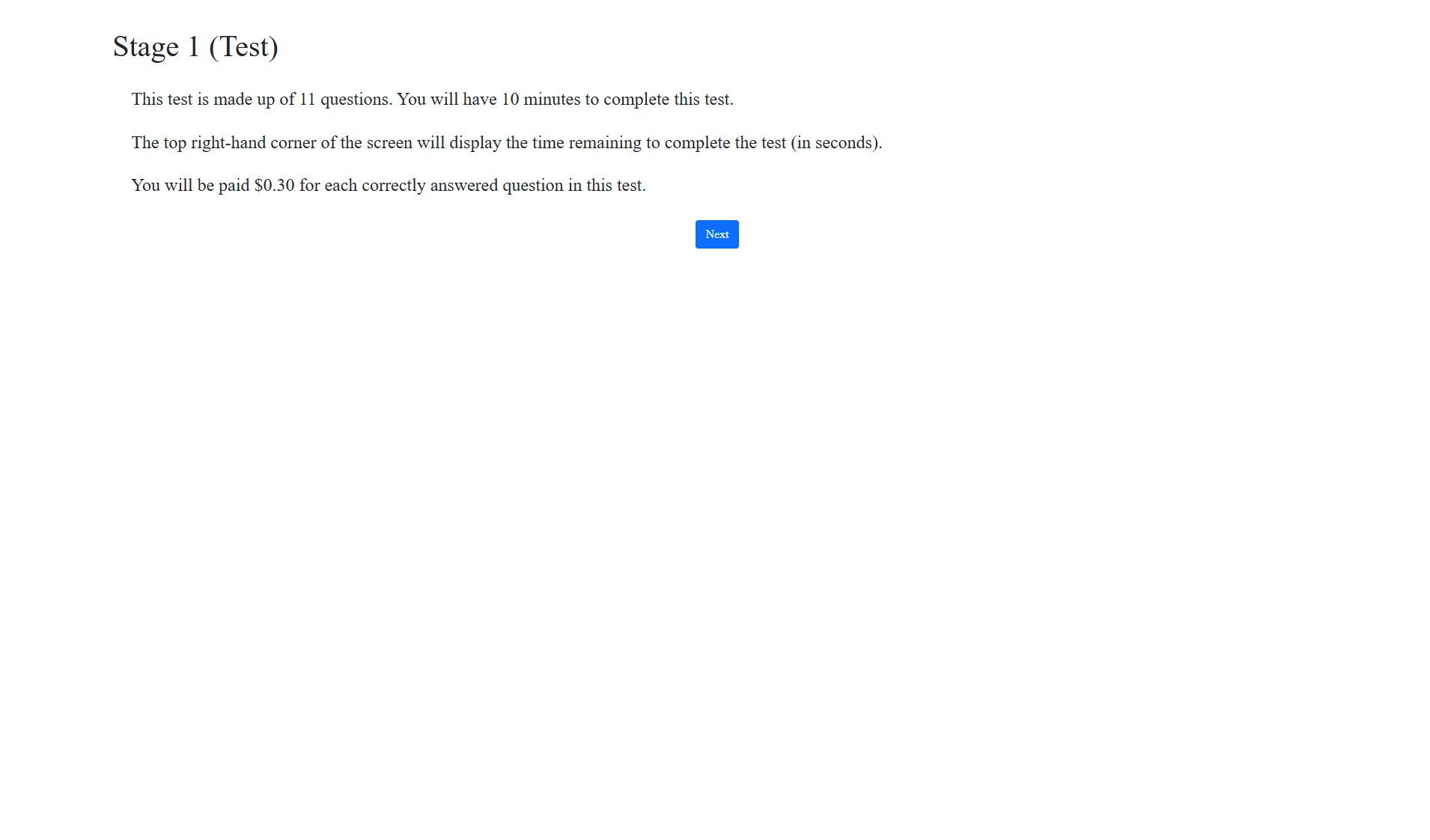}
    }
\end{textblock*}

\clearpage

\section*{}\label{Screen:Section1b}

\begin{textblock*}{\textwidth}(0.8cm,-0.1cm)
 \rotatebox{90}{
    \includegraphics[width=1.85\textwidth]{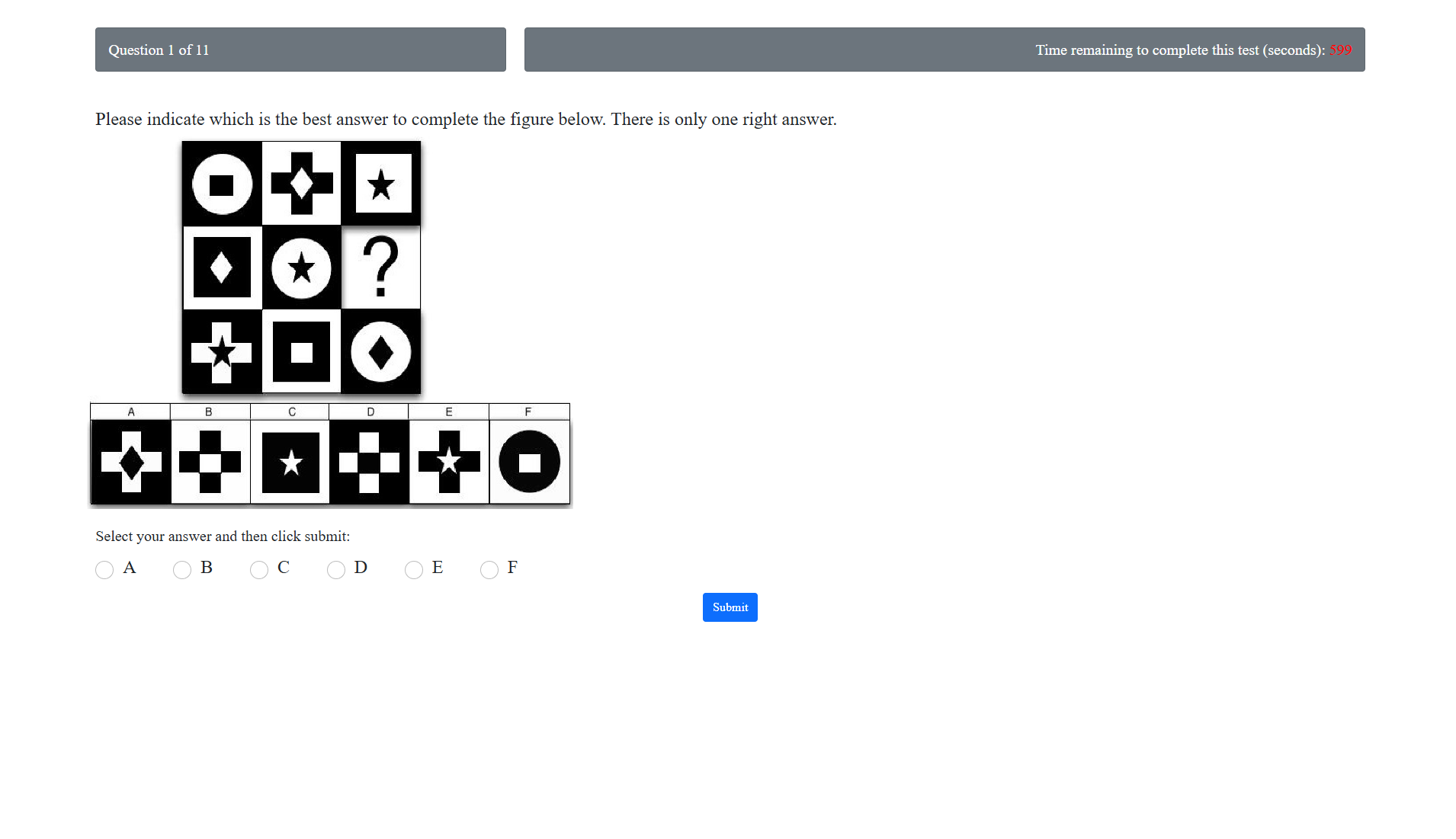}
    }
\end{textblock*}

\begin{textblock*}{\textwidth}(18cm, 18cm)
 \rotatebox{90}{
   \textit{Note: This is a sample item. See Section \ref{Sect:Tests} for details.}
    }
\end{textblock*}

\clearpage

\section*{}

\begin{textblock*}{\textwidth}(0.8cm,-0.1cm)
 \rotatebox{90}{
    \includegraphics[width=1.85\textwidth]{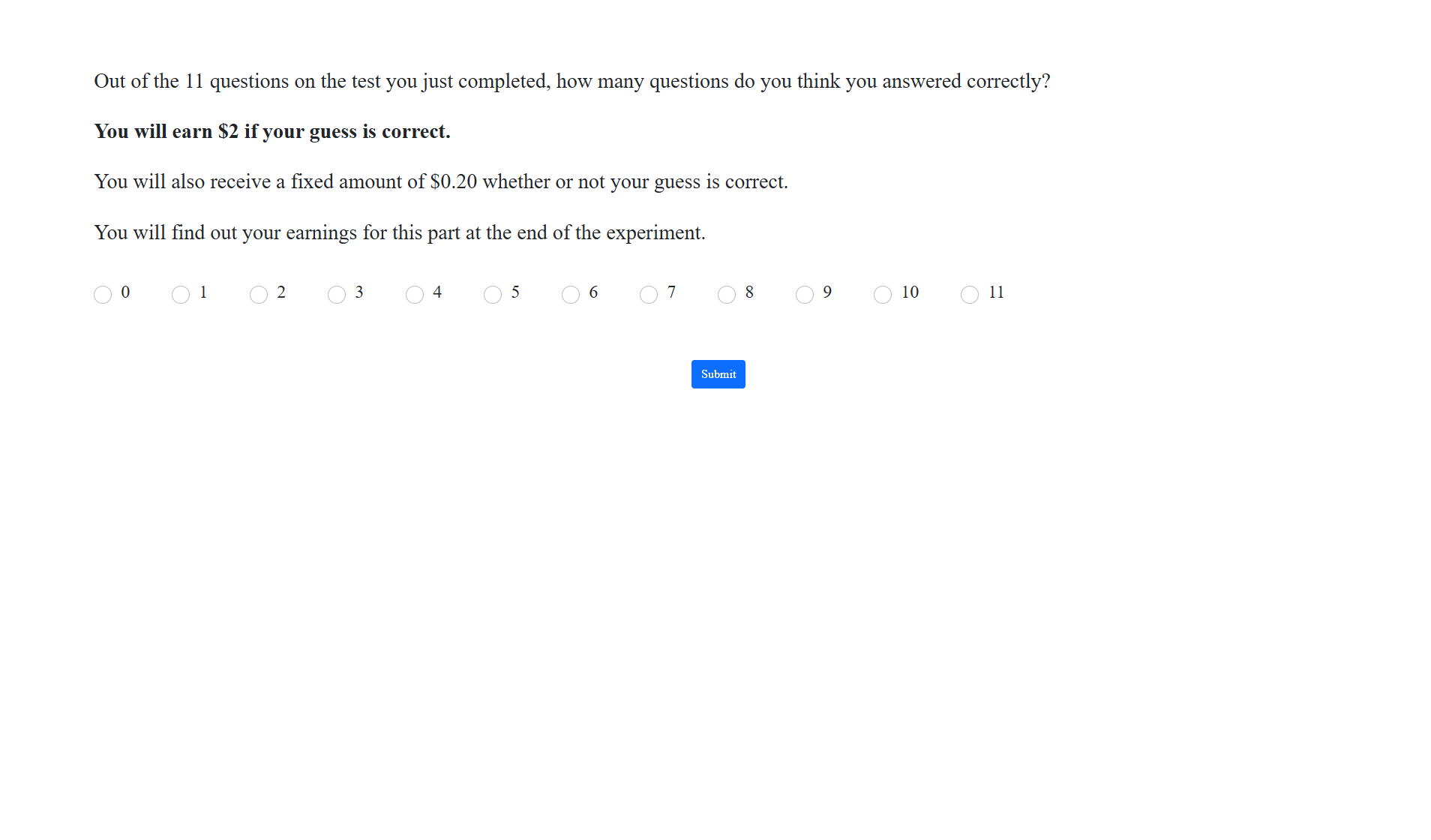}
    }
\end{textblock*}

\clearpage

\section*{}\label{Screen:Section2}

\begin{textblock*}{\textwidth}(0.8cm,-0.1cm)
 \rotatebox{90}{
    \includegraphics[width=1.85\textwidth]{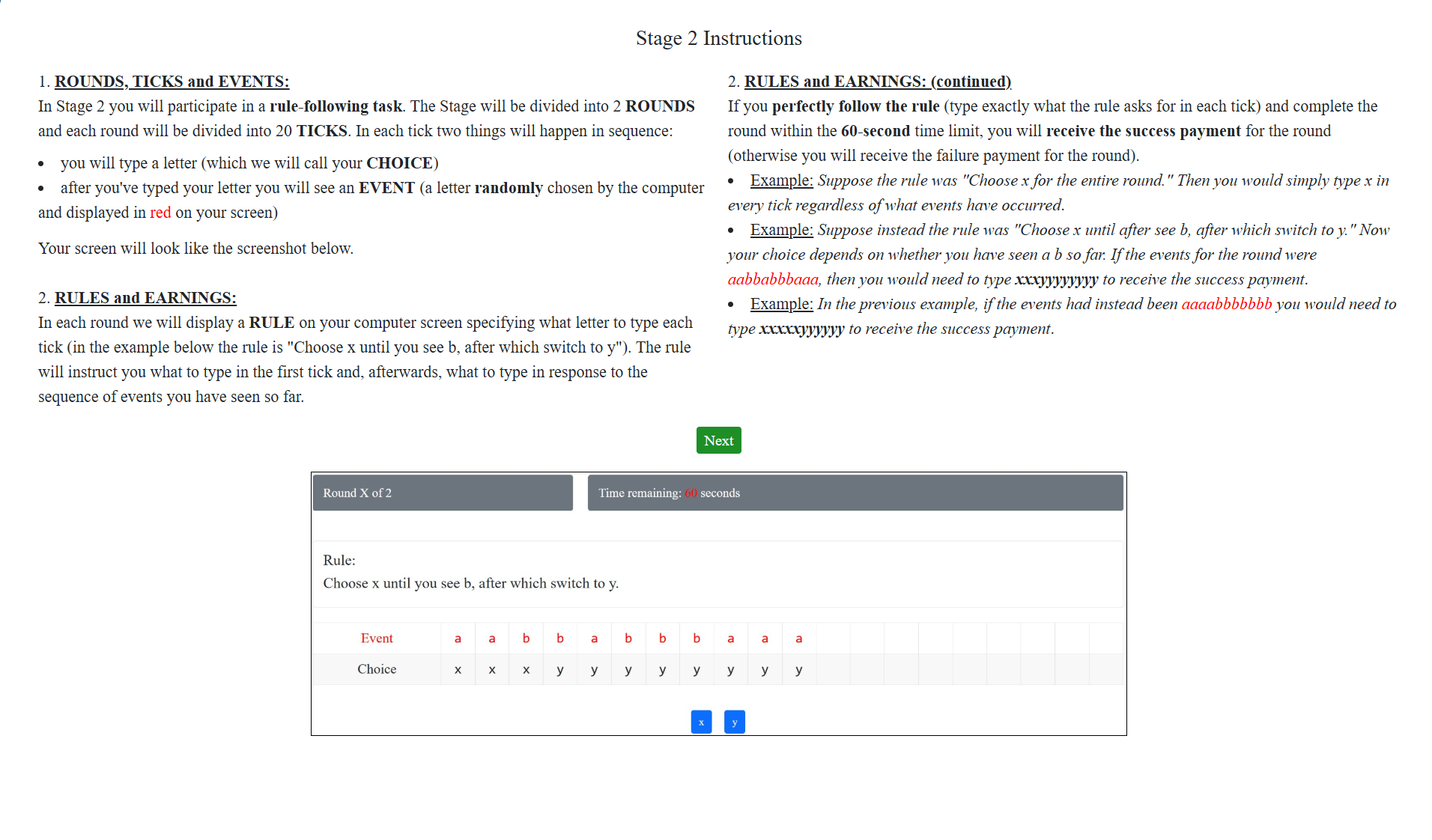}
    }
\end{textblock*}

\clearpage

\section*{}

\begin{textblock*}{\textwidth}(0.8cm,-0.1cm)
 \rotatebox{90}{
    \includegraphics[width=1.85\textwidth]{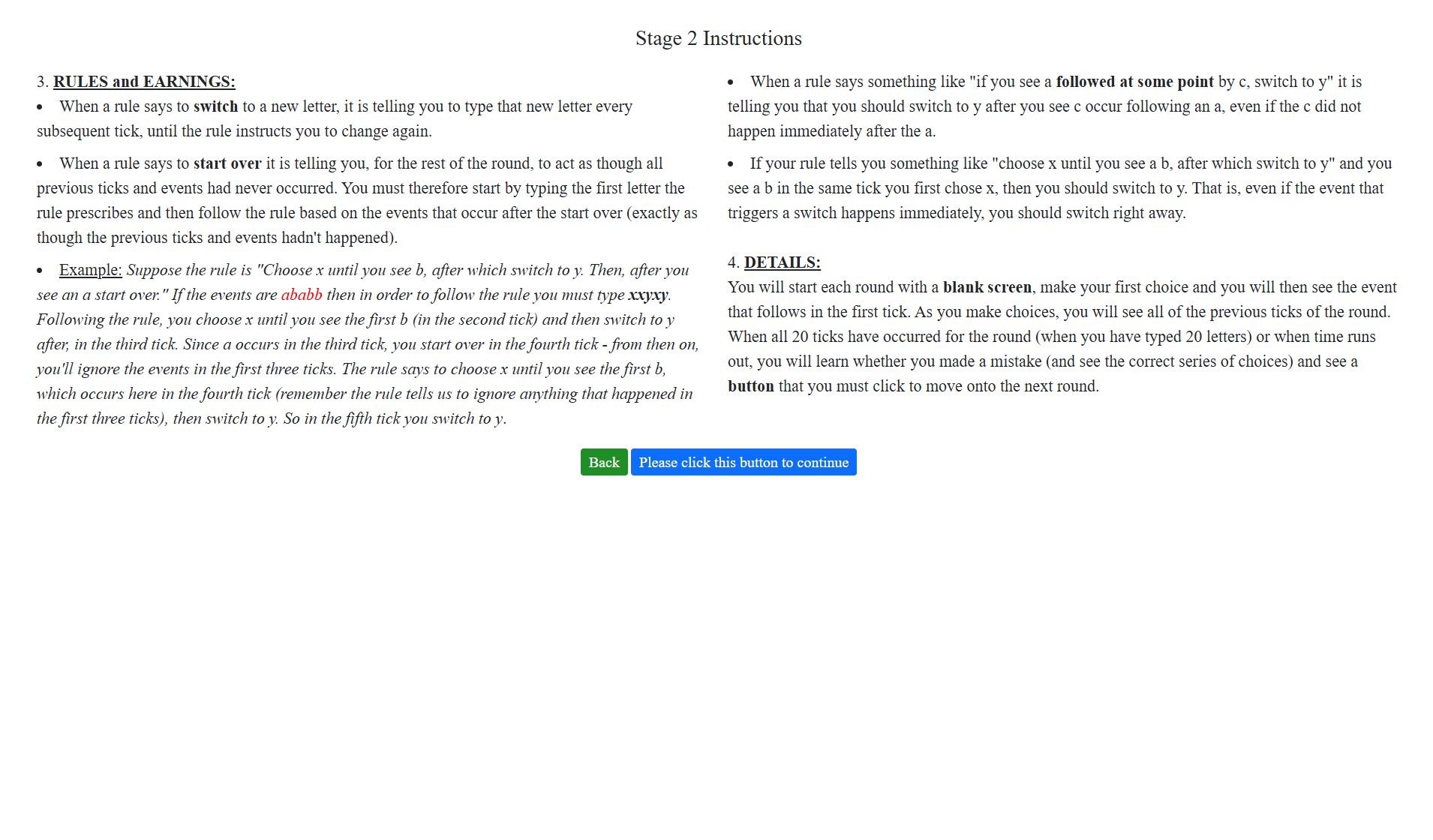}
    }
\end{textblock*}

\clearpage

\section*{}

\begin{textblock*}{\textwidth}(0.8cm,-0.1cm)
 \rotatebox{90}{
    \includegraphics[width=1.85\textwidth]{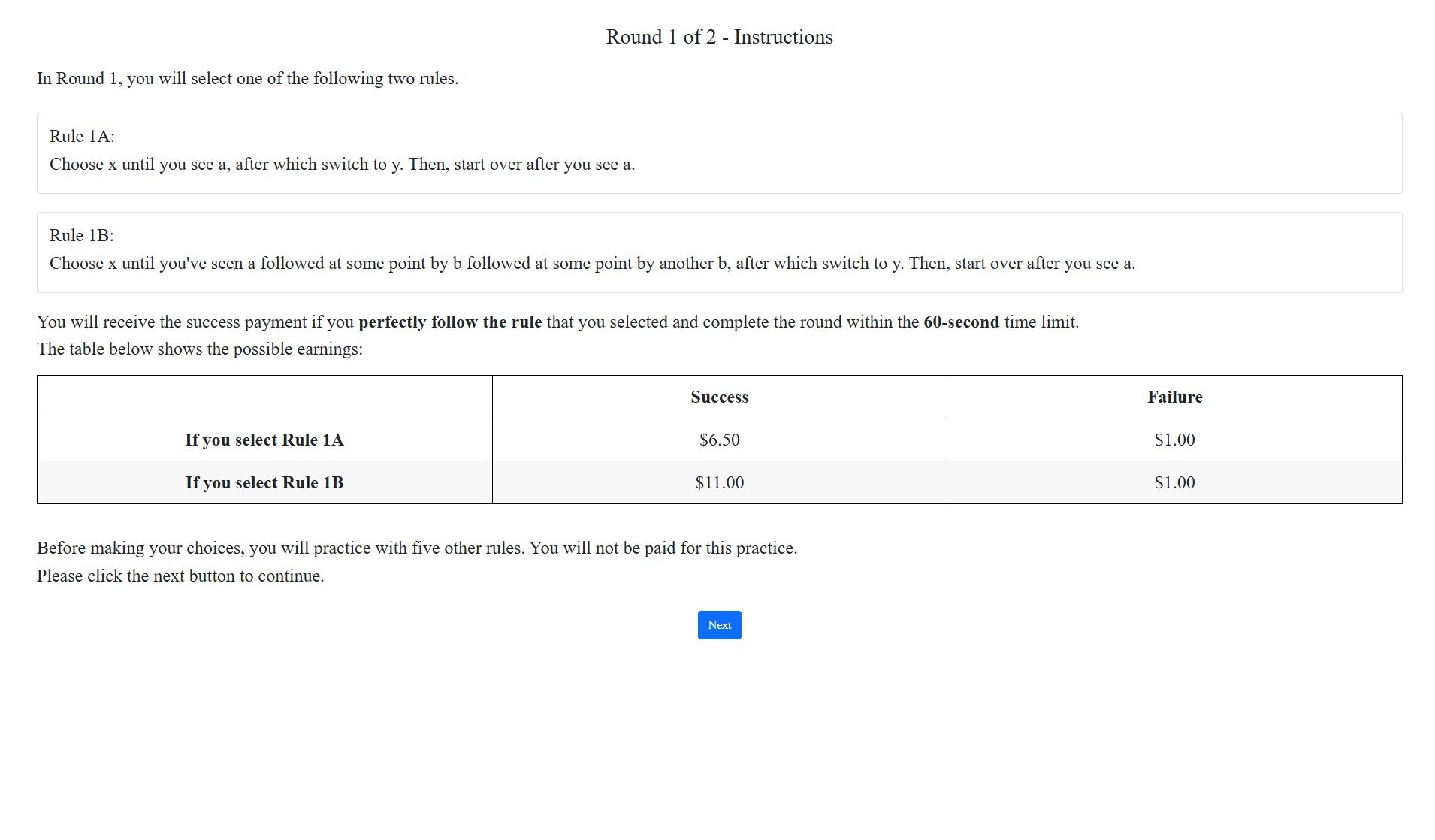}
    }
\end{textblock*}

\clearpage

\section*{}

\begin{textblock*}{\textwidth}(0.8cm,-0.1cm)
 \rotatebox{90}{
    \includegraphics[width=1.85\textwidth]{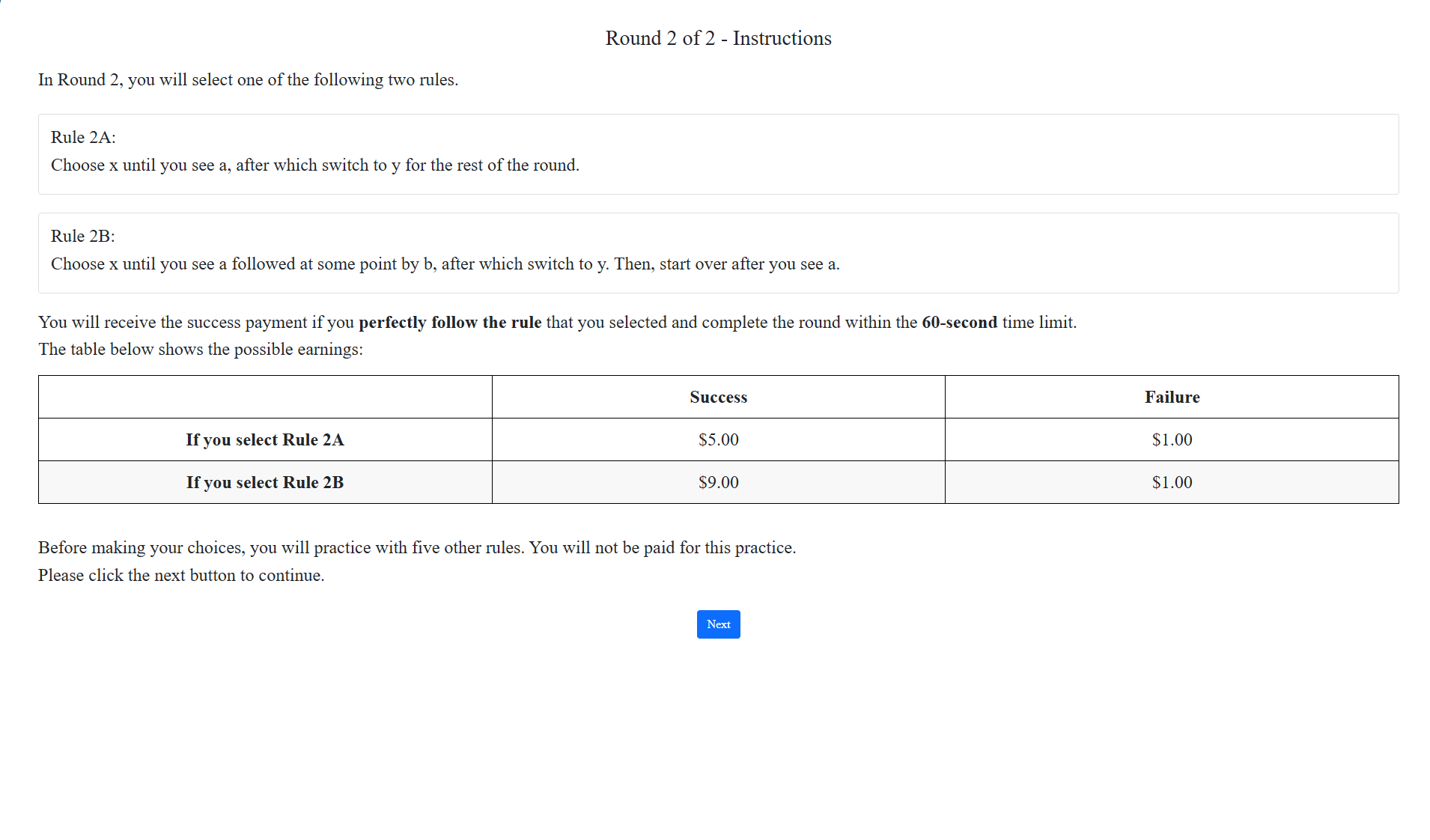}
    }
\end{textblock*}

\clearpage

\section*{}

\begin{textblock*}{\textwidth}(0.8cm,-0.1cm)
 \rotatebox{90}{
    \includegraphics[width=1.85\textwidth]{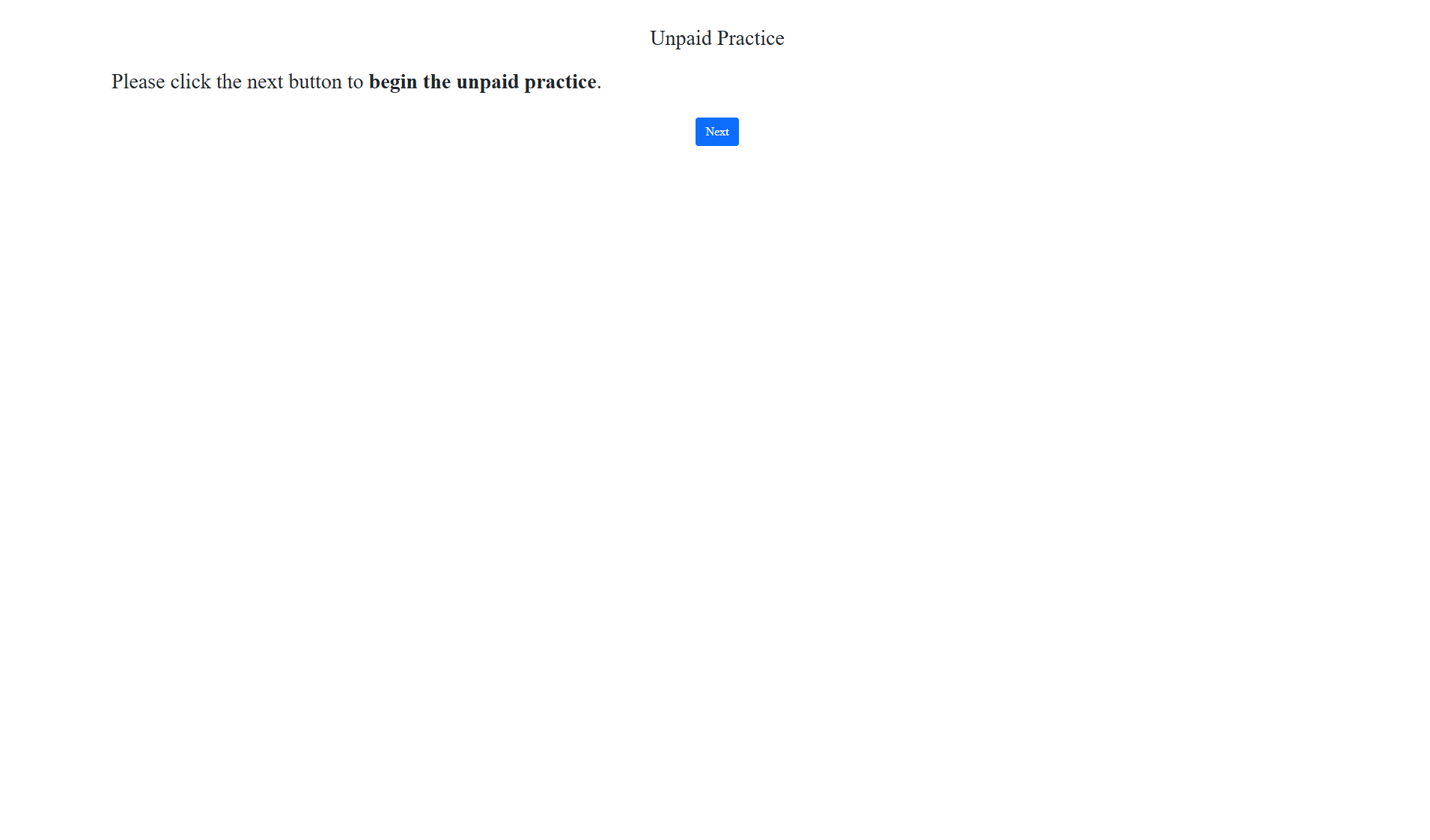}
    }
\end{textblock*}

\clearpage

\section*{}

\begin{textblock*}{\textwidth}(0.8cm,-0.1cm)
 \rotatebox{90}{
    \includegraphics[width=1.85\textwidth]{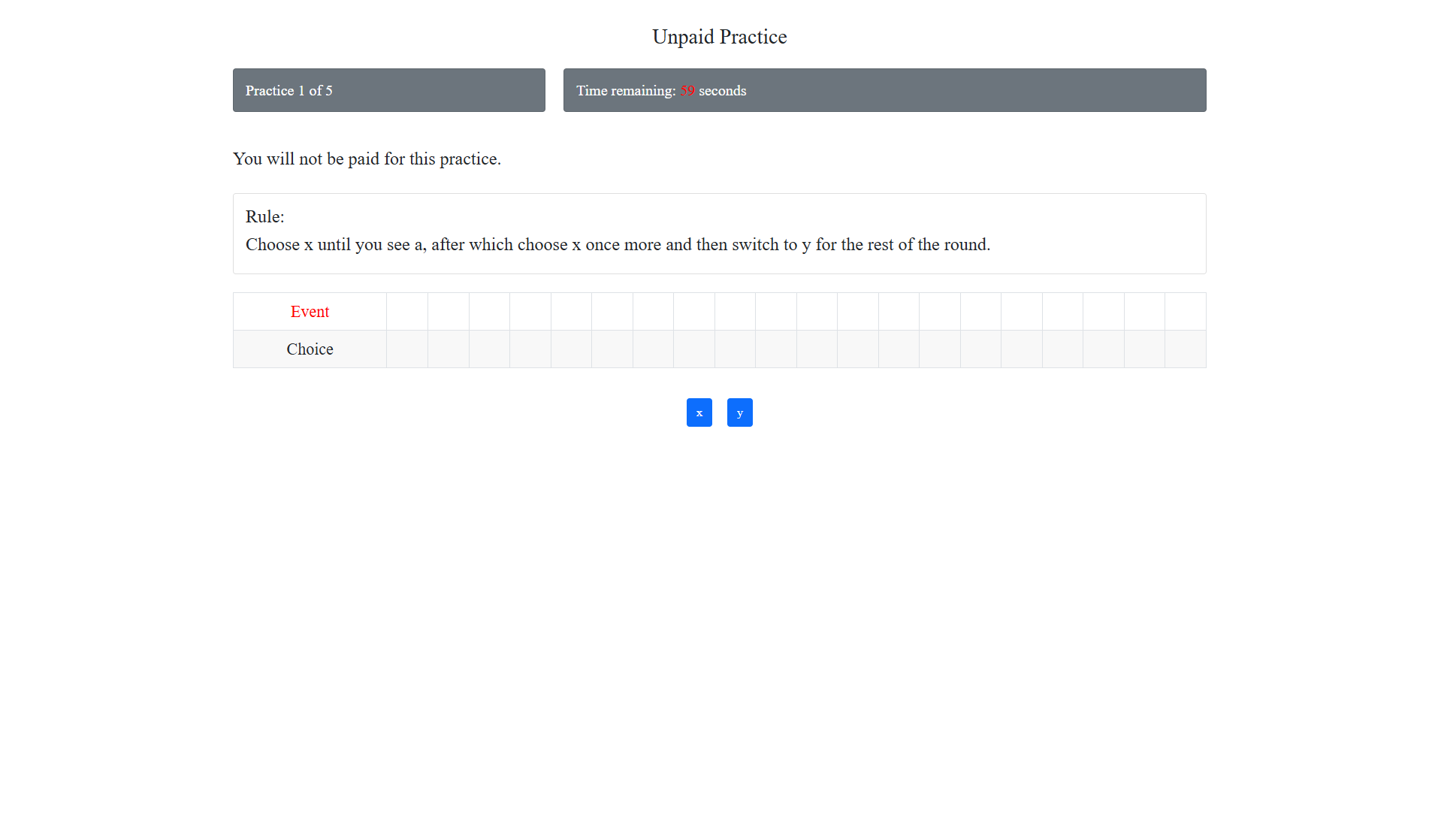}
    }
\end{textblock*}

\clearpage

\section*{}

\begin{textblock*}{\textwidth}(0.8cm,-0.1cm)
 \rotatebox{90}{
    \includegraphics[width=1.85\textwidth]{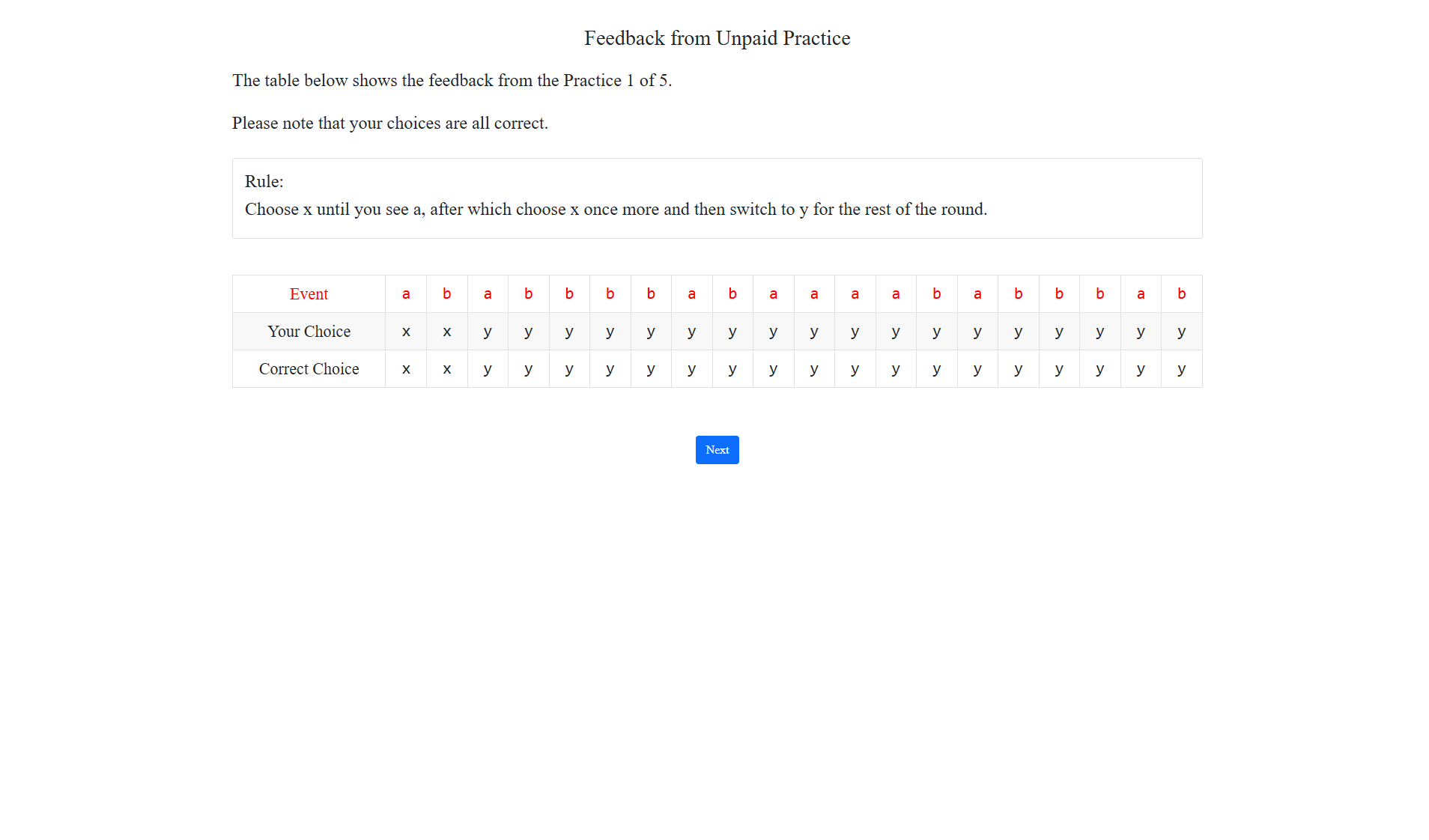}
    }
\end{textblock*}

\clearpage

\section*{}

\begin{textblock*}{\textwidth}(0.8cm,-0.1cm)
 \rotatebox{90}{
    \includegraphics[width=1.85\textwidth]{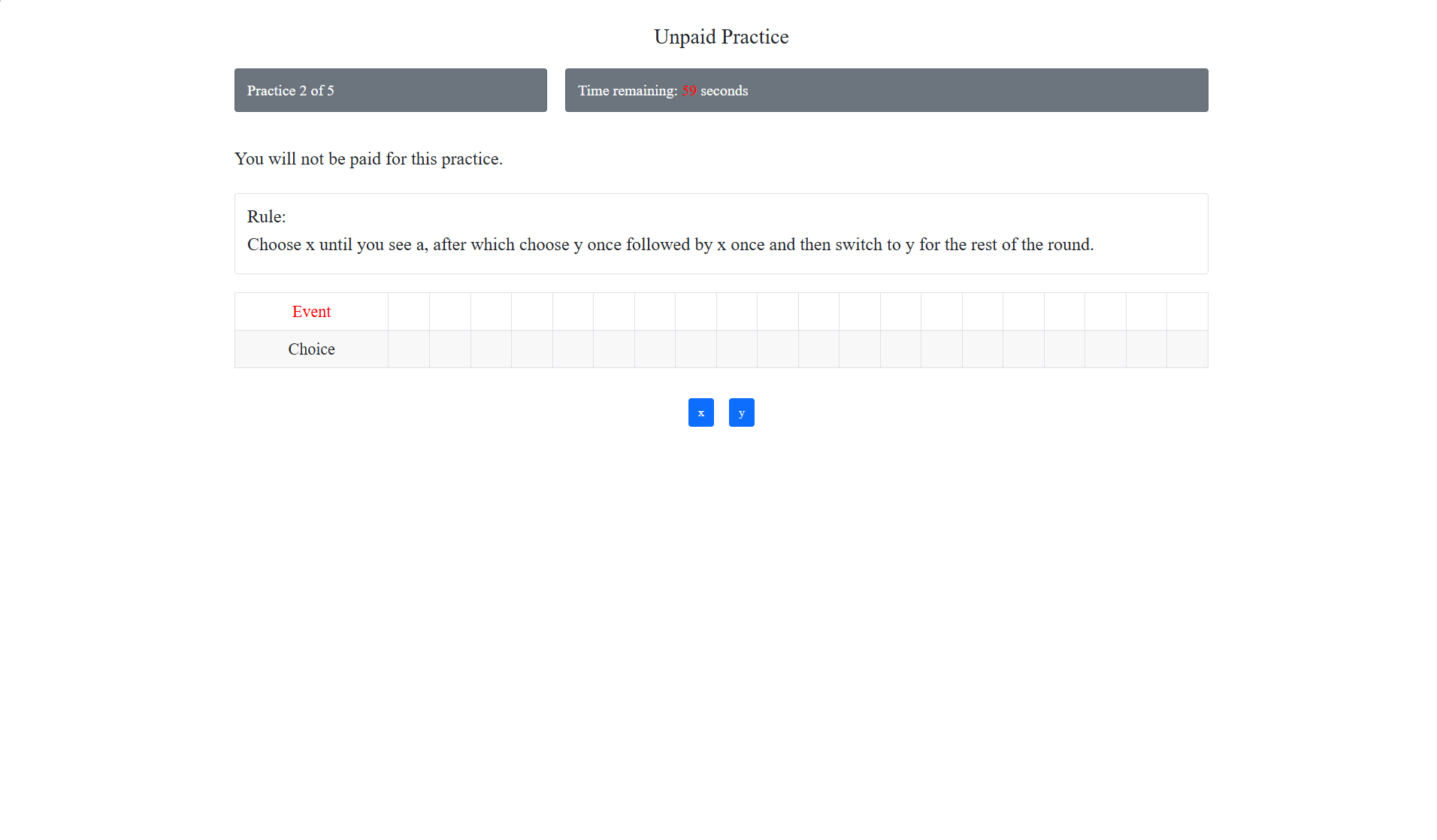}
    }
\end{textblock*}

\clearpage

\section*{}

\begin{textblock*}{\textwidth}(0.8cm,-0.1cm)
 \rotatebox{90}{
    \includegraphics[width=1.85\textwidth]{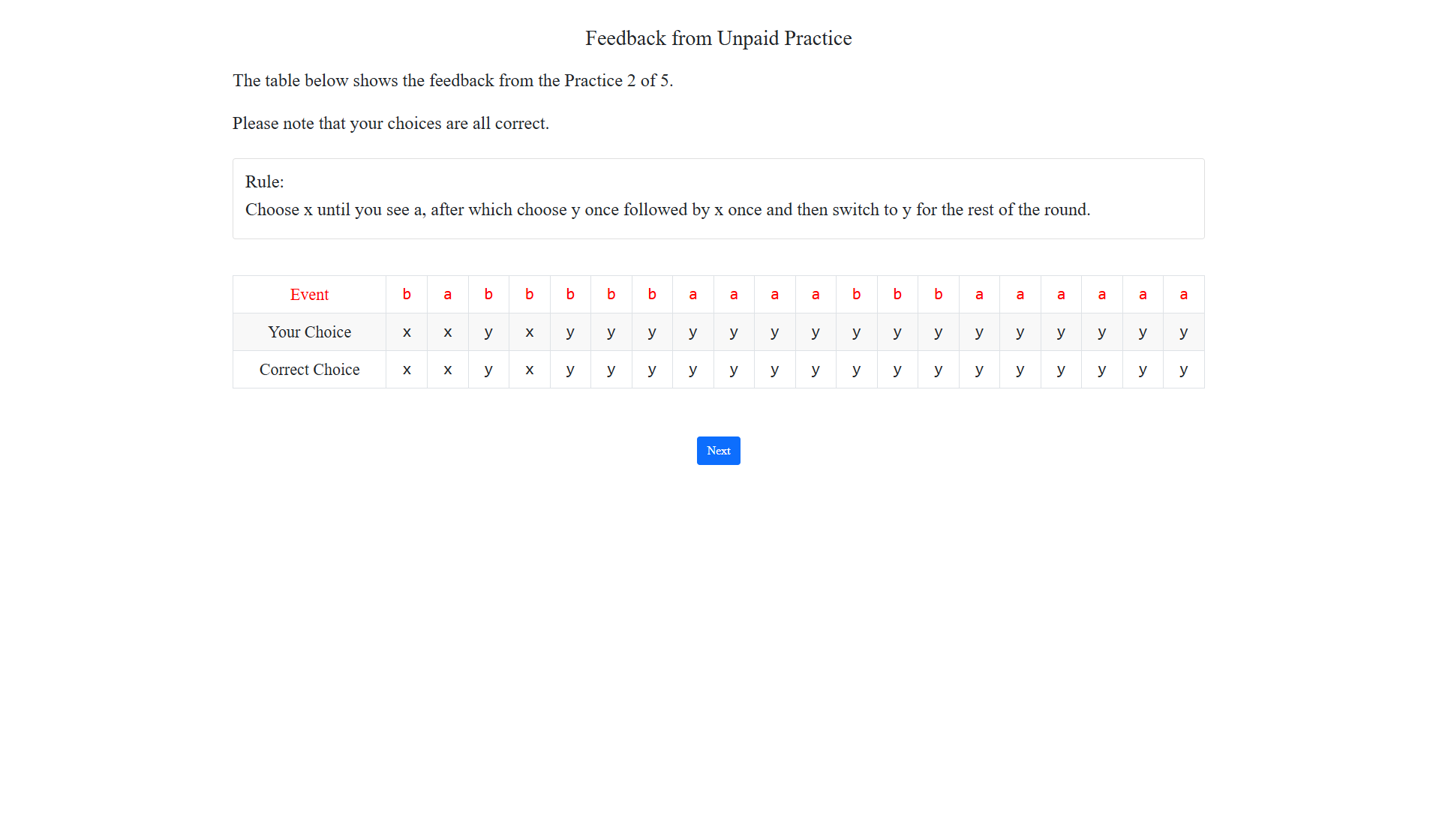}
    }
\end{textblock*}

\clearpage

\section*{}

\begin{textblock*}{\textwidth}(0.8cm,-0.1cm)
 \rotatebox{90}{
    \includegraphics[width=1.85\textwidth]{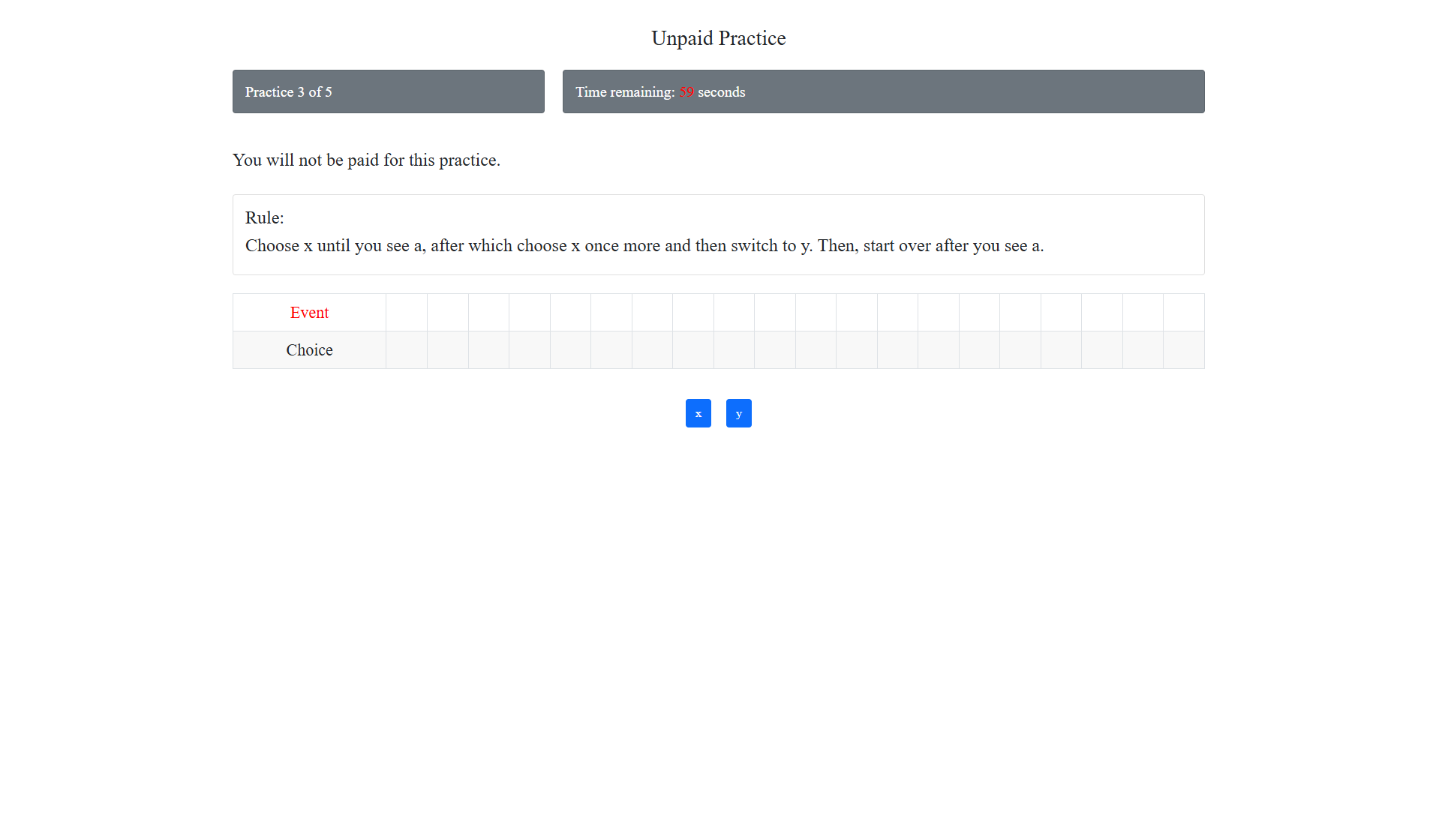}
    }
\end{textblock*}

\clearpage

\section*{}

\begin{textblock*}{\textwidth}(0.8cm,-0.1cm)
 \rotatebox{90}{
    \includegraphics[width=1.85\textwidth]{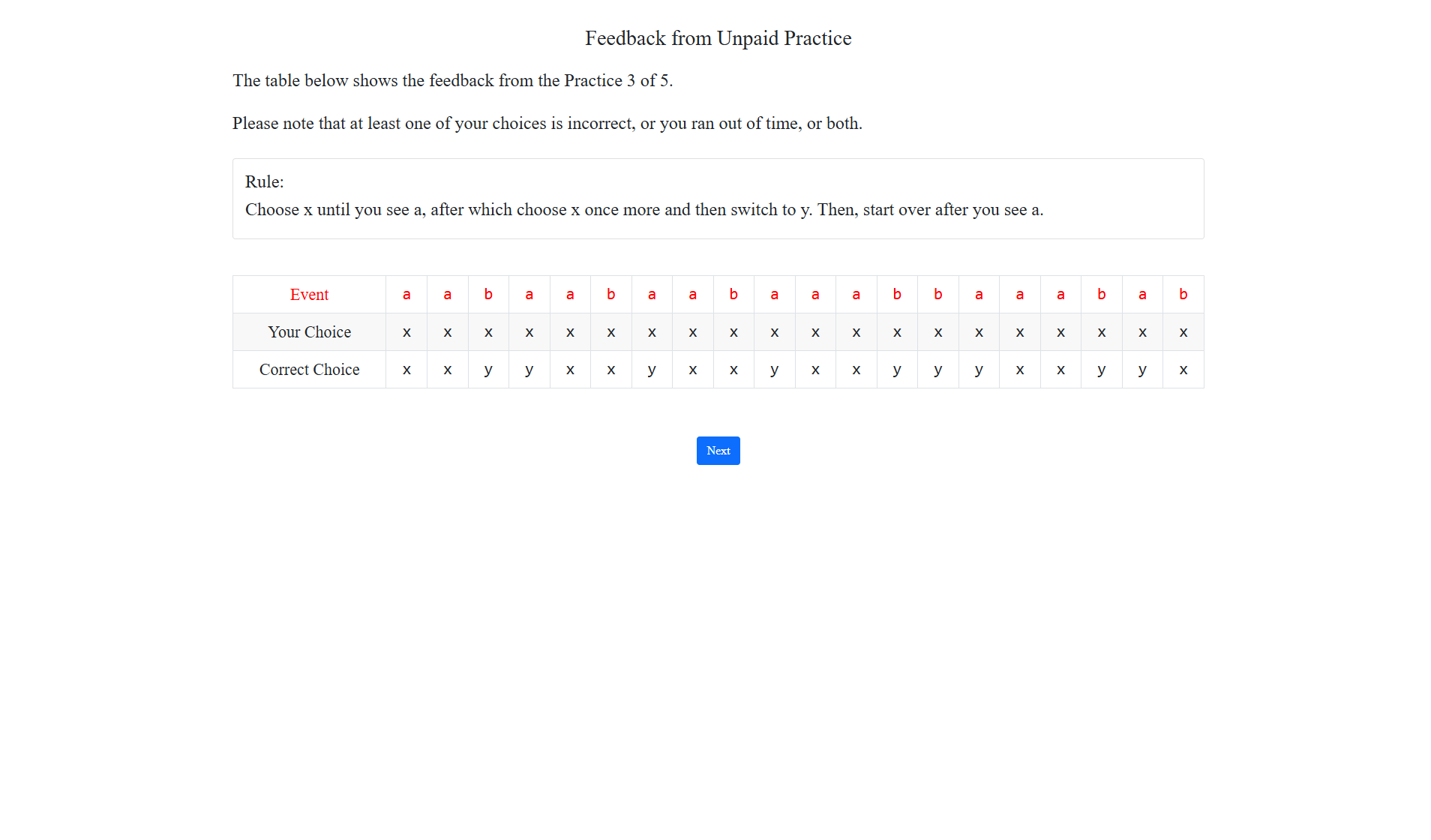}
    }
\end{textblock*}

\clearpage

\section*{}

\begin{textblock*}{\textwidth}(0.8cm,-0.1cm)
 \rotatebox{90}{
    \includegraphics[width=1.85\textwidth]{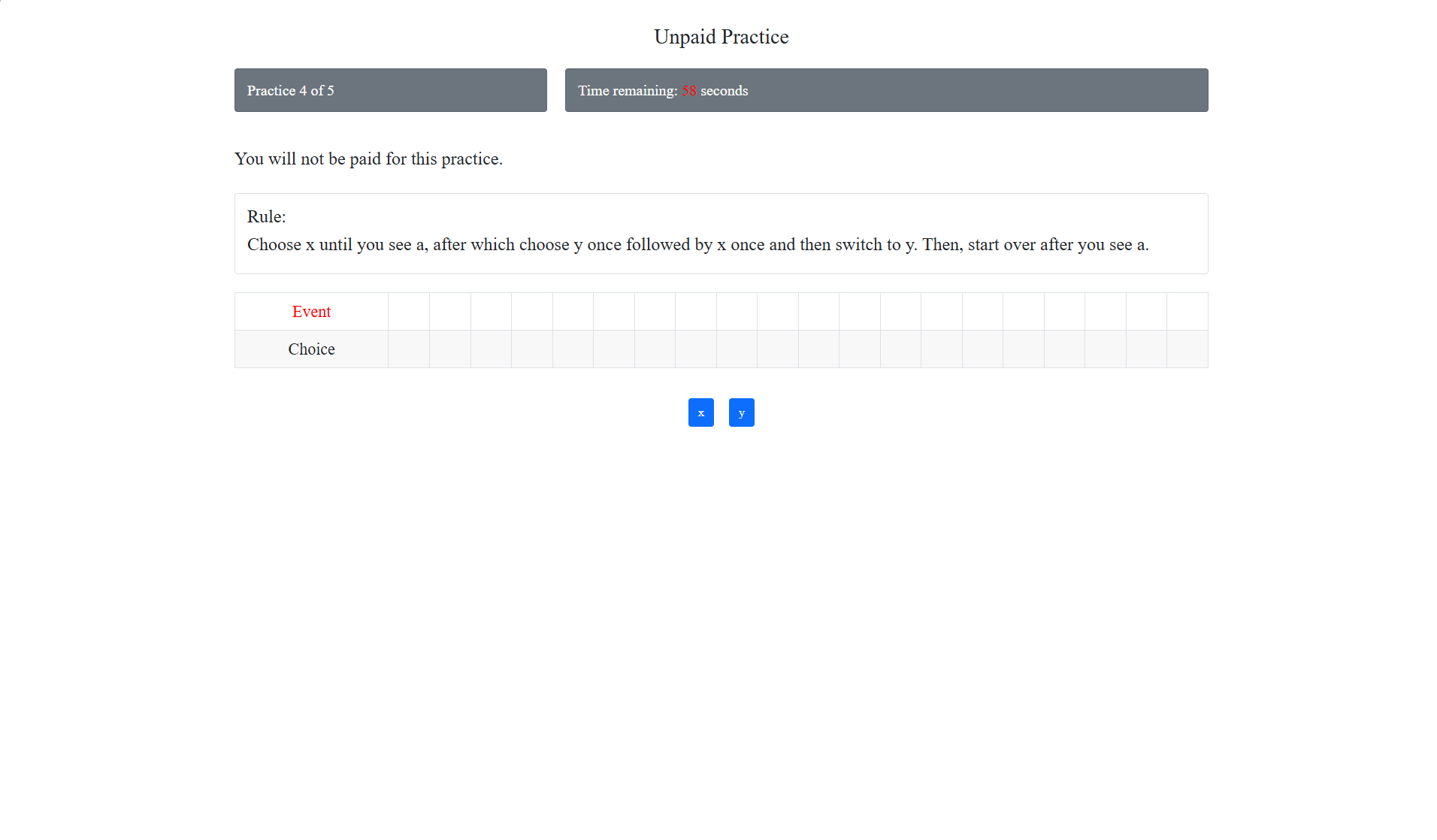}
    }
\end{textblock*}

\clearpage

\section*{}

\begin{textblock*}{\textwidth}(0.8cm,-0.1cm)
 \rotatebox{90}{
    \includegraphics[width=1.85\textwidth]{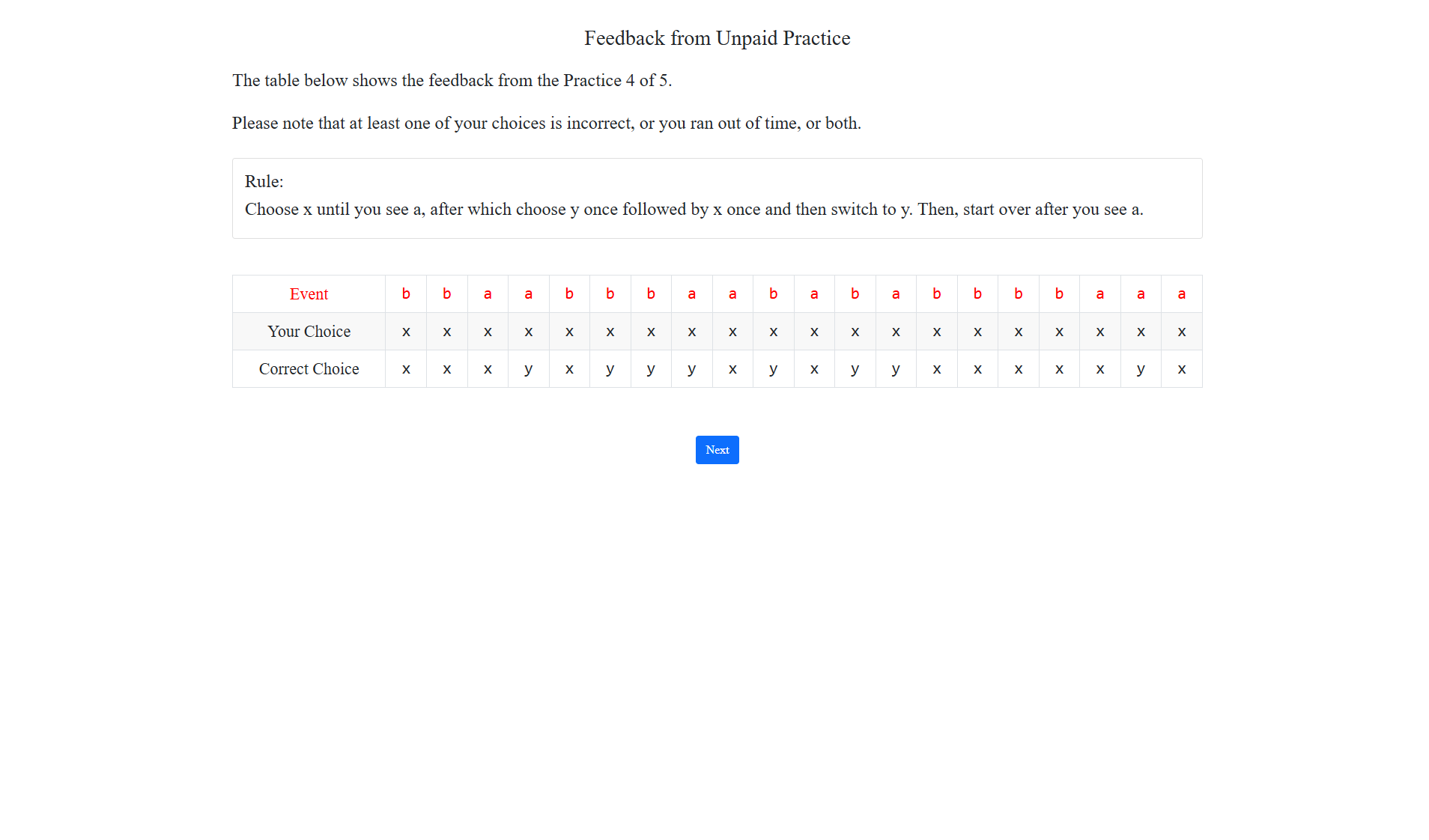}
    }
\end{textblock*}

\clearpage

\section*{}

\begin{textblock*}{\textwidth}(0.8cm,-0.1cm)
 \rotatebox{90}{
    \includegraphics[width=1.85\textwidth]{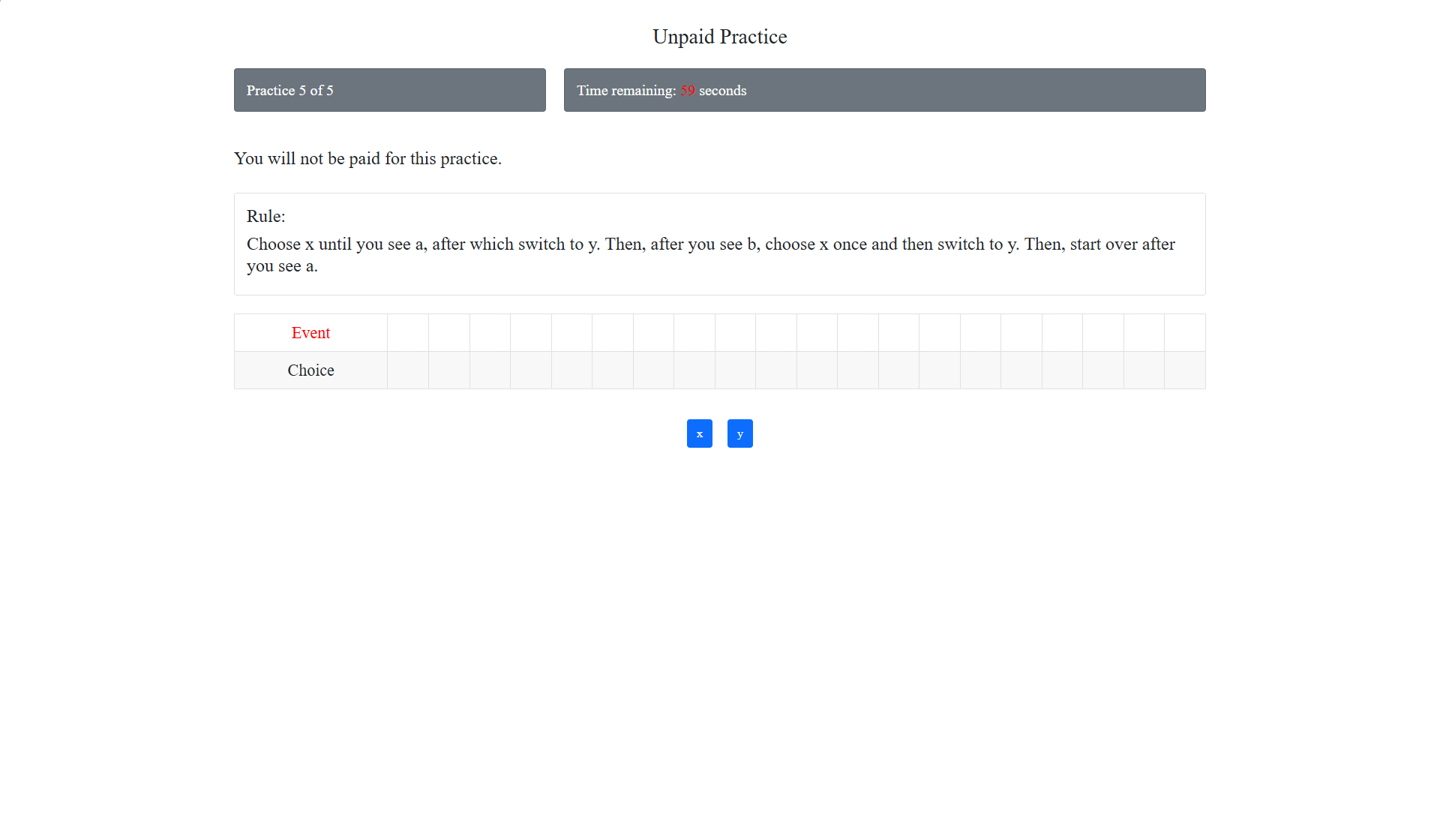}
    }
\end{textblock*}

\clearpage

\section*{}

\begin{textblock*}{\textwidth}(0.8cm,-0.1cm)
 \rotatebox{90}{
    \includegraphics[width=1.85\textwidth]{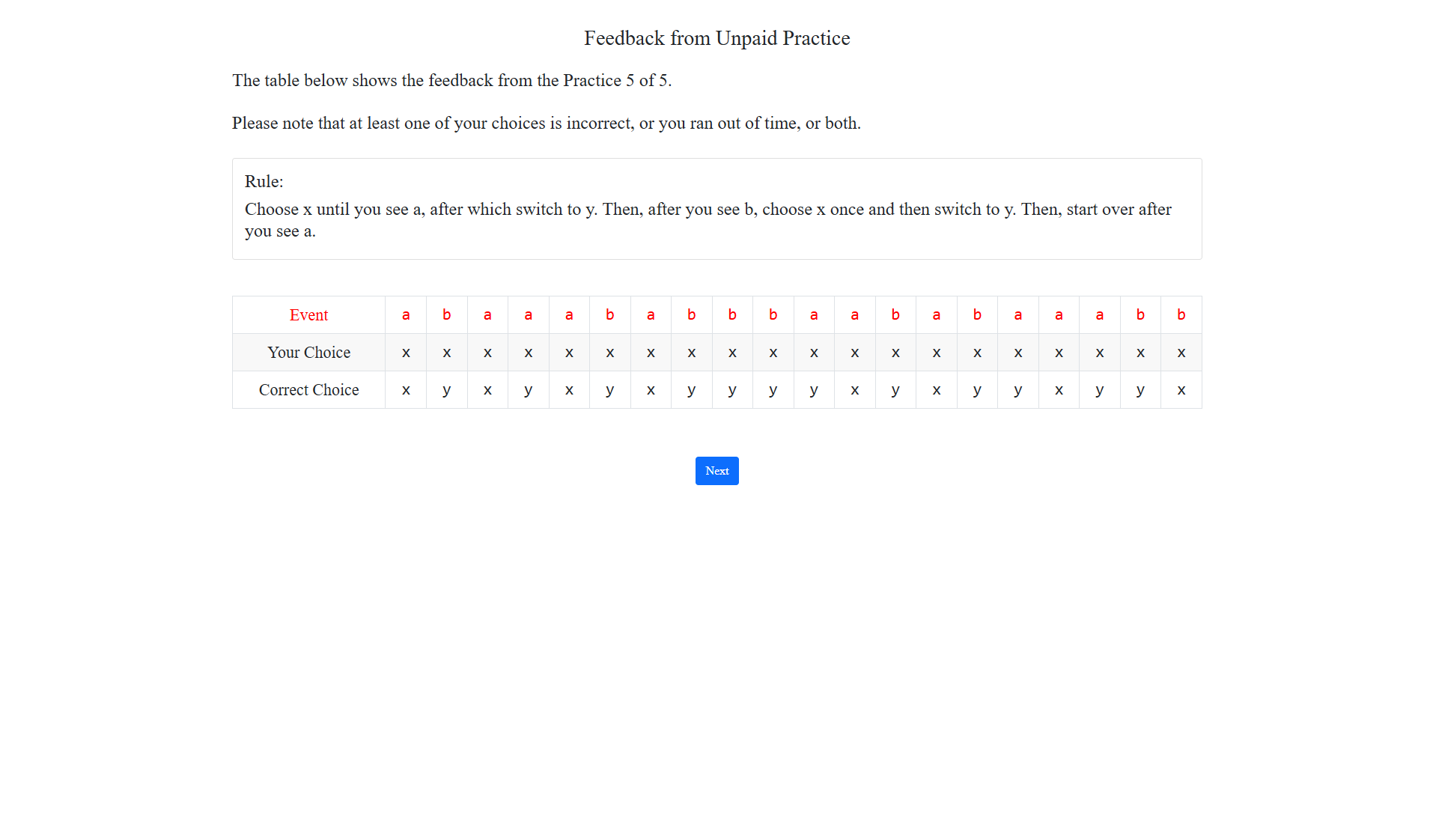}
    }
\end{textblock*}

\clearpage

\section*{}

\begin{textblock*}{\textwidth}(0.8cm,-0.1cm)
 \rotatebox{90}{
    \includegraphics[width=1.85\textwidth]{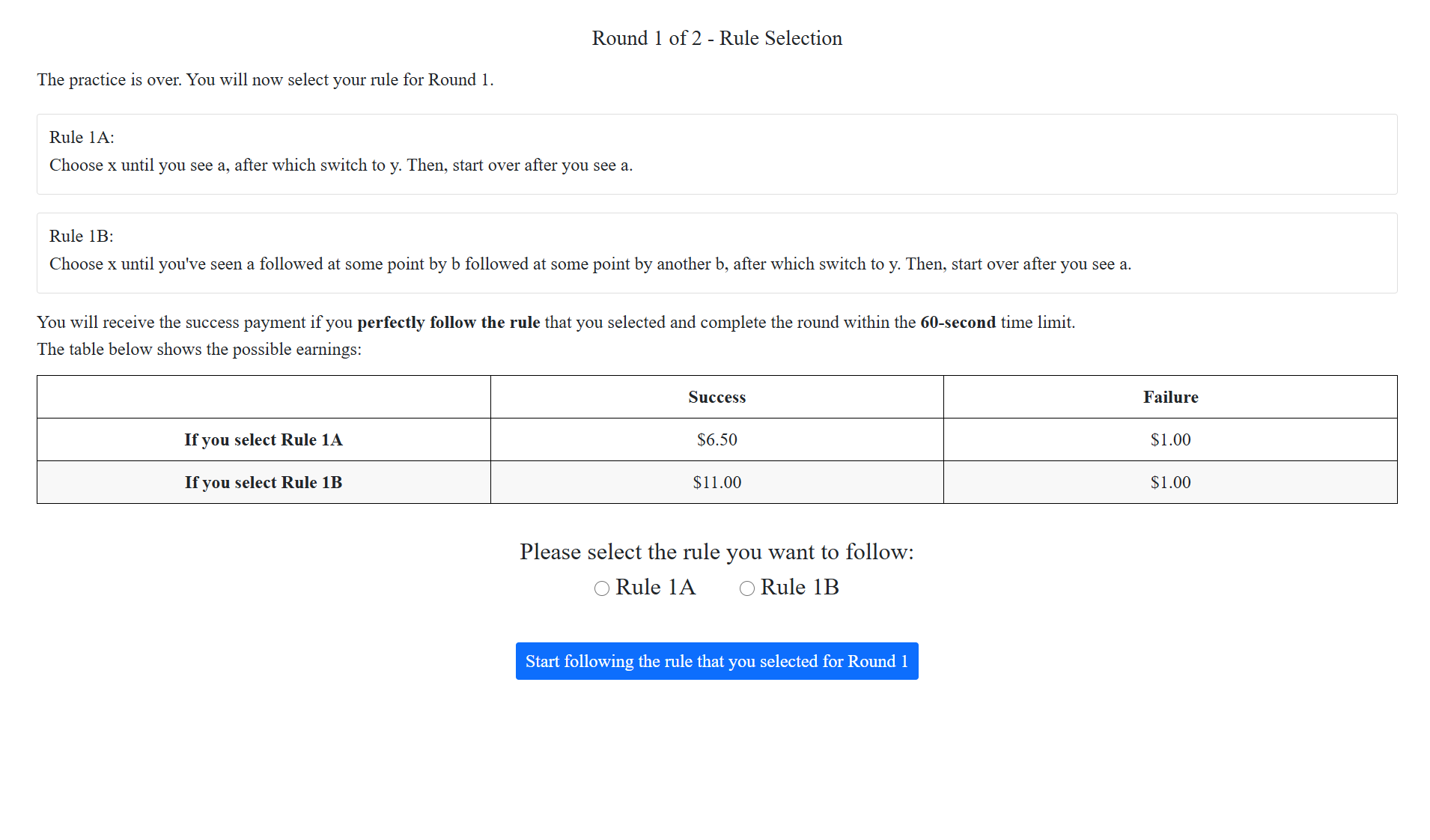}
    }
\end{textblock*}

\clearpage

\section*{}

\begin{textblock*}{\textwidth}(0.8cm,-0.1cm)
 \rotatebox{90}{
    \includegraphics[width=1.85\textwidth]{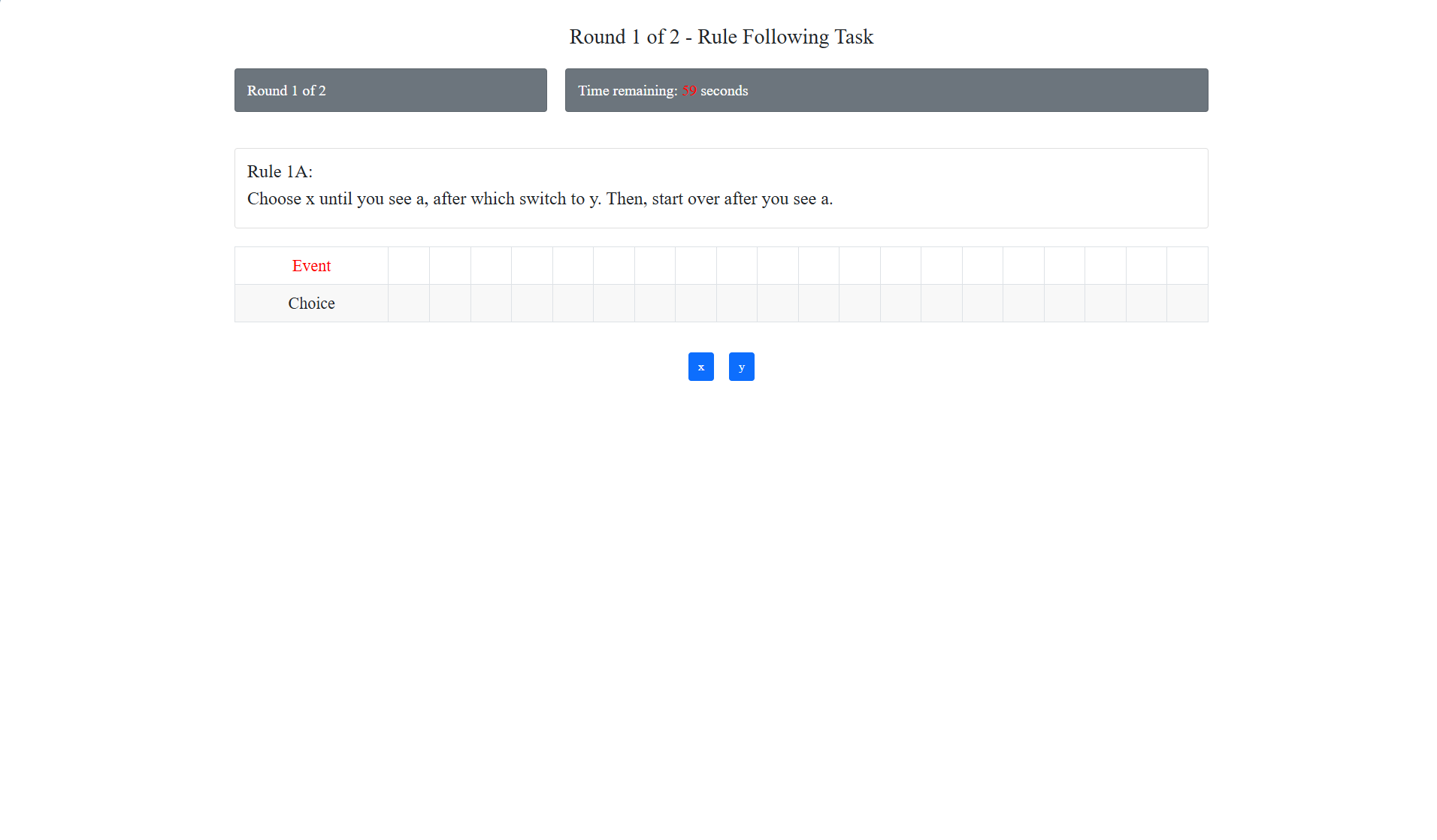}
    }
\end{textblock*}

\clearpage

\section*{}

\begin{textblock*}{\textwidth}(0.8cm,-0.1cm)
 \rotatebox{90}{
    \includegraphics[width=1.85\textwidth]{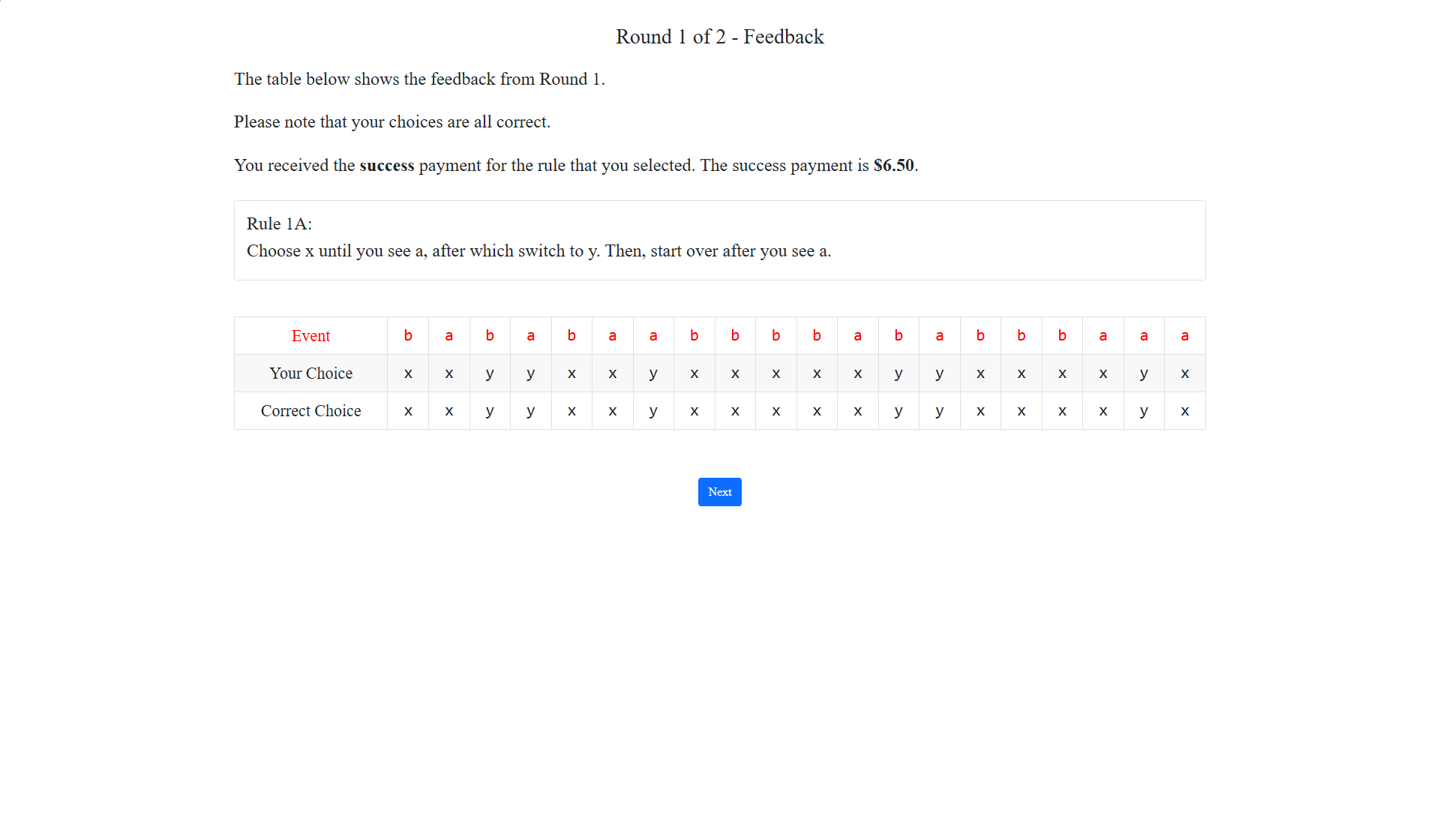}
    }
\end{textblock*}

\clearpage

\section*{}

\begin{textblock*}{\textwidth}(0.8cm,-0.1cm)
 \rotatebox{90}{
    \includegraphics[width=1.85\textwidth]{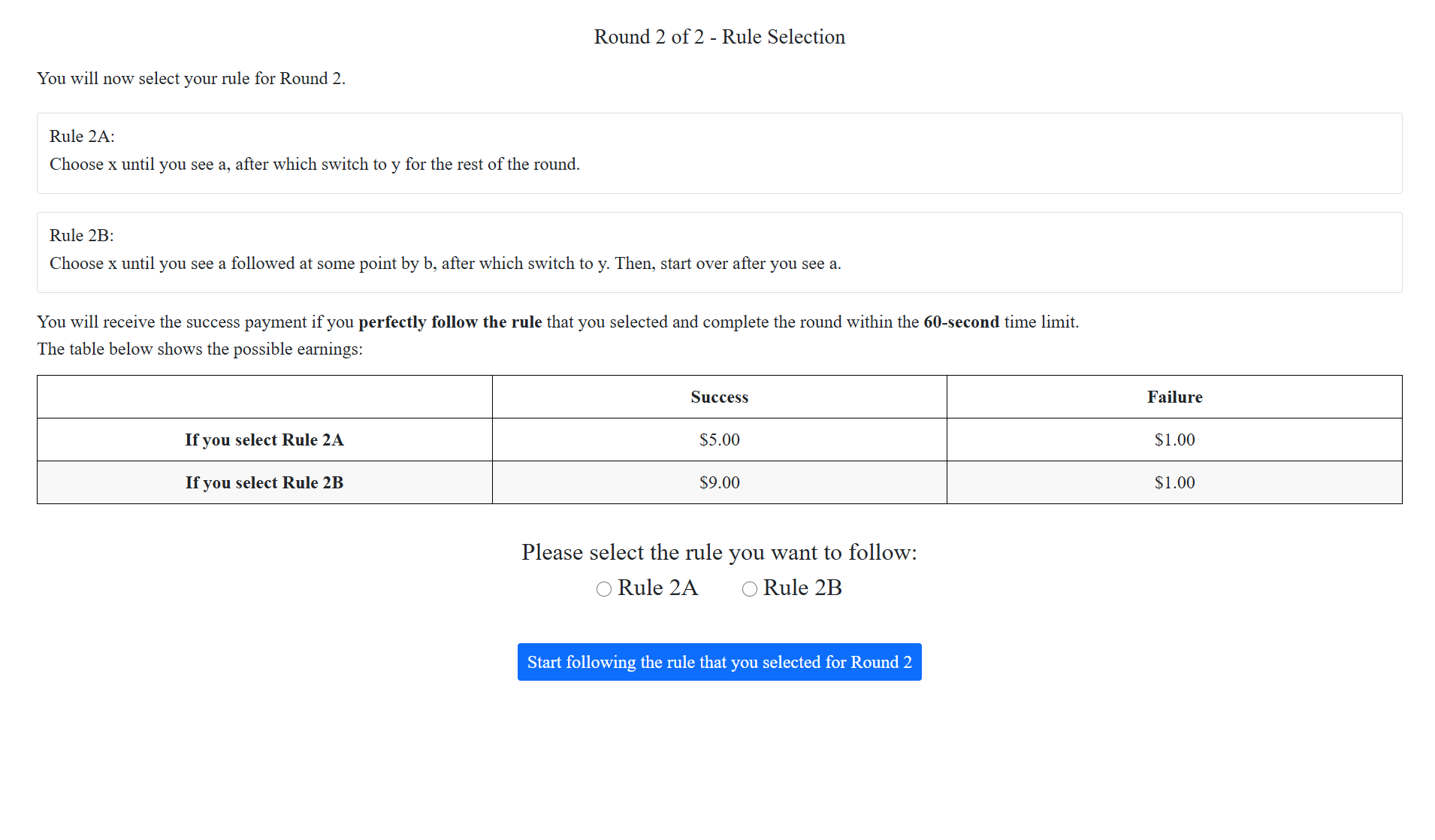}
    }
\end{textblock*}

\clearpage

\section*{}

\begin{textblock*}{\textwidth}(0.8cm,-0.1cm)
 \rotatebox{90}{
    \includegraphics[width=1.85\textwidth]{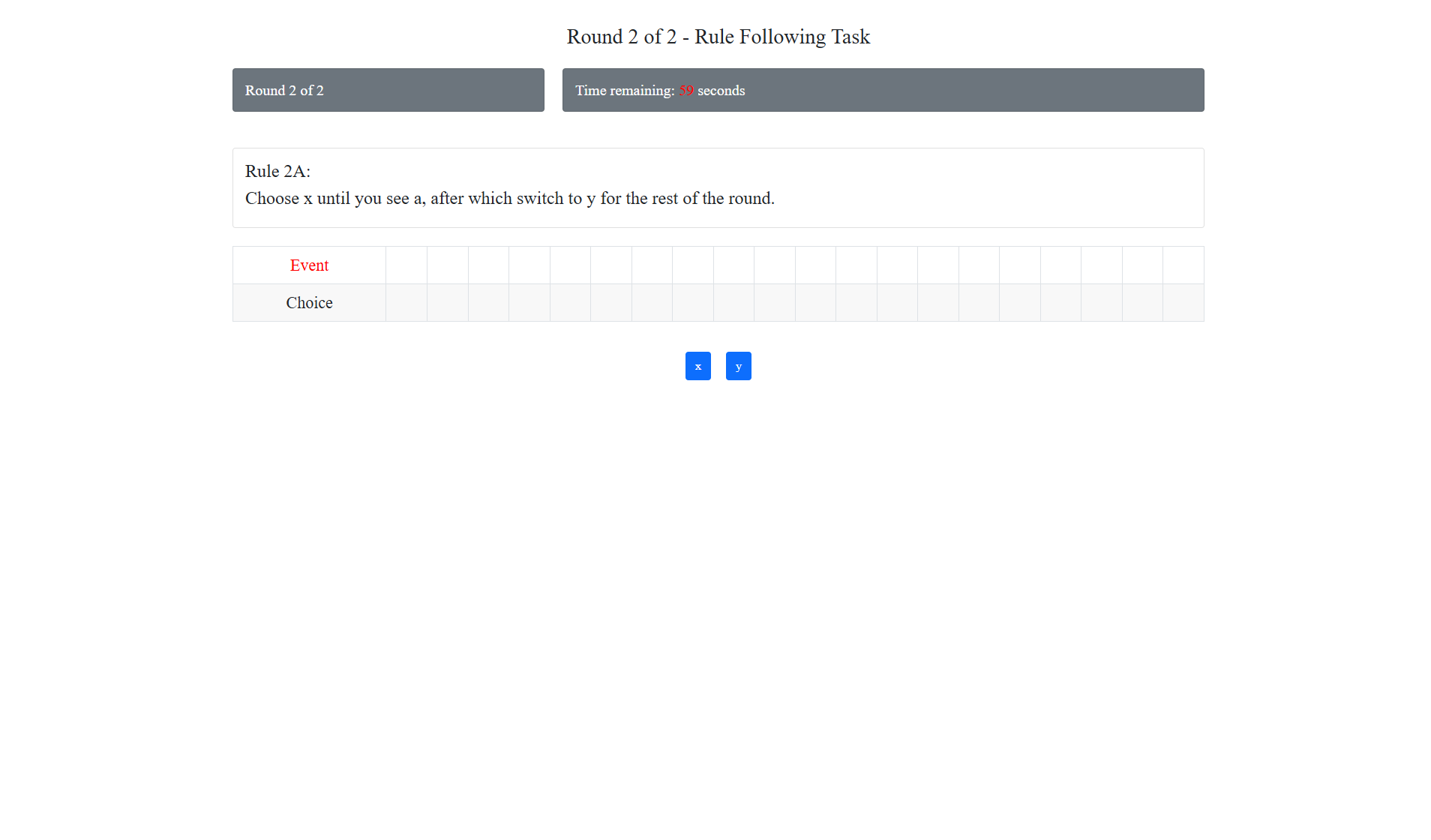}
    }
\end{textblock*}

\clearpage

\section*{}

\begin{textblock*}{\textwidth}(0.8cm,-0.1cm)
 \rotatebox{90}{
    \includegraphics[width=1.85\textwidth]{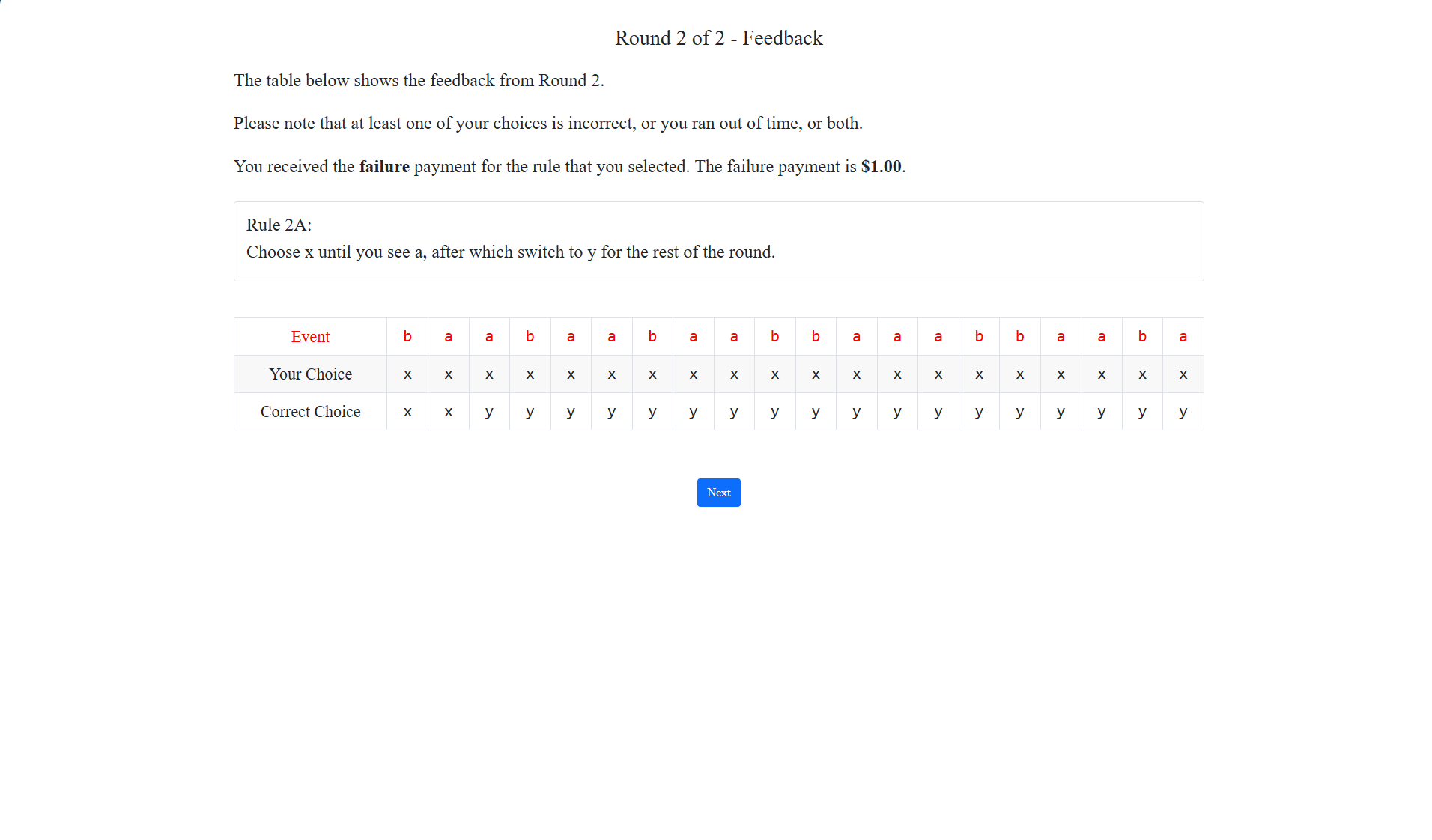}
    }
\end{textblock*}

\clearpage

\section*{}

\begin{textblock*}{\textwidth}(0.8cm,-0.1cm)
 \rotatebox{90}{
    \includegraphics[width=1.85\textwidth]{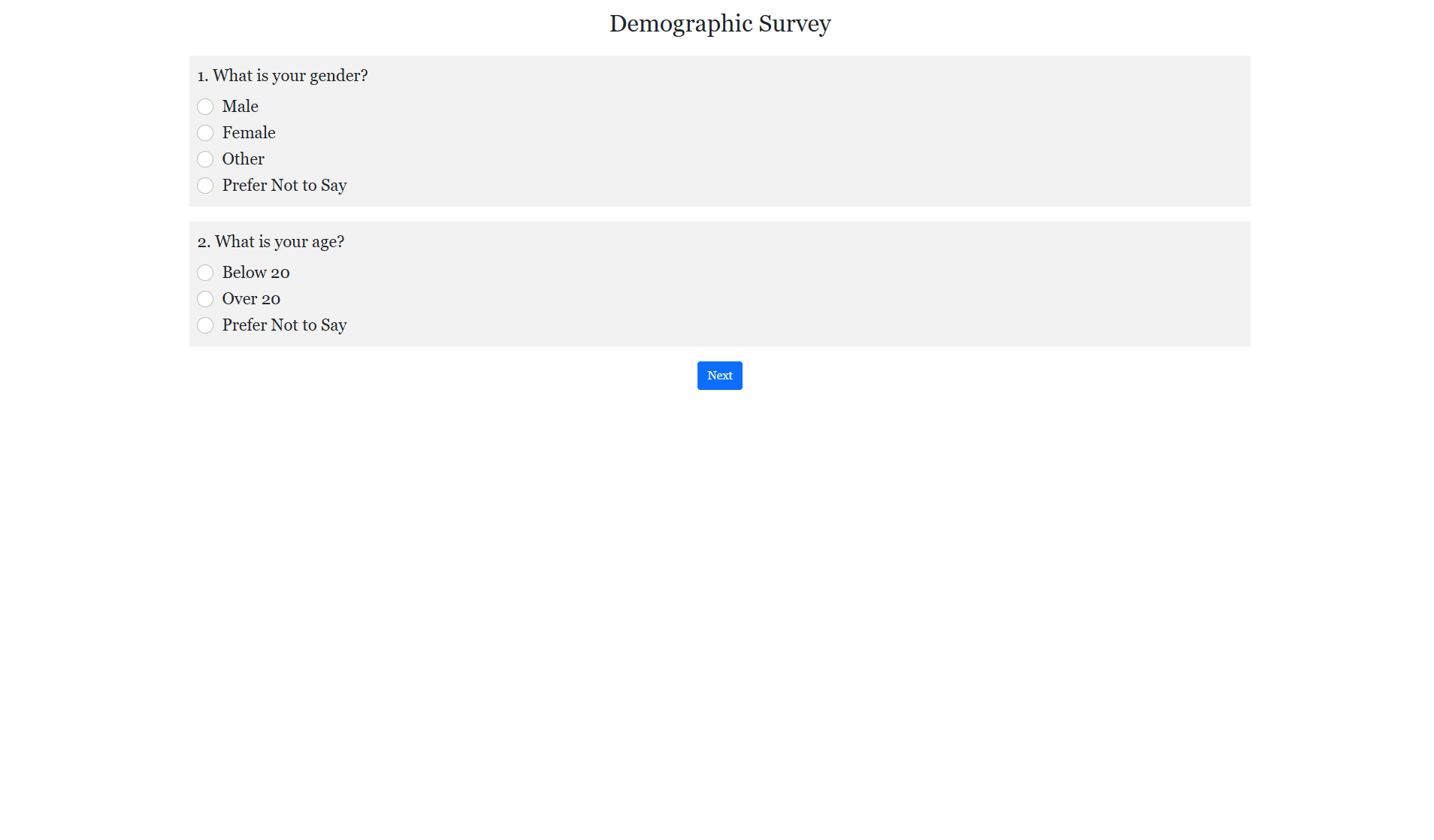}
    }
\end{textblock*}

\clearpage

\section*{}

\begin{textblock*}{\textwidth}(0.8cm,-0.1cm)
 \rotatebox{90}{
    \includegraphics[width=1.85\textwidth]{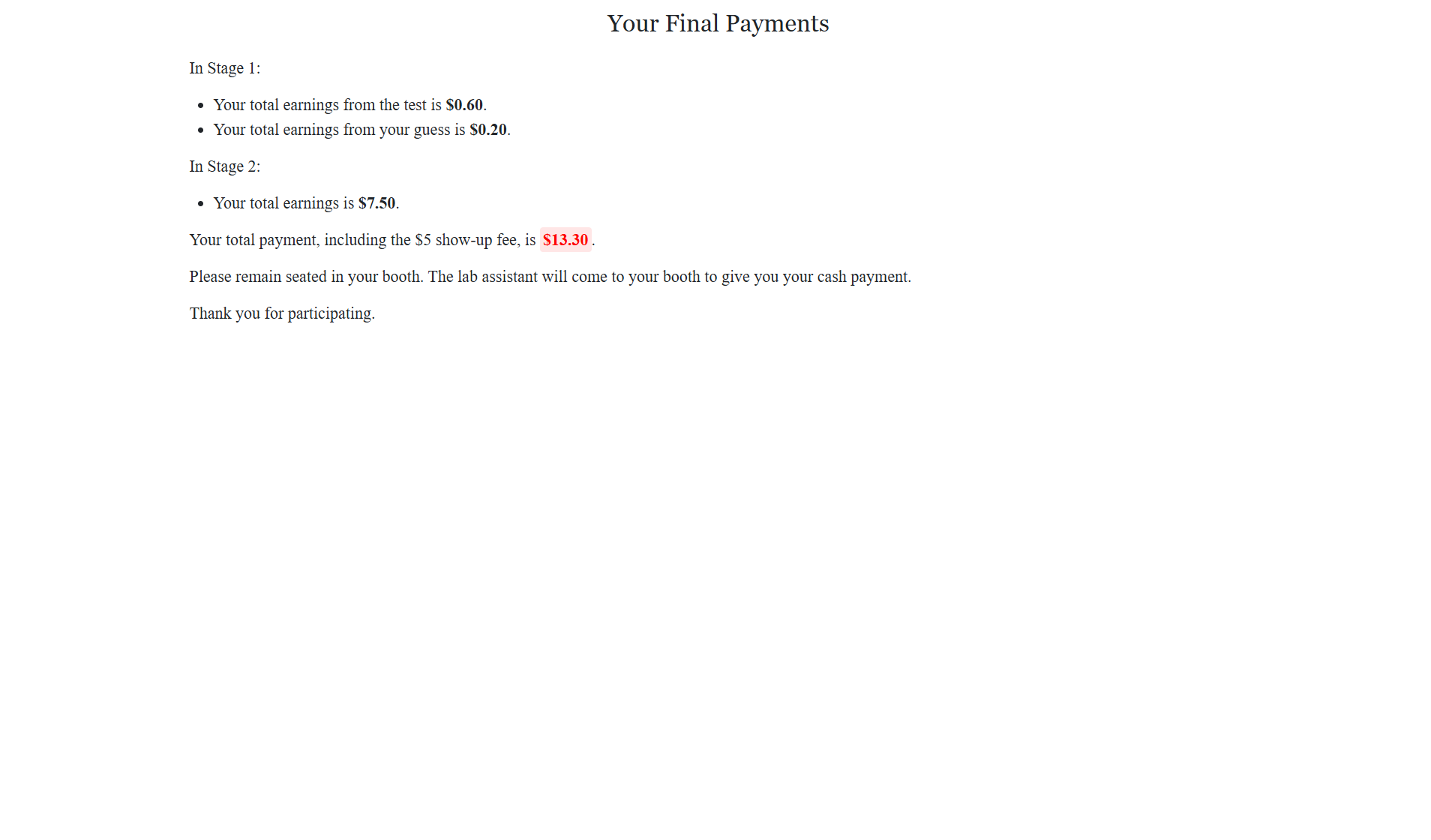}
    }
\end{textblock*}

\clearpage

\end{landscape}

\end{document}